# Parametric wave spectrum models for prediction of Stokes drift profiles

W. Erick Rogers, Yalin Fan

Naval Research Laboratory, Stennis Space Center, MS, USA

Luis Zamudio

Florida State University, Tallahassee, FL, USA

Elizabeth M. Douglass

Naval Research Laboratory, Stennis Space Center, MS, USA

Corresponding author: W. Erick Rogers (w.e.rogers.civ@us.navy.mil)

# CONTENTS

**LIST OF FIGURES**

**Executive Summary**

In this report, 15 variants of nine parametric models for non-directional ocean wave spectra are reviewed. They are compared against observation-based spectra created by taking spectra measured by moored buoys, binning by wind speed, and averaging. Surface Stokes drift (SSD) and Stokes drift profiles are computed using the model and observation-based spectra, and these are compared. This ‘parametric wave model’ strategy for estimating Stokes drift is contrasted with other methods for estimating Stokes drift, such as using spectra from a dynamical wave model. Advantages and shortcomings are reviewed.

One practical application for Stokes drift profiles is in ocean circulation modeling. Here, the information is used to compute wave-enhanced mixing (‘Langmuir turbulence’). We present examples of such cases, in the context of a global ocean model, tested in two modes (coupled and uncoupled) and with two types of forcing (Stokes drift approximated using wind speed and Stokes drift computed using a full ocean wave model). A more comprehensive review of the uncoupled tests with the full ocean wave model can be found in a separate report, Fan et al. (2026).

We evaluate simple Stokes drift models that use wind speed and/or wind stress but do not require a dynamical wave model. We find some are sufficiently accurate for routine prediction of SSD. A subset of these models can also predict Stokes drift profiles (via parametric spectra), but none are accurate in this role.

The SSD values computed from a very large drifting buoy dataset are compared with the total drift calculated using positional change of the same buoys. The SSD is roughly 25% of the total drift, consistent with expectations based on “rules of thumb”. However, the scatter is large, indicating that such rules have predictive skill only in the mean sense.

# 1. Introduction

Building on the theoretical framework for Langmuir roll cells proposed by Craik and Leibovich (1976), it has long been recognized—and subsequently analyzed as 'Langmuir turbulence' by McWilliams et al. (1997)—that the vertical shear in the wave-induced Stokes drift interacts with the vertical shear of the wind-driven Eulerian current to enhance mixing. This mechanism was later parameterized for practical ocean models by researchers such as Kantha and Clayson (2004).

By including this mechanism, it is hoped that the skill of ocean models can be improved; for example, by correcting tendencies to underpredict the mixed layer depth (MLD) and overpredict summer sea surface temperatures (SST) when/where waves are relatively energetic (Belcher et al. 2012; Li et al. 2016).

## 1.1. Motivation

Stokes drift was introduced in an ocean circulation model during a separate, prior effort. The original purpose of this study was to evaluate methods used in that study. There, the Stokes drift profile was approximated using only wind speed (Section 6.1). The wind speed is used to compute a one-dimensional wave spectrum, $E(f)$, which is then used to compute a Stokes drift profile, $U(z)$, which is then used in routines in the ocean model to compute wave-enhanced mixing.

Finding shortcomings with the details of this approach (Section 6.3), we expanded our scope to consider other, comparable methods for computing the Stokes drift profile (Section 5) based on many parametric forms of $E(f)$ (Section 4). All these parametric models use wind speed, but three additionally require information about wave age, i.e., a quantification of the stage of development of a wind sea.

## 1.2. Contents

This report is organized as follows. In Section 2, we review and contrast many potential strategies for introducing Stokes drift in an ocean circulation model. In Section 3, we introduce an observational dataset that is employed in the next two sections of the report. In Section 4, we present 15 variants of nine parametric models for non-directional wave spectra, $E(f)$, and compare against the observational dataset at two different wind speeds. We also compare methods for estimating friction velocity used in the parametric models (Section 4.3), methods for computing frequency at which the spectral slope increases in the high-frequency face of the spectrum (i.e., the transition from equilibrium range to saturation range) (Section 4.4), and methods for estimating the low-frequency limit of the equilibrium range (Section 4.5).

In Section 5, we compare Stokes drift from $E(f)$ given by the parametric models against Stokes drift computed from the observation-based $E(f)$. We also discuss the impact of directionality on Stokes drift, methods for computation of Stokes drift from buoy and WW3 spectra and investigate the sensitivity to the high-frequency limits on the integration used to calculate Stokes drift. In Section 6, we present applications from a global circulation model. This includes results with and without wave-enhanced mixing, and results in both coupled and uncoupled mode of operation. In Section 7, we show results from a wave model run within a coupled regional modeling system, and compare Stokes drift calculated by that numerical model against Stokes drift computed using simple parametric methods. In Section 8, we use wave spectra from drifting buoys to compare surface Stokes drift (SSD) computed from the buoy spectra, accounting for directional spread, against simple estimates of SSD. The observation-based SSD is also compared with the drift computed from buoy position (i.e., total drift) to quantify the contribution of the former to the latter.

## 2. Strategies for including Stokes drift in an ocean wave model

Here we introduce a taxonomy for methods that one might use to provide Stokes drift profiles to an ocean wave model. While considering these criteria of taxonomy, we discuss the advantages and disadvantages of different methods.

Criterion #1. Is it based on a monochromatic wave, a wave spectrum, a "Stokes spectrum", or some other approach?

In classical oceanography, the monochromatic wave approximation is used to describe depth-dependent Stokes drift velocity as:

$$U(z) = U(z=0)e^{2k_r z}$$

where $U(z=0)$ is the surface drift and $k_r$ is a representative wavenumber.

The representative wavenumber can be taken from a wave model, say as the mean wavenumber. Alternatively, to force consistency with bulk properties, some models, such as Skyllingstad and Denbo (1995), back-calculate an effective, representative wavenumber $k_r$ using the surface velocity $v_0$ and the depth-integrated transport $V_s$: $k_r = U(z=0)/(2V_s)$
Using this back-calculated $k_r$ guarantees that the depth-integrated transport of the monochromatic profile matches the true physical transport.

While mathematically simple, this profile decays too slowly compared to observations and spectral models (Breivik et al. 2014).

Another approach is to compute or import a wave spectrum and use that to compute the Stokes drift profiles.

A third, intermediate, approach is to develop a parametric Stokes profile model and calibrate it to provide profiles that are more realistic than the monochromatic approach, but without requiring explicit definition of the wave spectrum. This is the approach used by Breivik et al. (2014).

A fourth approach is to import a "Stokes spectrum". This is an output option for the spectral ocean wave model WAVEWATCH III ("WW3", Tolman (1991), WW3DG (2018)) and is a description of the Stokes profiles in frequency space, which are converted to vertical space for use by the ocean model.  All information that is computed by the wave model relevant to Stokes drift is thereby passed to the ocean model. It is the most concise method for passing the "full" Stokes information: it is slightly more concise than passing the "1.5-dimensional" spectrum from WW3 which we introduce below. However, if there are other (non-Stokes) uses for the 1.5-dimensional spectra, that method might be preferred over the Stokes spectrum.

<u>Criterion #2</u>. If based on a wave spectrum, how does it describe that spectrum?

Some potential methods describe the non-directional wave spectrum only. Others provide the full two-dimensional wave spectra. Others provide a "1.5-dimensional spectrum" (1.5-D), defined as the energy level, mean direction, and directional spread at each frequency, $E(f)$, $\theta_m(f)$, and $\sigma_\theta(f)$. With the latter, all information that is computed by the wave model relevant to Stokes drift is passed to the ocean model.

<u>Criterion #3</u>. How does it account for the effect of the wave directionality on Stokes drift?

Directionality is relevant to Langmuir mixing: it affects how the Stokes drift interacts with the Eulerian current. In some cases, all information from the wave model related to Stokes drift is retained: this is true if the wave model provides the full 2-dimensional spectrum, the 1.5-dimensional spectrum, or the Stokes spectrum. In other cases, the directionality may be completely ignored, e.g. by setting all wave directions as "toward east". In other cases, the directionality may be described by taking a single direction to represent that time/location—say the mean wave direction, the peak wave direction[1], or the direction of Stokes transport, $V_s$—and applying it uniformly at all depths.

<u>Criterion #4</u>. How does it account for the directional spread on Stokes drift profiles?

Like with Criterion #3, in some cases, all information from the wave model related to Stokes drift is retained: this is true if the wave model provides the full 2-dimensional spectrum, the 1.5-dimensional spectrum, or the Stokes spectrum.

Other methods might use a "rule of thumb" to give a rough approximation of the impact of directional spread. Or the directional spread might be disregarded altogether. See Breivik et al. (2014), Webb and Fox-Kemper (2015) and Clarke and van Gorder (2018) for more discussion.

<u>Criterion #5</u>. How does it handle the high frequency tail of the wave spectrum?

In the case of a monochromatic spectrum, the answer is always "not at all".

[1] The definitions for mean wave direction and peak wave direction are not important here, but can be found in WW3DG (2018).

In the case of a wave spectrum or Stokes spectrum defined on discrete frequencies, the tail is truncated at the highest discrete frequency. In principle, it is possible to attach a high-frequency "diagnostic tail".

Criterion #6. Does it require information from an external wave model?

In the most simple cases, the strategy is to "get by" without any information from an external wave model. The ocean model already imports information about wind speed, so it is very simple (measured by lines of new code required) to rely on this exclusively and build a parametric wave spectrum from that. This has an obvious drawback: it omits all information about the stage of development of a wind sea (fetch, duration) and all information about swell.

Criterion #7. How many numbers are provided to the ocean model per geographic grid point?

This is largely determined by the prior criteria. Examples are given below.

Examples

Some examples of possible ways to describe Stokes Drift profiles in an ocean model are given in Table 1.

Table 1. Taxonomy of Stokes Drift Profile Implementations

| Approach | Physical Basis | Directionality | High-Frequency (HF) Tail | Requires Wave Model? | Input Parameters (per grid point) | Key Weaknesses / Assumptions |
|---|---|---|---|---|---|---|
| Wind-Based Parametric | Monochromatic (Implicitly) | Assumed aligned with wind | Ignored | No (Uses wind only) | 2 parameters: *u10* and *v10*. | Completely ignores swell, fetch, and wave development physics; highly idealized. |
| "Classic" Monochromatic | Monochromatic (Explicit) | Single direction (e.g., mean wave direction, MWD) | Ignored | Yes | 3 parameters: Surface drift magnitude (SSD), bulk wavenumber ($k_r$), and bulk direction ($\theta_r$). | Profile decays too quickly at depth; lacks realistic shear profile shape. |
| Skyllingstad & Denbo (1995) | Monochromatic (Transport-Matched) | Highly simplified: all Stokes drift is eastward. | Ignored | Yes | 2 parameters: Surface drift magnitude (SSD) and Stokes transport magnitude (\|Vs\|). | Deep-water profile shape remains physically incorrect; directionality is overly simplified (unidirectional). |
| Breivik et al. (2014) | Parametric (Exponential Integral) | Single direction: mean wave direction, MWD | Included in analytical integration of Phillips (1958) spectrum. | Yes | 3 parameters: Surface drift magnitude (SSD), Stokes transport magnitude (\|Vs\|), and Mean Wave Direction (MWD). | Profile is still an approximation; direction is kept uniform with depth (no direction shear). Invocation of Phillips (1958)†. |

| Approach | Physical Basis | Directionality | High-Frequency (HF) Tail | Requires Wave Model? | Input Parameters (per grid point) | Key Weaknesses / Assumptions |
|---|---|---|---|---|---|---|
| WAVEWATCH III "Stokes Spectrum" | Spectral | Full directionality computed per frequency bin | Not included but can be attached. | Yes | $nf \times 2$ parameters: Zonal (*us(f)*) and Meridional (*vs(f)*) drift components for each frequency bin. | Significant computing burden to pass and store. Requires reconstruction of vertical profiles in ocean model or a pre-processor. |
| WAVEWATCH III "1.5-dimensional" spectrum | Spectral | Full directionality computed per frequency bin | Not included but can be attached. | Yes | $nf \times 3$ parameters: energy level $E(f)$, mean direction $\theta_m(f)$, and directional spread $\sigma_\theta(f)$. | Significant computing burden to pass and store. |
| Full 2D Spectrum | Spectral | Full directionality computed per frequency and direction bin | Not included but can be attached. | Yes | $nf \times n\theta$ parameters: Wave energy spectrum $E(f,\theta)$ across all bins. | Highest data-passing and memory requirements; generally impractical for global coupled operational models. |

**Table Abbreviations & Notes:**
**SSD**: Magnitude of the surface Stokes drift velocity
**$|V_s|$**: Magnitude of the depth-integrated Stokes transport.
**$k_r$, $\theta_r$**: Representative wavenumber and wave direction.
**MWD**: Mean Wave Direction (standard bulk output from wave models): this is one possible definition of $\theta_r$.
***nf*, $n\theta$**: Number of discrete frequency and direction bins used in the spectral wave model
***u10, v10**:* Zonal and Meridional components of the 10-m wind speed $U_{10}$

***us(f), vs(f)***: Zonal and Meridional components of the Stokes drift, output by WW3 per frequency bin.

†: The mathematical foundation of the Breivik et al. (2014) exponential integral profile ($E_1$) is strictly derived from an idealized $f^{-5}$ Phillips spectrum integrated from $f_p$ to ∞. While these spectral assumptions are physically outdated—ignoring the widely accepted $f^{-4}$ equilibrium range (Section 4.4) and overestimating energy near the spectral peak—the final operational method is largely insulated from these errors. By dynamically tuning the depth-scale parameter $k_e$ using an estimate of surface Stokes drift and integrated transport, and integrated transport from a wave model, the formulation acts as an effective "stretch-to-fit" approximation that mimics realistic profile behavior despite its outdated spectral origins.

## 3. CDIP observational dataset for ocean wave spectra

In later sections, we compare parametric spectra against wave spectra derived from buoy measurements. The buoy data are provided by the Coastal Data Information Program (CDIP). CDIP is operated by the Ocean Engineering Research Group (OERG), part of the Integrative Oceanography Division (IOD) at Scripps Institution of Oceanography (SIO), which is affiliated with the University of California, San Diego. We use data from 13 buoys as described in Rogers and Thomson (2026) and Rogers et al. (2026), with up to 32 months duration per buoy, depending on outages, May 2022 to December 2024. Data were downloaded from the CDIP website. These 13 buoy positions are shown in Figure 1.

Rogers and Thomson (2026) and Rogers et al. (2026) also used drifting buoys, but those are not used in our primary analysis (Sections 4 and 5), since spectra from these buoys present an ambiguity with respect to the frame of reference: ideally, individual spectra from drifting buoys should be Doppler-shifted to either an intrinsic (relative to the medium that the waves travel in) or absolute (fixed relative to earth) frequency space (see Rogers et al. (2026)). The CDIP buoys are all moored and so all observed spectra are in the fixed frame of reference.

It is useful to organize the buoy data by local wind speed, but the CDIP buoys do not measure or report wind speed, so our 10-m wind speeds are taken from the ECMWF (European Centre for Medium-Range Weather Forecast) Reanalysis v5 (ERA5) reanalysis (Hersbach et al. 2020), at 1-hourly intervals and 1/4° geographic resolution. Co-locations are created using nearest neighbor in time and bilinear interpolation in space. The native dataset is in the form of 10-m neutral wind vectors; we only use the magnitudes here.

The spectra are provided at 64 frequencies with irregular spacing (0.005 Hz at low frequencies and 0.01 Hz at higher frequencies) from 0.025 to 0.58 Hz. We organize the spectra into wind speed bins with width of 1 m/s. In most tables and figures herein, we use two wind speeds as examples. The first is relatively low but occurs very often in our co-location dataset $U_{10} = 6$ m/s, being 12.2% of total population ($n = 4.55 \times 10^5$). The second example is for a wind speed that is higher, $U_{10} = 12$ m/s, being 2.94% of total population. Using a near-gale-force or higher wind speed is not practical, since it would result in a small population, e.g., 0.7% with $U_{10} = 15$ m/s.

Within a wind speed bin, the mean spectrum and quartile spectra (25%, 50%, and 75% quantiles, with 50% being the median) are computed for each frequency of the CDIP spectra.

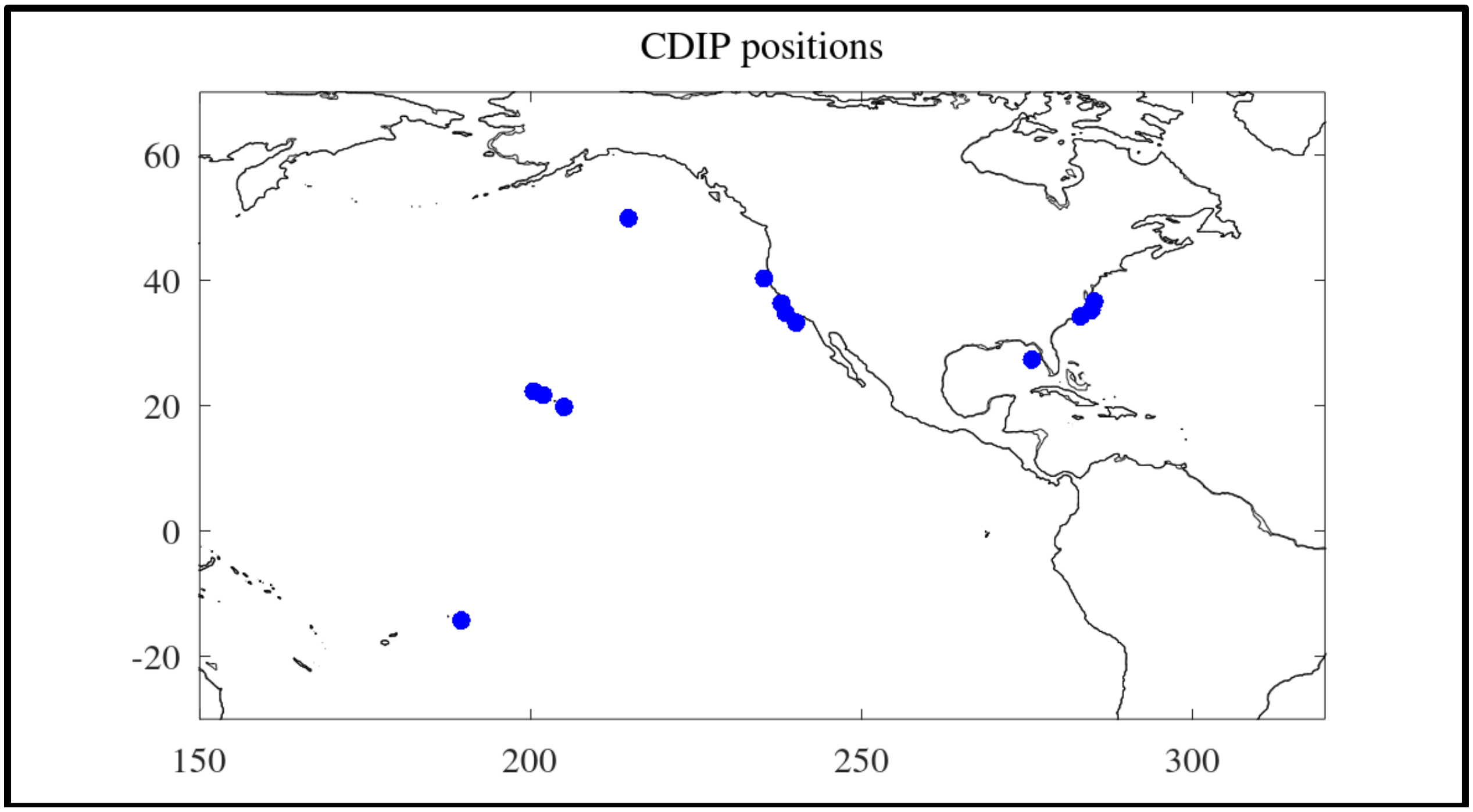

Figure 1. Positions of the 13 CDIP buoys used to evaluate the parametric spectra. Data are from May 2022 to December 2024.

## 4. Candidate parametric models of ocean wave spectra: introduction and evaluation against observational dataset

The non-directional ocean wave spectrum is denoted here as $E(f)$. It is the spectral density of the sea surface elevation $\eta$ with time (SI units: $m^2$s). Integration over frequency $f$ yields the variance of the sea surface elevation, $\eta_{rms}^2 \equiv m_0 = \int E(f)df$. Since this quantity is proportional to the energy per surface area (SI units: J $m^{-2}$) via $E = \rho_w g \eta_{rms}^2$, it is often referred to as the wave energy spectrum. (Here, $\rho_w$ is the density of water and $g$ is gravitational acceleration.)

### 4.1. Phillips spectrum: high frequencies, $f^{-5}$

The simplest parametric model is that of Phillips (1958). It is found by dimensional analysis, assuming that, given large fetch and duration, the wave spectrum reaches an "equilibrium" where it is only governed by frequency and gravity: $E(f) \propto g^2 f^{-5}$. This is, of course, a non-sensical assumption, as it implies that the spectrum level is the same for low and high wind speeds, say $U = 1$ m/s and $U = 20$ m/s. This error was acknowledged by Phillips (1985), but the formula remains as a useful reference point, and has had a "second life" since its simplicity makes otherwise unsolvable analytical equations solvable (e.g., Breivik et al. 2014), and it has been used as the basis for a breaking threshold (e.g., Babanin et al. 2007, 2010[2]). This allows the spectrum level to increase with wind speed in a modern wave model, like a spring: dissipation pulls the spectrum down toward a "rest state" (the breaking threshold), while wind pulls the spectrum upward; whenever wind speed changes, these two forces quickly reach a new equilibrium.

[2] In its simplest form, ignoring directional spreading ($A$=1 in their notation), the breaking threshold of Babanin et al. (2007, 2010) is the Phillips spectrum divided by 3.3.

Further, it is often acknowledged that a $f^{-5}$ spectral slope (meaning, $E(f) \propto f^{-5}$) *does* exist for individual spectra, i.e., without the universal constant, at high frequencies (Forristall 1981, Kahma and Calkoen 1992, Tsagareli 2009). This is sometimes called the saturation range or dissipation range (Resio et al. 2004). Nowadays, it is generally not referred to as the equilibrium range[3], even though this term was used in the title of Phillips (1958).

The original form of the Phillips spectrum is $F(\omega) = \alpha_P g^2 \omega^{-5}$, where $\omega = 2\pi f$ is the circular frequency, $F(\omega)$ is the spectral density, $F(\omega) = E(f)/2\pi$, $\alpha_P$ is a coefficient of proportionality. Thus,

$$E(f) = \frac{\alpha_P g^2}{\omega^4 f} = \frac{\alpha_P g^2}{(2\pi)^4 f^5} \qquad (1)$$

Following Pierson and Moskowitz (1964), we use $\alpha_P = 0.0081$. Other values can be found in the literature, e.g., Breivik et al. (2014) use $\alpha_P = 0.0083$ and Phillips (1958) originally suggested $\alpha_P = 0.0074$.

### 4.2. Toba spectrum: high frequencies, $f^{-4}$

Toba (1973) proposed an alternative spectrum, also on dimensional grounds: $E(f) \propto g u_* f^{-4}$. Toba denoted the proportionality coefficient as "$\alpha_g$" (his equation 2.12), but we adopt the more modern notation, "$\beta$", following Battjes et al. (1987). The native formula is then: $F(\omega) = \beta g u_* \omega^{-4}$, which is equivalent to $E(f) = \beta g u_* \omega^{-3} f^{-1} = \beta g u_* (2\pi)^{-3} f^{-4}$. Toba proposed $\beta$ =0.02 based on laboratory experiments. This value is very low relative to later estimates (Battjes et al. 1987) and is not used herein.

Besides the value for $\beta$, a second ambiguity exists with the Toba formula: the friction velocity $u_*$ must be defined. This is addressed in the next section.

### 4.3. Formulae for friction velocity

For estimation of $u_*$, we use Hwang (2011). His equation 5 is $C_{10} = 10^{-5}(-0.160 U_{10}^2 + 9.67 U_{10} + 80.58)$, valid for $U_{10} \leq 50$ m/s, and $u_* = U_{10}\sqrt{C_{10}}$. Here, $U_{10}$ is the 10-meter wind speed and $C_{10}$ is the corresponding drag coefficient. Following Smith et al. (2013), we include a small modification to extend into extreme wind speeds: we set the value of $u_*$ at its maximum ($u_*$=2.02558 m/s at $U_{10}$ =50.33 m/s in the original formula) when $U_{10} > 50.33$ m/s.

Prior to Hwang (2011), other simple formulae existed for estimating $u_*$ from $U_{10}$: see Garratt (1977), Wu (1980, 1982), and WAMDIG (1988). Forristall (1981) and Battjes et al. (1987) use the Garratt formula.

---

[3] The term is used for another part of the spectrum, as we will discuss in Section 4.4.

Hwang (2011) is compared against earlier formulae in Figure 2. "Garratt (1977) direct" computes $U_{10}$ from $u_*$ using a logarithmic wind profile (with a fixed von Karman parameter $\kappa = 0.41$ and computed roughness length $z_0$), while "Garratt (1977) indirect" is a linear fit proposed by Garratt to compute $u_*$ from $U_{10}$. WAMDIG (1988) is identical to Wu (1980, 1982) except at $U_{10} < 7.5$ m/s, where $C_{10}$ is a constant, $1.2875 \times 10^{-3}$. Jarosz et al. (2007) is a fit using ocean-current observations during Hurricane Ivan in 2004. The primary feature of the two newer formulae is that $u_*$ saturates at high wind speeds, which implies that $C_{10}$ "rolls over" to decrease at hurricane wind speeds. Comparisons to other datasets, such as Powell (2007), can be found in Smith et al. (2013).

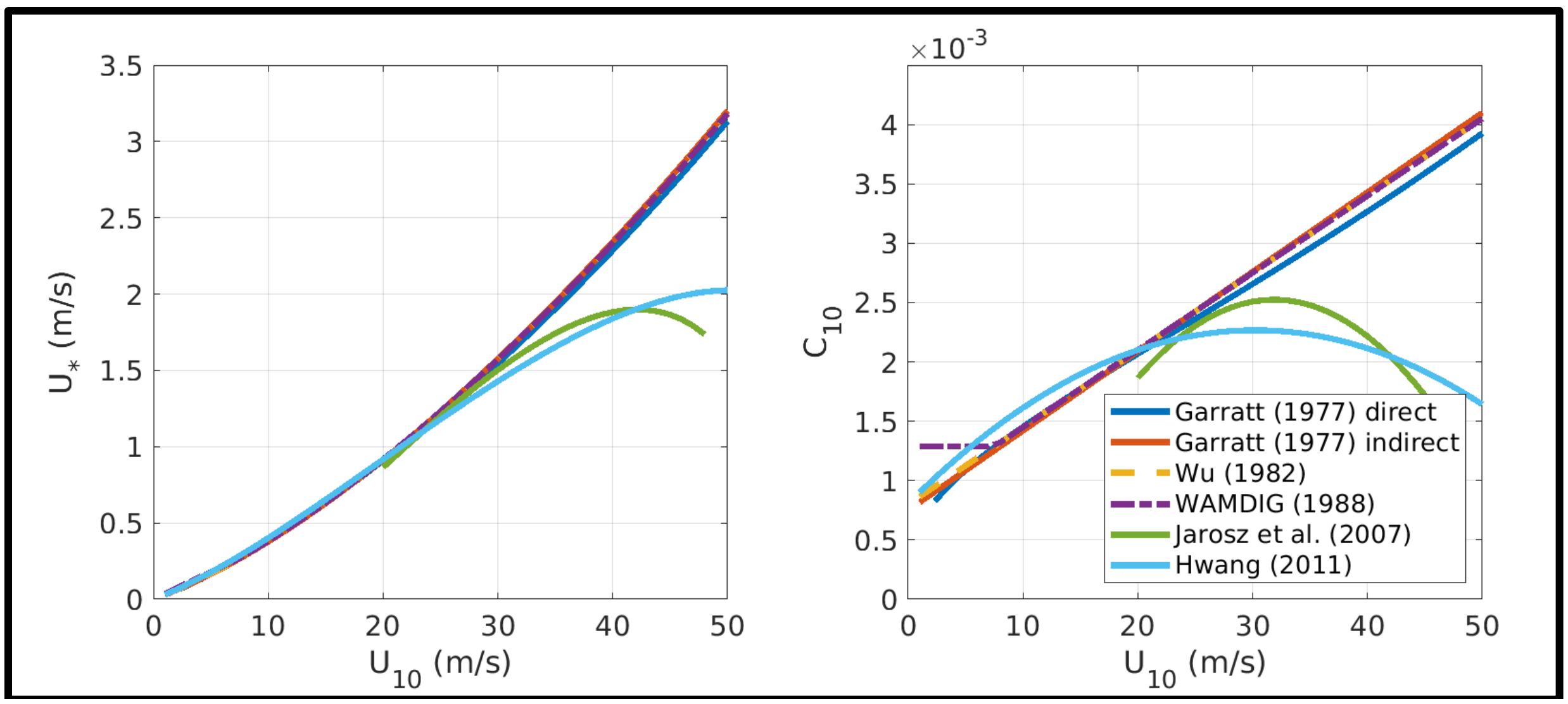

Figure 2. Comparison of Hwang (2011) against earlier formulae for $u_*$ (left panel) and $C_{10}$ (right panel).

### 4.4. Transition frequency

We have already introduced the concept of the saturation range (called the "dissipation range" in Resio et al. 2004), where many models apply a $f^{-5}$ spectral slope, $E_5(f) = E(f) \propto f^{-5}$. At lower frequencies, a $f^{-4}$ spectral slope is commonly assumed. According to most sources, this typically is bound at a lowest frequency around $1.5f_p$; we will discuss this low frequency limit in Section 4.5. The region of $E_4(f) = E(f) \propto f^{-4}$ is—since the mid-1980s, at least—referred to as the "equilibrium range".

Forristall (1981) proposed that there is a frequency at which there is a transition from the $f^{-4}$ equilibrium range to the $f^{-5}$ saturation range. This concept is also seen in later works, e.g., Kitaigorodskii (1983), Kahma and Calkoen (1992), Tsagareli (2009), Lenain and Melville (2017), and Vincent et al. (2019).

There are three essential components of a parametric model that includes a transition from $f^{-4}$ to $f^{-5}$:

1. The energy level in the equilibrium range, $E_4(f)$, e.g., a calibration of the Toba formula.
2. The energy level in the saturation range, $E_5(f)$, e.g., a calibration of the Phillips formula.
3. The transition frequency $f_t$.

However, one only needs to specify two of the three components. The third naturally follows. Thus, there are three possible strategies available to represent the transition.

The first strategy is to prescribe (1) and (2). Then, (3) is simply the frequency of intersection.

The second strategy is to prescribe (1) and (3), e.g., as described in Vincent et al. (2019). The third strategy is to prescribe (2) and (3), but we know of no examples of this in the literature.

In the second and third strategies, the transition frequency $f_t$ is estimated using an empirical method such as the three listed here:

- Forristall (1981) studied measured spectra and found a transition frequency $f_t = 0.0275g/u_*$, i.e., $\sim 0.27/u_*$ in S.I. units.
- Kahma and Calkoen (1992), in a grand average of spectra, found $f_t = 5g/(2\pi U_{10})$, i.e., $7.8/U_{10}$ in S.I. units. At $U_{10}$=7 m/s and applying the Hwang formula, this is $f_t \sim 0.29/u_*$ (S.I. units).
- Lenain and Melville (2017) find a transition at $r = ku_*^2/g \approx 1-2.5 \times 10^{-2}$ (using their Figure 13 in lieu of the numbers given in their abstract). Taking a value of $2 \times 10^{-2}$, and assuming deep water, we have $f_t = \sqrt{2 \times 10^{-2}}g/(2\pi u_*)$. In S.I. units, this is $0.22/u_*$.

Example calculations are shown in Figure 3 (with $r = 2 \times 10^{-2}$ for the Lenain and Melville formula).

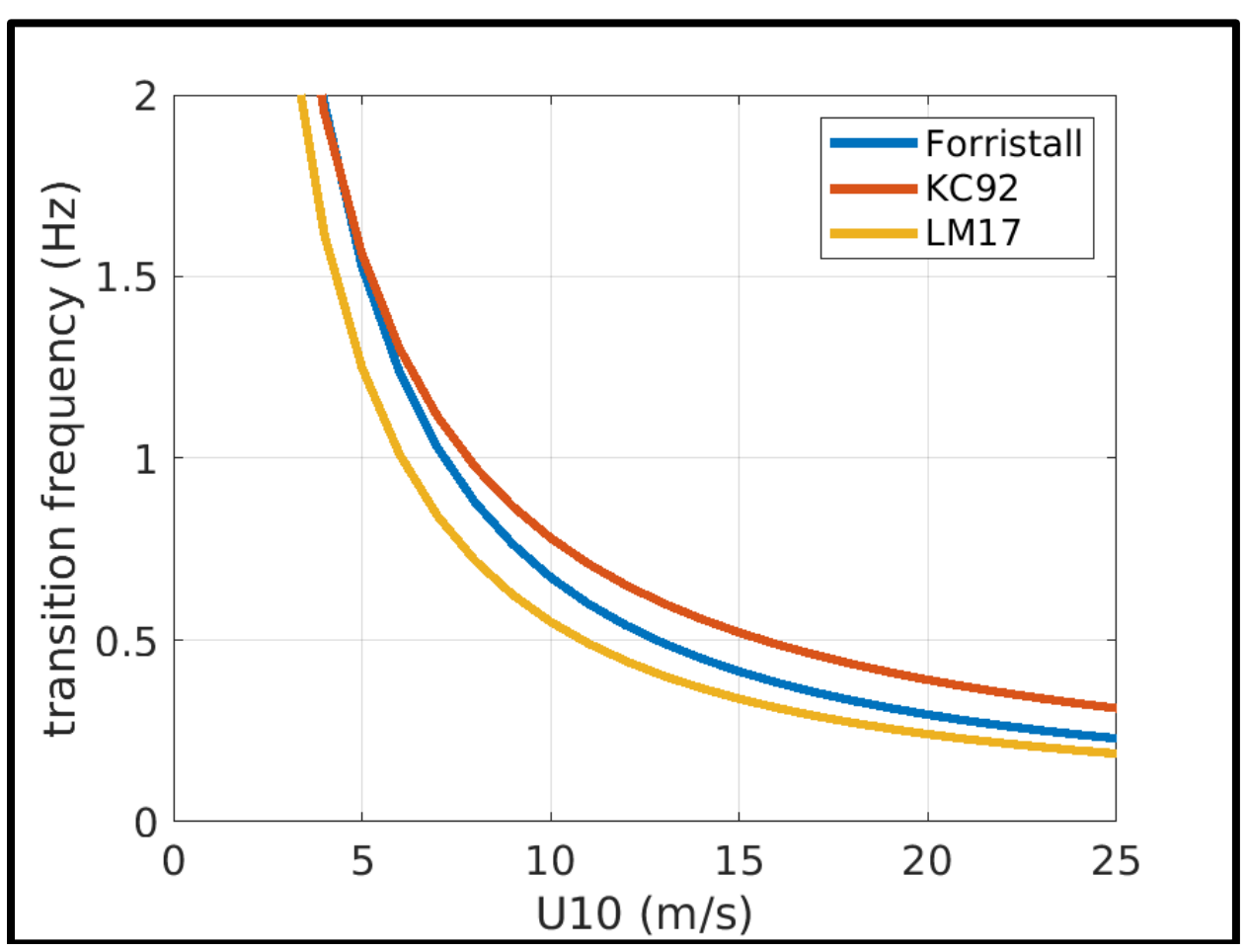

Figure 3. Transition frequencies as a function of $U_{10}$ according to Forristall (1981), Kahma and Calkoen (1992), and Lenain and Melville (2017). The Hwang formula is used to convert from $U_{10}$ to $u_*$.

### 4.5. Low-frequency limit on the equilibrium spectrum

The low-frequency limit on the equilibrium spectrum is commonly given as $1.5f_p$, where $f_p$ is the peak frequency. Donelan et al. (1985) state: "*To illustrate the slope of the rear face, the spectra have been multiplied by* $\omega^4$ *... in the frequency region* $1.5\omega_p < \omega < 3.0\omega_p$ *(see figure 12). It is clear that an* $\omega^{-4}$ *power law is a good description of the rear face of the spectrum in the energy containing region*". Notably, their Figure 12 has little data beyond $3.0\omega_p$, so there is no indication that $3.0\omega_p$ is a high-frequency limit on the equilibrium range; only the low-frequency limit $1.5\omega_p$is clearly shown. Resio et al. (2004), Tsagareli (2009), and Vincent et al. (2019) also use $1.5f_p$ as a rule of thumb for the low-frequency limit of the equilibrium range.

For the definition of $f_p$ used to determine this limit, herein we use $f_p$ calculated as $f_p = \nu g/U_{10}$, where $\nu$ is the non-dimensional peak frequency. The fully developed Pierson-Moskowitz value, $\nu = 0.13$, is used unless the $\nu$ value is explicitly stated.

### 4.6. Battjes/JONSWAP spectrum, $f^{-4}$ in high frequencies

Given the Toba form, $E(f) = \beta g u_*(2\pi)^{-3}f^{-4}$ , Battjes et al. (1987) reviewed $\beta$ values inferred from the literature to that point (e.g., $\beta = 0.11$ from Forristall (1981)) and calibrated the formula to spectral data from the JONSWAP field experiments in 1968-1969 (Hasselmann et al. 1973), finding $\beta = 0.13$.

Battjes et al. use the Garratt (1977) formula for converting from $U_{10}$ to $u_*$. We use the Hwang formula instead.

We treat this model as valid only in the equilibrium range, with a low frequency limit of $1.5f_{p,PM}$ (i.e. 1.5 times the peak frequency of a "fully-developed" spectrum) and a high frequency limit according to Forristall (1981). Example spectra are shown in Figure 4, using the calibration to the data of JONSWAP ($\beta = 0.13$) and Mitsuyasu et al. (1980) ($\beta = 0.087$). These are the highest $\beta$ value and a mid-range $\beta$ value from Table 2 of Battjes et al. (1987), who performed the calibration.

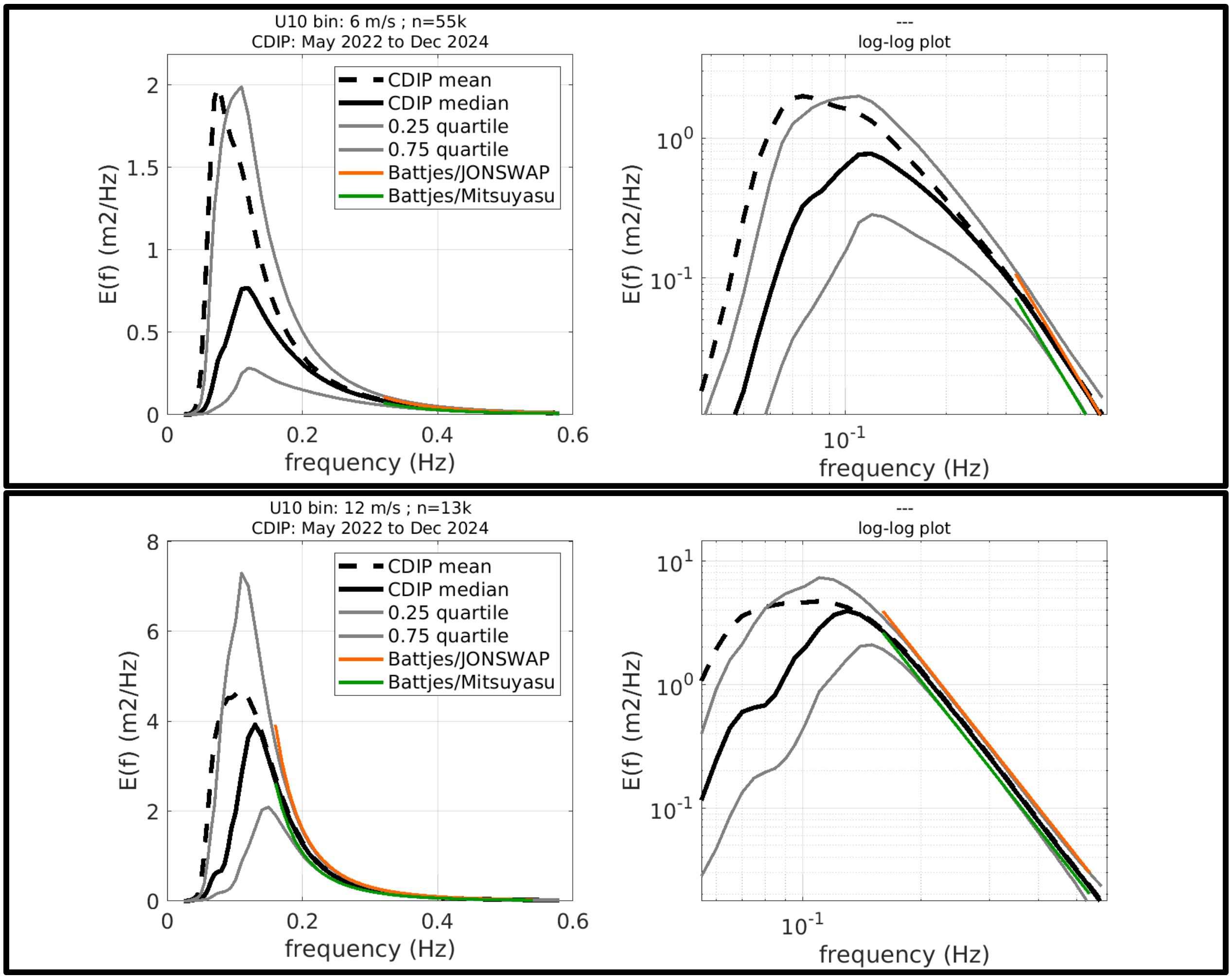

Figure 4. Comparison of two calibrations of the Battjes et al. (1987) parametric spectrum (described in this section, 4.6) against mean and quartile spectra computed from CDIP data, described in Section 3. Upper panels: $U_{10} = 6$ m/s. Lower panels, $U_{10} = 12$ m/s. Left panels: plots in linear scale. Right panels: plots in log-log scale.

### 4.7. Pierson-Moskowitz spectrum: full spectrum, $f^{-5}$ in high frequencies

Thus far, we have introduced two parametric spectra, the Phillips spectrum and the Toba spectrum. These two models are only intended for the high frequencies. They do not include frequencies lower than the equilibrium range ($f^{-4}$ range) and certainly do not attempt to describe the shape of the spectral peak. One of the earliest parametric spectra was proposed by Pierson and Moskowitz (1964)[4]. We refer to it here as the "PM spectrum". This was notionally for "fully developed seas", i.e., not limited by fetch or duration. The spectrum is:

$$F(\omega) = \left(\frac{\alpha g^2}{\omega^5}\right) e^{-\beta\left(\frac{\omega_p}{\omega}\right)^4} \tag{2}$$

[4] For a review of earlier proposed forms, see the *History* section in Moskowitz (1964).

where $\alpha = 8.10 \times 10^{-3}$, $\beta = 0.74$, and $\omega_p = g/U_{19.5}$, with $U_{19.5}$ being the wind speed reported by the weather ships at 19.5 m elevation. Note that without the exp() term, it is equivalent to the Phillips spectrum, $F(\omega) = \alpha_P g^2 \omega^{-5}$.

We use:

$$E(f) = E_P(f) e^{-1.25\left(\frac{f_p}{f}\right)^4} = \left(\frac{\alpha g^2}{f^5 (2\pi)^4}\right) e^{-1.25\left(\frac{f_p}{f}\right)^4} \tag{3}$$

where $E_P(f)$ is the Phillips spectrum and $f_p$ is the peak frequency estimated using $f_p = 0.13 g/U_{10}$. See also Young (1999), equation 5.51. The non-dimensional peak frequency is therefore $\nu \equiv f_p U_{10}/g = 0.13$, and $U_{10}/C_p = \nu(2\pi) = 0.817$. With $C_p$ being the phase velocity at the peak frequency, this implies that the waves at that frequency are outrunning the wind.

A useful feature of the PM spectrum is that it requires specification of only one parameter, the wind speed. A disadvantage of the spectrum is that it uses $f^{-5}$ in the equilibrium range, where we have known for decades that $f^{-4}$ is more appropriate; see Toba (1973), Forristall (1981), Kitaigorodskii (1983), Donelan et al. (1985), and Phillips (1985).

### 4.8. JONSWAP spectrum: full spectrum, $f^{-5}$ in high frequencies

The JONSWAP field experiments in 1968-1969 led to a new parametric model (Hasselmann et al. 1973). This parametric model is an extension of the PM spectrum, including an additional shape factor $S$ to make it applicable to fetch-limited conditions: $E(f) = E_{PM}(f)S$. We refer the reader to Hasselmann et al. (1973), Young (1999), and Holthuijsen (2007) for details of this shape factor and provide a short summary here: it includes three empirical shape parameters, $\gamma$, $\sigma_a$, and $\sigma_b$, which are a function of wave age. The first is a "peak enhancement" factor that controls the maximum of $E(f)$, while the latter two control the width of the peak to the left and right, respectively. The $\alpha$ parameter, assumed constant in the PM spectrum, is also a function of wave age. For large fetch and duration, the shape factor approaches unity and the $\alpha$ approaches the PM value.

In the literature, one can find different approaches to applying the JONSWAP spectrum, but it is our opinion that the most sensible method is to start with two of three parameters: $U_{10}$, $f_p$, and $U_{10}/C_p$, with the latter being one form of the "inverse wave age" and $C_p$ is the phase velocity at the peak. From the two starting parameters, the third can be computed. Next, the shape parameters can be estimated using empirical fits such as those described in Young (1999). For example, with $\nu = 0.2$, we get $\gamma \approx 3$, $\sigma_a \approx \sigma_b \approx 0.08$, and $\alpha \approx 0.009$.

### 4.9. Donelan et al. spectrum: full spectrum, $f^{-4}$ in high frequencies

Donelan et al. (1985) proposed an alternative to the JONSWAP spectrum, based on new field and laboratory observations. The non-directional spectrum is given as:

$$F(\omega) = \frac{\alpha_D g^2}{\omega^4 \omega_p} e^{-\left(\frac{\omega_p}{\omega}\right)^4} S \tag{4}$$

or

$$E(f) = \frac{\alpha_D g^2}{(2\pi)^4 f^4 f_p} e^{-\left(\frac{f_p}{f}\right)^4} S \tag{5}$$

where $S$ is a shape factor that follows that of JONSWAP, but with a different method for prescribing the shape parameters that compose it.

As with JONSWAP, $\alpha_D$ and the shape parameters used in $S$ are determinable from wave age using empirical fits provided by Donelan et al., so like with JONSWAP, it is customary to start with two of three parameters: $U_{10}$, $f_p$, and a metric for inverse wave age.

The Donelan spectrum also provides a description of the directional distribution of the spectrum, $E(f,\theta) = E(f)D(f,\theta)$, where $\int_0^{2\pi} D(f,\theta)d\theta = 1$. We refer the reader to Donelan et al. (1985) for a description of the formula for $D(f,\theta)$. The Donelan spectrum is popular with engineers and scientists that require a parametric model for the full directional spectrum.

Figure 5 compares the Donelan spectrum described in this section (4.9) and the JONSWAP spectrum described in the prior section (4.8) against the CDIP-based mean and quartile spectra. For the parametric spectra, we use $\nu = 0.2$, which corresponds to developing wind sea.

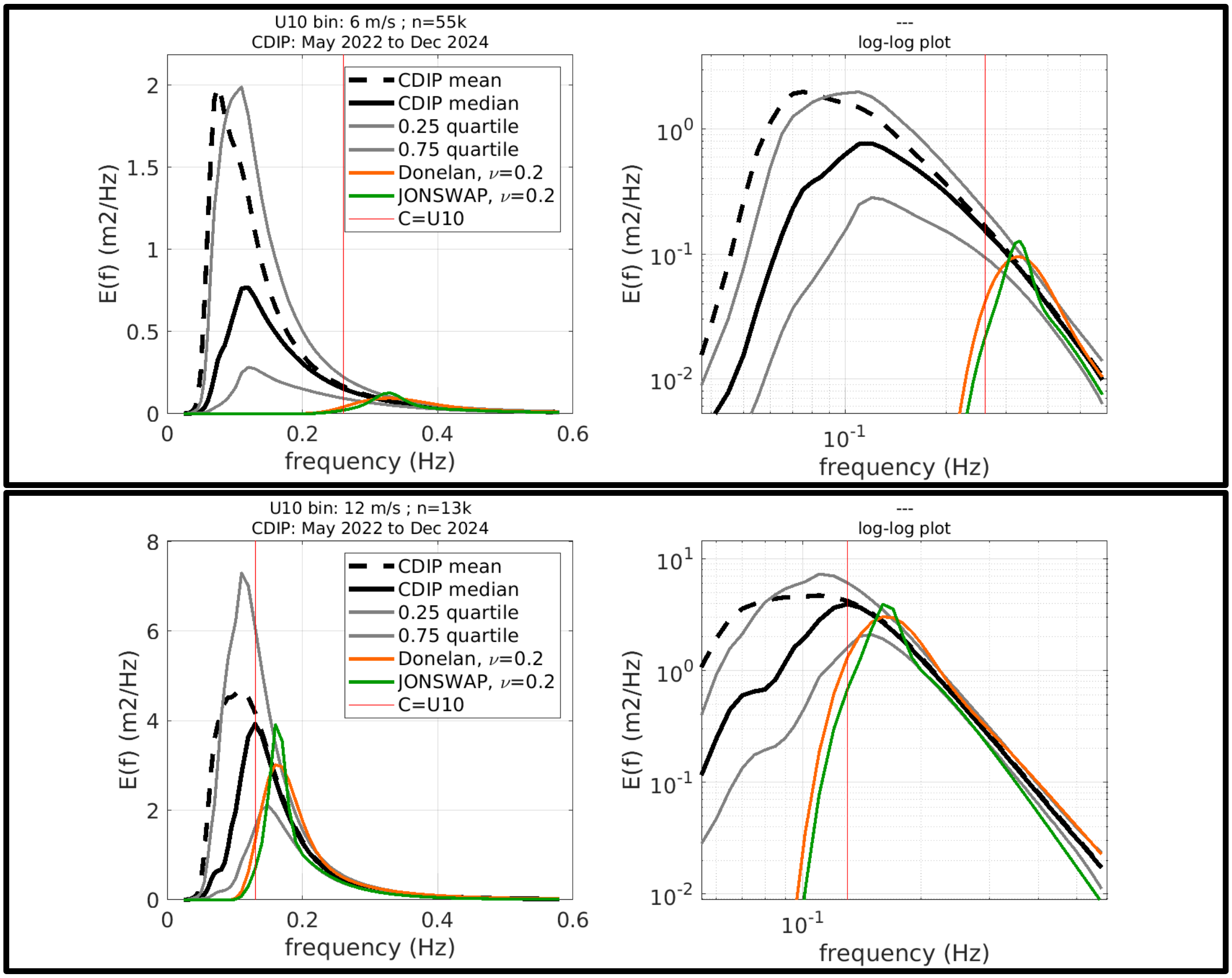

Figure 5. Like Figure 4, except that the two parametric spectra are Donelan et al. (1985) and Hasselmann (et al. 1973). The vertical red lines indicate $U_{10}/C = 1$.

### 4.10. Alves et al. PM spectrum, merging Pierson-Moskowitz and Donelan spectra

Alves et al. (2003) recognized the problem just mentioned (namely: the use of $f^{-5}$ in the equilibrium range by PM) and set out to correct the problem by merging the PM spectrum with that of Donelan et al. (1985), which uses a $f^{-4}$ slope. They also reanalyzed the original spectra of Moskowitz (1964) and Pierson and Moskowitz (1964) using technical reports provided by Dr. Pierson. Their proposed new PM formula is, in their equation 19:

$$E(f) = \left(\frac{\alpha_D g^2}{f_p f^4 (2\pi)^4}\right) e^{-\left(\frac{f_p}{f}\right)^4} \tag{6}$$

where $\alpha_D = 6.15 \times 10^{-3}$. The peak frequency is not changed (see their equation 12): $f_p = 0.13g/U_{10}$.

The authors claim that "these values of [$\alpha_D$] for a spectral form with high-frequency tail proportional to $f^{-4}$ were consistent with Donelan et al. (1985), who fitted a variant of [their equation 19] to fetch-limited data collected in Lake Ontario, Canada". If one applies the Donelan et al. formula (their equation 5.2) with a small inverse wave age ($U/C$ =1 or $\nu = f_p U_{10}/g = 0.16$), one gets $\alpha_D = 6 \times 10^{-3}$, so it is indeed consistent.

Unfortunately, Alves et al. (2003) did not account for the fact that the peak frequency strongly affects the tail level predicted by the Donelan formula. The Toba (1973) and Donelan et al. (1985) both use the $f^{-4}$ spectrum, so it is possible to relate the $\alpha_D$ of the Alves spectrum with the $\beta$ of the Toba spectrum. This was worked out by Battjes et al. (1987) (their equation B2):

$$\beta = \alpha_D \frac{g}{\omega_m u_*}$$

If we use $\omega_m = (2\pi) f_{p,PM} = (2\pi) 0.13 g/U_{10}$, and relate $u_*$ to $U_{10}$ using the Hwang formula, we get the $\beta$ values shown in the solid black line in Figure 6. These are not at all consistent with the median $\beta$ from the Donelan et al. dataset as evaluated by Battjes et al. (1987): they have $\beta = 0.11$ in their Table 2, which we indicate with a red line in Figure 6. The discrepancy appears to be introduced via $\omega_m$ (or $f_p$ in (6)). The Alves et al. (2003) value of $\alpha_D$ is consistent with Donelan et al., but the corresponding value of $\beta$, which more directly controls the tail level, is *not* consistent with Donelan et al., since Alves et al. are addressing the case of fully developed seas, e.g., $U/C$ =1 or $\nu = f_p U_{10}/g = 0.16$, while the typical data point in the Donelan dataset is for fetch-limited seas, $U/C$ =2 or $\nu = 0.32$. The dashed black line is the result if we convert from $\alpha_D$ to $\beta$ using $\nu = 0.32$. This illustrates the sensitivity to $f_p$ in (6).

The bottom line is that, because of a discrepancy in wave age between source material (Donelan et al.) and application (fully developed spectra), the tail level from the Alves et al. model, equation (6), will be much higher than that from other $f^{-4}$ parametric models.

Figure 7 compares the Alves et al. (2003) parametric model with the original PM spectrum. As expected, the tail level of the former is much higher than the latter.

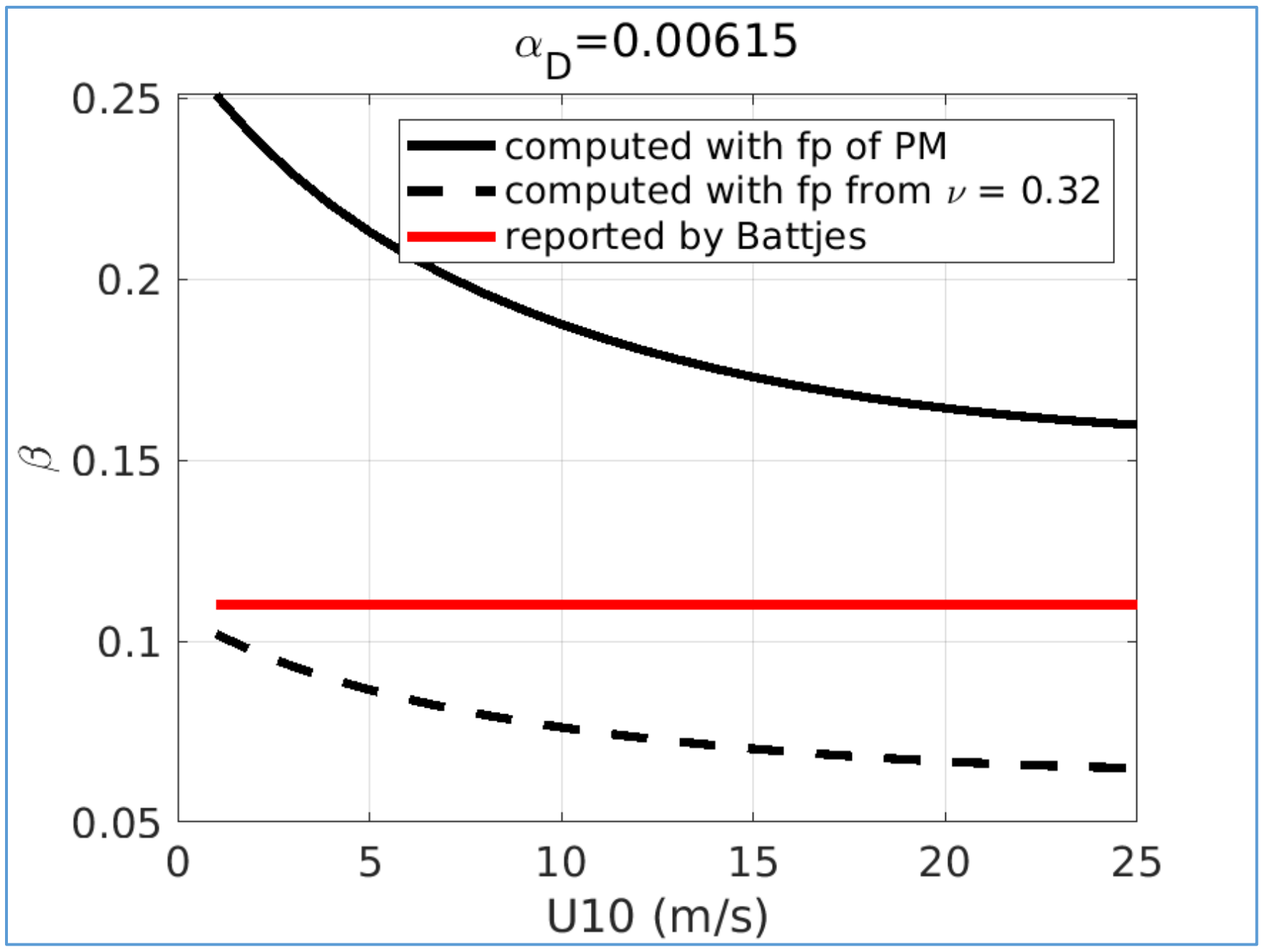

Figure 6. Black lines: using the Alves et al. (2003) value $\alpha_D = 6.15 \times 10^{-3}$ and converting to $\beta$. This illustrates how $\beta$ and thus the tail level is strongly dependent on the peak frequency used. The solid black line shows the case where $f_p$ is for fully developed seas. The dashed black line shows the case where $f_p$ is appropriate to the Donelan et al. dataset instead. The red line indicates the mean $\beta$ of Donelan et al. (1985) as reported by Battjes et al. (1987).

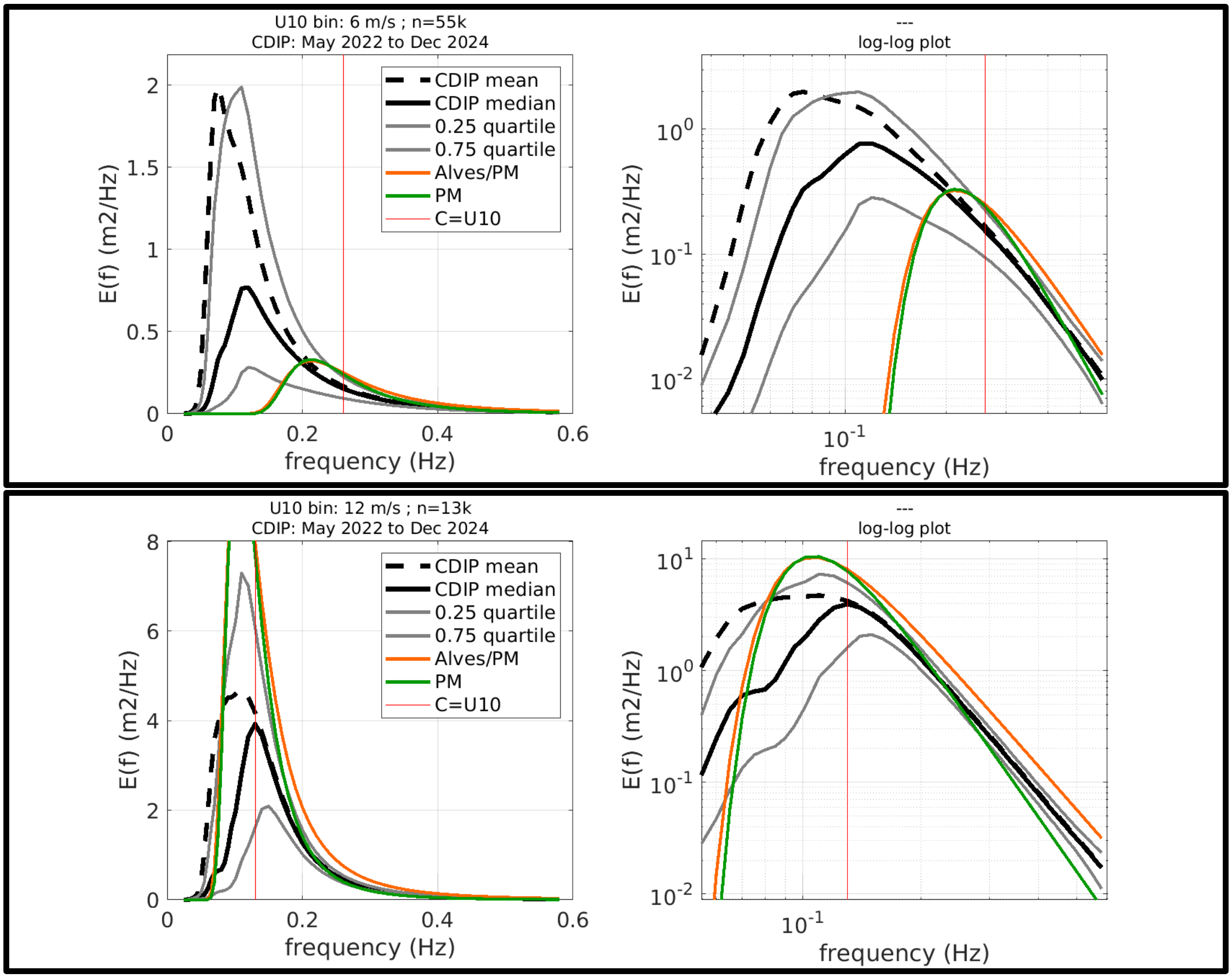

Figure 7. Like Figure 4, except that the two parametric spectra are the PM spectrum and the Alves et al. (2003) spectrum. The vertical red lines indicate $U_{10}/C = 1$.

### 4.11. Vincent spectrum, high frequencies, $f^{-4}$ and $f^{-5}$

Though it is not their primary aim, Vincent et al. (2019) present an empirical spectrum within their introduction. Within the equilibrium range, they have $E(f) = \alpha_0 U_{10} g (2\pi)^{-3} f^{-4}$ (their equations 1 and 8). Since their equation is based on $U_{10}$, it cannot be converted to/from the Toba form. They cite values of $\alpha_0 \approx 0.003 - 0.005$.

Within the saturation range, they simply attach an $f^{-5}$ tail: $E(f) = \alpha_0 U_{10} g (2\pi)^{-3} f_t f^{-5}$. In principle, they could have proceeded to formulate an $f_t$ that would cause this to conform to a calibration of the Phillips spectrum: $\alpha_0 U_{10} g (2\pi)^{-3} f_t f^{-5} = \alpha_P g^2 (2\pi)^{-4} f^{-5}$ or $f_t = \alpha_P g / (2\pi \alpha_0 U_{10})$. However, instead they suggest computation of $f_t$ using a form of Lenain and Melville (2017): $k_t = 0.015 g u_*^{-2}$, or in deep water, $f_t = \sqrt{0.015} g / (2\pi u_*)$. Since their formula retains some small dependence on wind speed via the drag coefficient $C_{10}$, it cannot be converted to/from the Phillips spectrum. Also, counter-intuitively, though this model uses a non-standard form of the Toba spectrum (based on $U_{10}$ instead of $u_*$), the model still requires a conversion from $U_{10}$ to $u_*$, for the $f_t$ calculation.

As with the Battjes model in Section 4.6, we treat this model as invalid for frequencies below the start of the equilibrium range for a fully developed sea, $1.5 f_{p,PM}$.
The Vincent et al. parametric spectrum using the lower and upper values that they stated, $\alpha_0 \approx 0.003 - 0.005$, is illustrated in Figure 8.

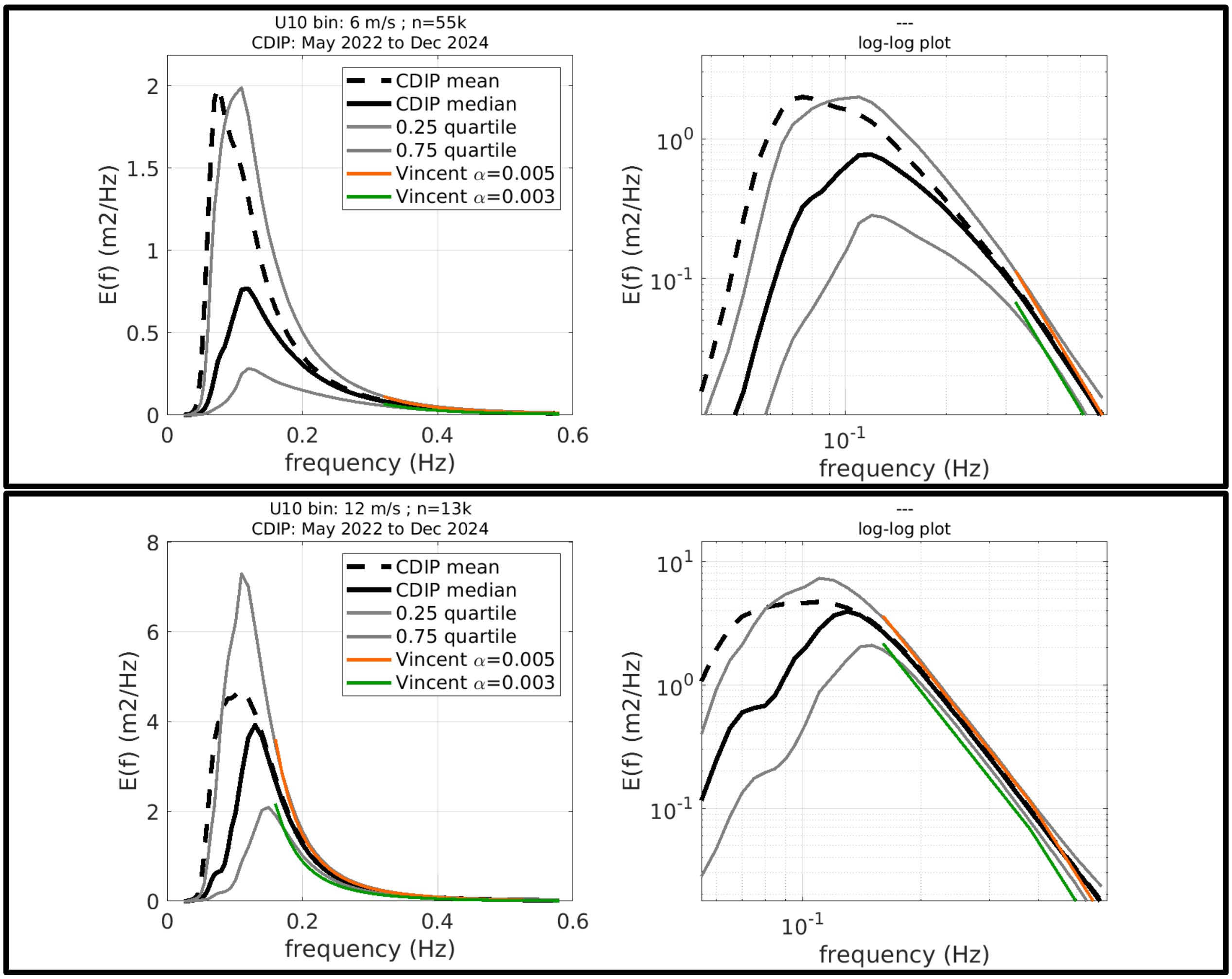

Figure 8. Like Figure 4, except that the parametric spectra are based on the model described by Vincent et al. (2019).

### 4.12. Forristall spectrum: high frequencies, $f^{-4}$ and $f^{-5}$

Within the equilibrium range, Forristall (1981) uses $E(f) = E_4(f) = \alpha_4 u_* g f^{-4}$. The "average" $\alpha_4$ reported in his Table 3 is $4.43\times 10^{-4}$. Since the Toba spectrum is $E(f) = \beta g u_* (2\pi)^{-3} f^{-4}$, the proportionality coefficient can be converted as $\beta = \alpha_4 (2\pi)^3$. Thus, the Forristall equilibrium range is the Toba spectrum with $\beta = 0.11$; in Battjes et al.'s Table 2, $\beta = 0.11$ is given for three sources: Kahma (1981), Forristall (1985), and Donelan et al. (1985).

As with the Battjes model in Section 4.6, we treat this model as invalid for frequencies below the start of the equilibrium range for a fully developed sea, $1.5f_{p,PM}$.

Within the saturation range, Forristall (1981) uses $E(f) = \alpha_5 g^2 f^{-5}$. The "average" $\alpha_5$ reported in his Table 3 is $12.2\times 10^{-6}$. Since the Phillips spectrum is $E(f) = \alpha_P g^2 (2\pi)^{-4} f^{-5}$, the proportionality coefficient can be converted as $\alpha_P = \alpha_5 (2\pi)^4$. Thus the Forristall saturation range is the Phillips spectrum with $\alpha_P = 0.019$, i.e., more than twice the PM level of $\alpha_P = 0.0081$.

Using the "average" coefficients of Forristall, we can work out the transition frequency:

$$E(f_t) = 4.43 \times 10^{-4} u_* g f_t^{-4} = 12.2 \times 10^{-6} g^2 f_t^{-5}$$

Thus, $f_t = g/(36.31 u_*)$ or $f_t = 0.27/u_*$ in S.I. units.

This is consistent with the formula given by Forristall (his equation 14): $f_t u_*/g = 0.0275$. If a different calibration is used, i.e., $\alpha_4 \neq 4.43\times 10^{-4}$ and/or $\alpha_5 \neq 12.2\times 10^{-6}$, then one should not expect $f_t u_*/g = 0.0275$ to hold.

Our application of the Forristall spectrum is slightly modified from the original: where Forristall (1981) used the Garratt formula to compute $u_*$ from $U_{10}$, we use the Hwang formula.

### 4.13. Battjes/Holthuijsen/Phillips spectrum: high frequencies, $f^{-4}$ and $f^{-5}$

As mentioned already, given the Toba form, $E(f) = \beta g u_* (2\pi)^{-3} f^{-4}$ , Battjes et al. (1987) reviewed $\beta$ values inferred from the literature to that point. They also found that the Toba form is more consistent with JONSWAP than the Phillips form ($f^{-5}$), and they found no correlation between $\beta$ and wave age. [This might be seen as contradicting the finding of Donelan et al. (1985) that $\alpha_D$ correlates with wave age. However, $\beta$ and $\alpha_D$ are not equivalent; see Battjes et al.'s Appendix B.]

As with the Battjes model in Section 4.6, we treat this model as invalid for frequencies below the start of the equilibrium range for a fully developed sea, $1.5f_{p,PM}$.

Holthuijsen (2007, p. 157), rather simplistically, arrive at a $\beta$ value by averaging the seven $\beta$ values given for field cases in Table 2 of Battjes et al.: $\beta = 0.096$. We can use this as a starting point for a new candidate model.

The new model shares features of the Forristall model. We have noted above that the energy level of the Forristall model in the saturation range, $E_5(f)$ is high relative to the Phillips tail, and the energy level of the Forristall model in the equilibrium range, $E_4(f)$ is high relative to that of the Holthuijsen calibration ($\beta = 0.11$ and $\beta = 0.096$, respectively). Thus, it makes sense to design the new model as a "lower energy" alternative to the Forristall model. To this end, we use $E_5(f) = \alpha_P g^2 (2\pi)^{-4} f^{-5}$ with $\alpha_P = 0.0081$ (i.e., the $\alpha_P$ used in Pierson and Moskowitz (1964)).

The transition frequency $f_t$ is the frequency of intersection:

$$E_4(f_t) = \beta g u_* (2\pi)^{-3} f_t^{-4} = E_5(f_t) = \alpha_P g^2 (2\pi)^{-4} f_t^{-5} \quad (7)$$

so $f_t = \alpha_P g/(2\pi\beta u_*) = 0.0134 g/u_*$.

In Figure 9, we compare this new "BHP" model with the Forristall model.

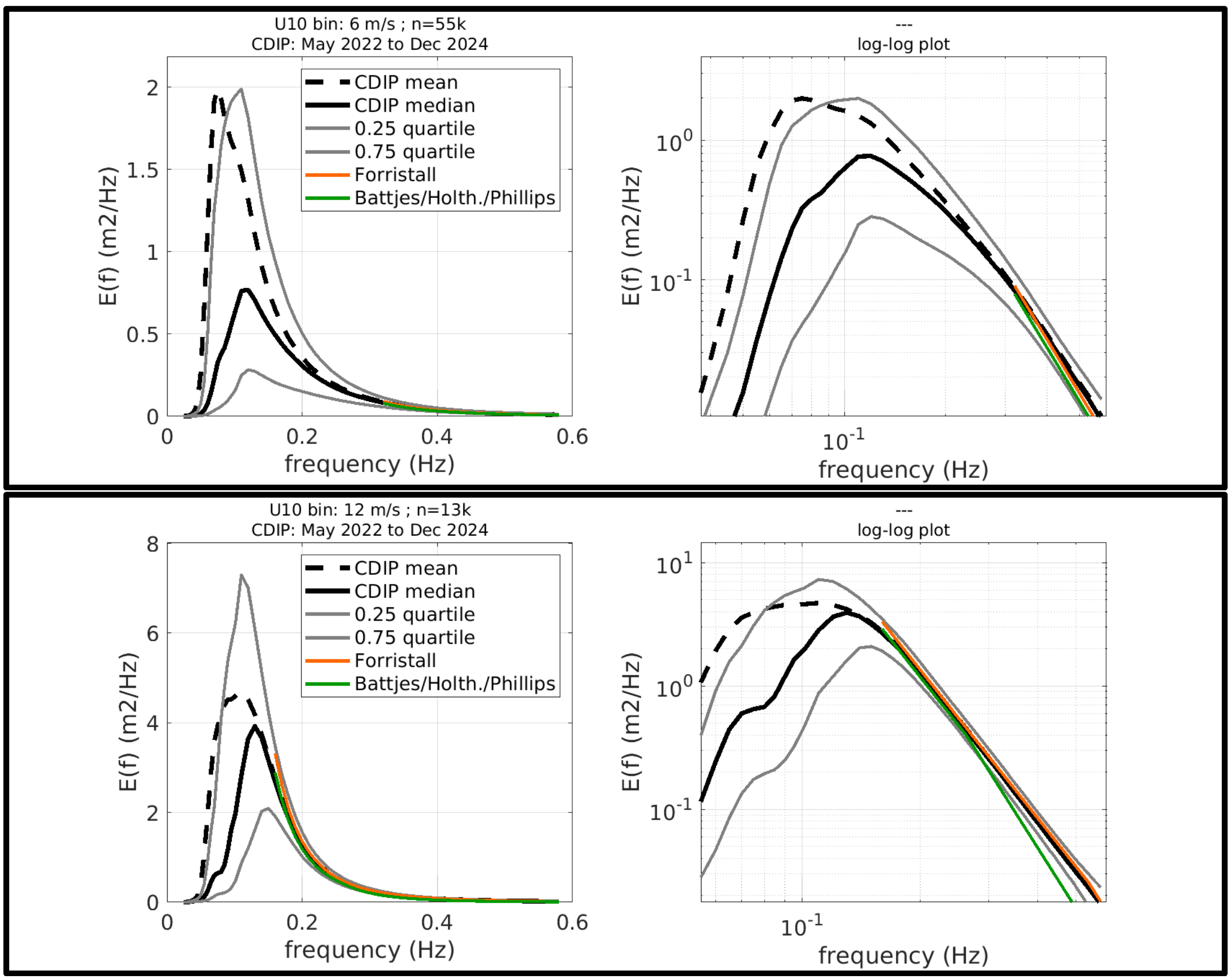

Figure 9. Like Figure 4, except that the parametric spectra are the Forristall model $\alpha_4 = 4.43 \times 10^{-4}$, i.e., $\beta = 0.11$, and the new "BHP" model, which uses $\beta = 0.096$.

### 4.14. Resio spectrum: high frequencies, $f^{-4}$

Resio et al. (2004) give a parametric model for the equilibrium range. Their equation (18), after assuming deep water and converting to frequency spectrum is:

$$E(f) = \alpha_{R4} u_m g f^{-4} (2\pi)^{-3} \tag{8}$$

where $u_m = \left(U_\lambda^2 C_p\right)^{\frac{1}{3}} - u_0$. Calibration parameters are: $\alpha_{R4} = 0.00553$ and $u_0 = 1.92$ m/s. $U_\lambda$ is the wind speed at an elevation of $0.065L_p$ (their paragraph 33). $C_p$ and $L_p$ are the phase velocity and wavelength at peak frequency $f_p$. The conversion to frequency spectrum is $E(f) = F(k)2\pi/C_g$, where $C_g$ is the group velocity.

Herein, the formula is assumed to be valid only in the range $1.5f_p < f < f_t$, where $f_t$ is computed using the Forristall formula.

Figure 10 compares the Resio spectrum with the Donelan spectrum. In both cases, we use $\nu = 0.2$, which corresponds to developing wind sea. Because of the restriction $f > 1.5f_p$, the Resio spectrum is only valid for a small range of frequencies in the $U_{10} = 6$ m/s case, where $1.5f_p = 0.49$ Hz.

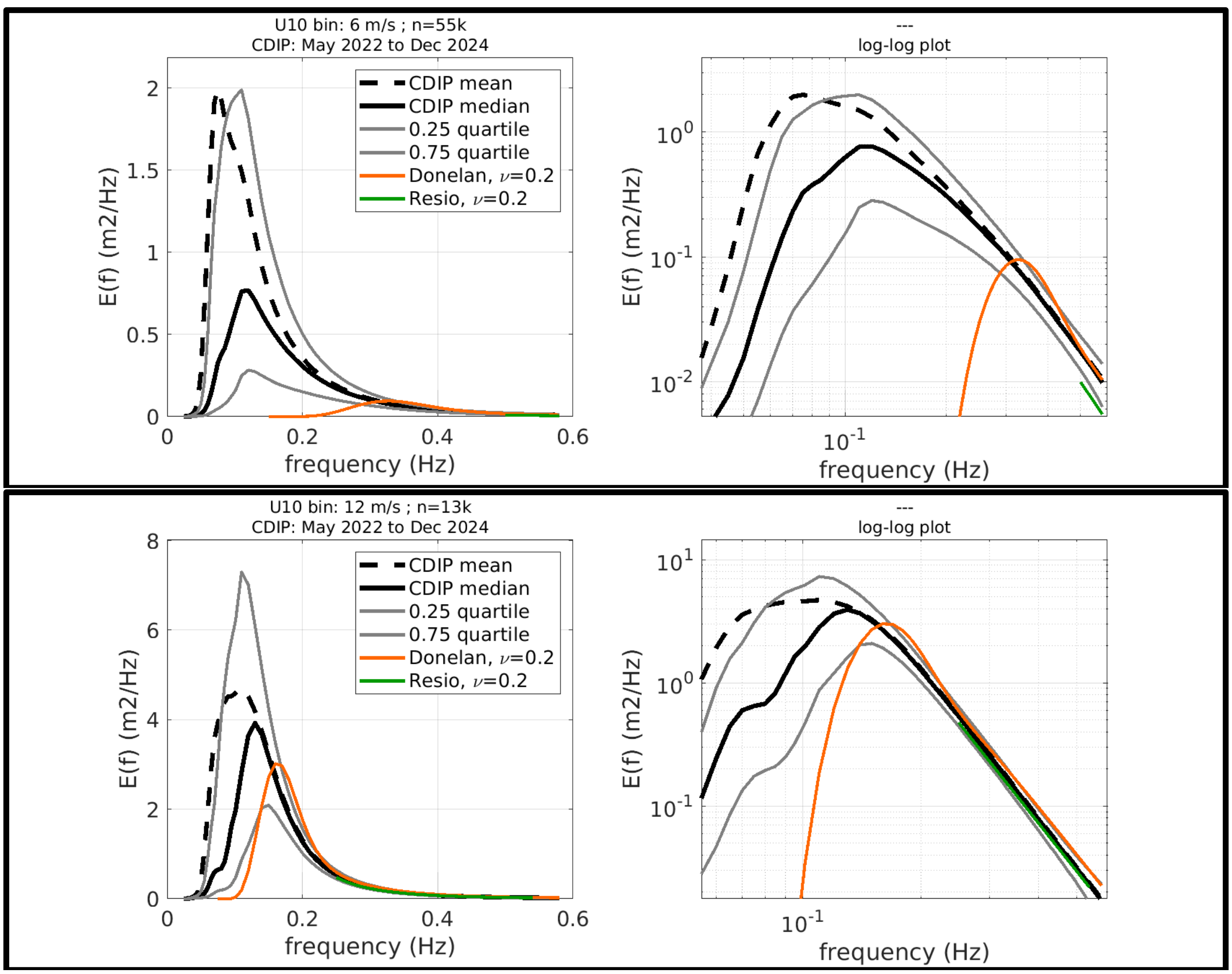

Figure 10. Like Figure 4, except that the two parametric spectra are Donelan et al. (1985) and Resio et al. (2004).

### 4.15. Additional comparisons

In Figure 9, we saw that for the $U_{10} = 12$ m/s case, the Forristall model is slightly high at the beginning of the equilibrium range, say $1.5f_p < f < 2f_p$. Thus, a non-standard calibration of the Forristall model, using $\alpha_4 = 4.00 \times 10^{-4}$ is tested. In Figure 11, it is compared with the mid-range calibration of the Vincent model, $\alpha = 0.004$. The comparison illustrates that the two models cannot be made to match via calibration, since a) the Vincent model uses a non-standard form of the Toba model that is in terms of $U_{10}$ rather than $u_*$ and b) the models use different formula for the transition frequency $f_t$. For the $U_{10} = 6$ m/s case, both models are slightly low at the highest frequencies, with the Vincent model being slightly better than the non-standard Forristall model. For the $U_{10} = 12$ m/s case, the non-standard Forristall calibration is a very good match to the observations, while the Vincent model underpredicts energy in the tail.

Based on comparisons to the CDIP spectra, we have seen that the Forristall model is one of the most—possibly the most—accurate parametric model considered here. However there does appear to be some residual bias that depends on wind speed. It may be possible to calibrate further by using a $\alpha_4$ that has some weak dependence on wind speed or drag coefficient.

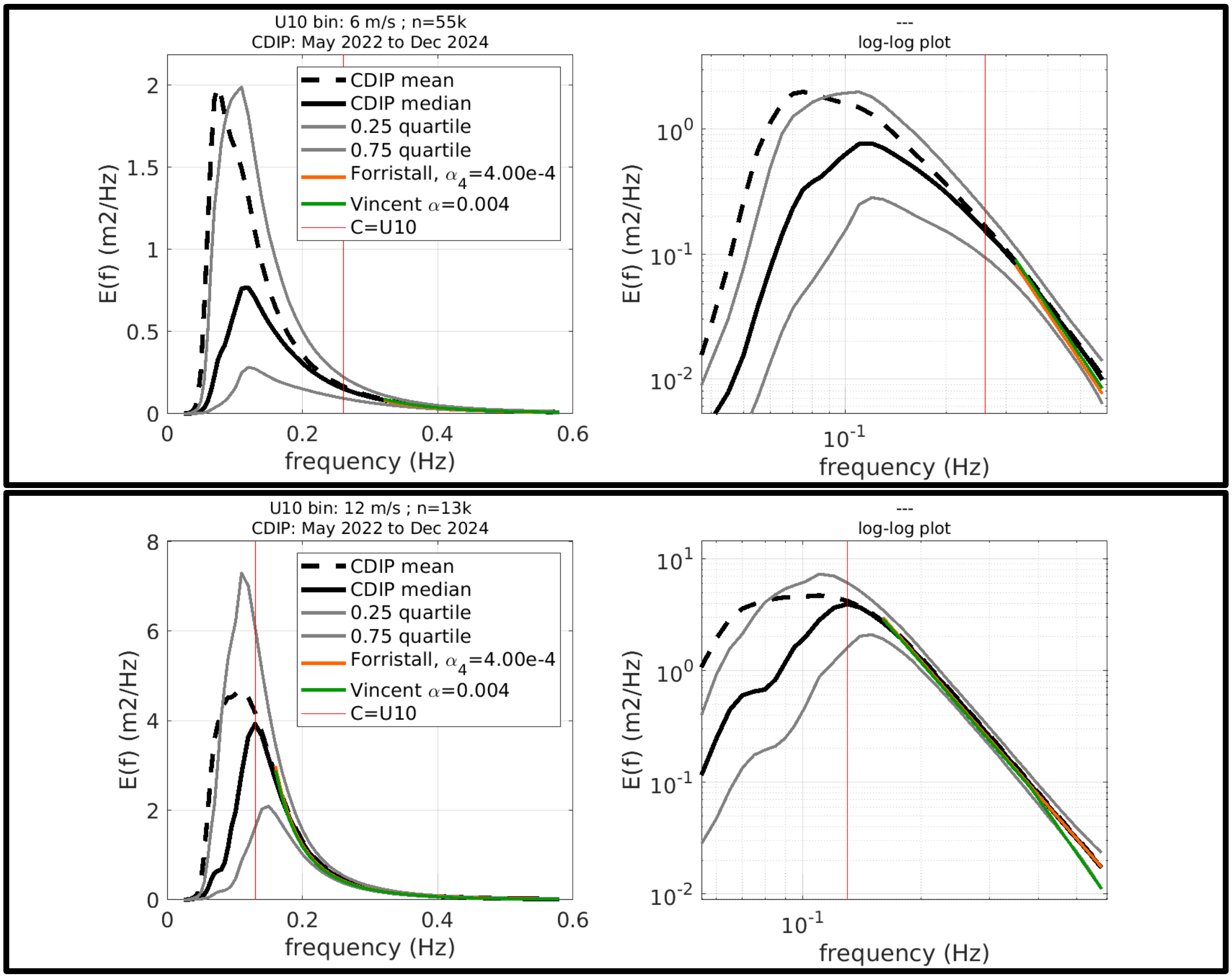

Figure 11. Like Figure 4, except that the parametric spectra are a non-standard (low) calibration of the Forristall model and a mid-range calibration of the Vincent model. The vertical red lines indicate $U_{10}/C = 1$.

In Figure 12, we compare nine of the parametric models at a relatively high wind speed ($U_{10} =$ 15 m/s), omitting the CDIP-based spectra for clarity. The nine models are:

1) the PM model, as previously shown in Figure 7 for lower wind speeds,
2) the Donelan model, with $\nu = 0.341$ (thus, a higher wind speed and younger wave age than was used in Figure 5 and Figure 10),
3) the Alves/PM model, as previously shown in Figure 7 for lower wind speeds,
4) the Battjes model with JONSWAP calibration, as previously shown in Figure 4 for lower wind speeds,
5) the BHP model, as previously shown in Figure 9 for lower wind speeds,
6) the Forristall model, as previously shown in Figure 9 for lower wind speeds,
7) the Vincent model with two values for α, 0.003 and 0.005, as previously shown in Figure 8 for lower wind speeds, and
8) the Resio model, with $\nu = 0.15$ (thus, a higher wind speed and older wave age than was used in Figure 10).

In Figure 13, we look at the same nine parametric models along a different dimensional axis: we look at only a single frequency (0.4 Hz, in the high frequencies), plotting the spectral level as a function of wind speed.

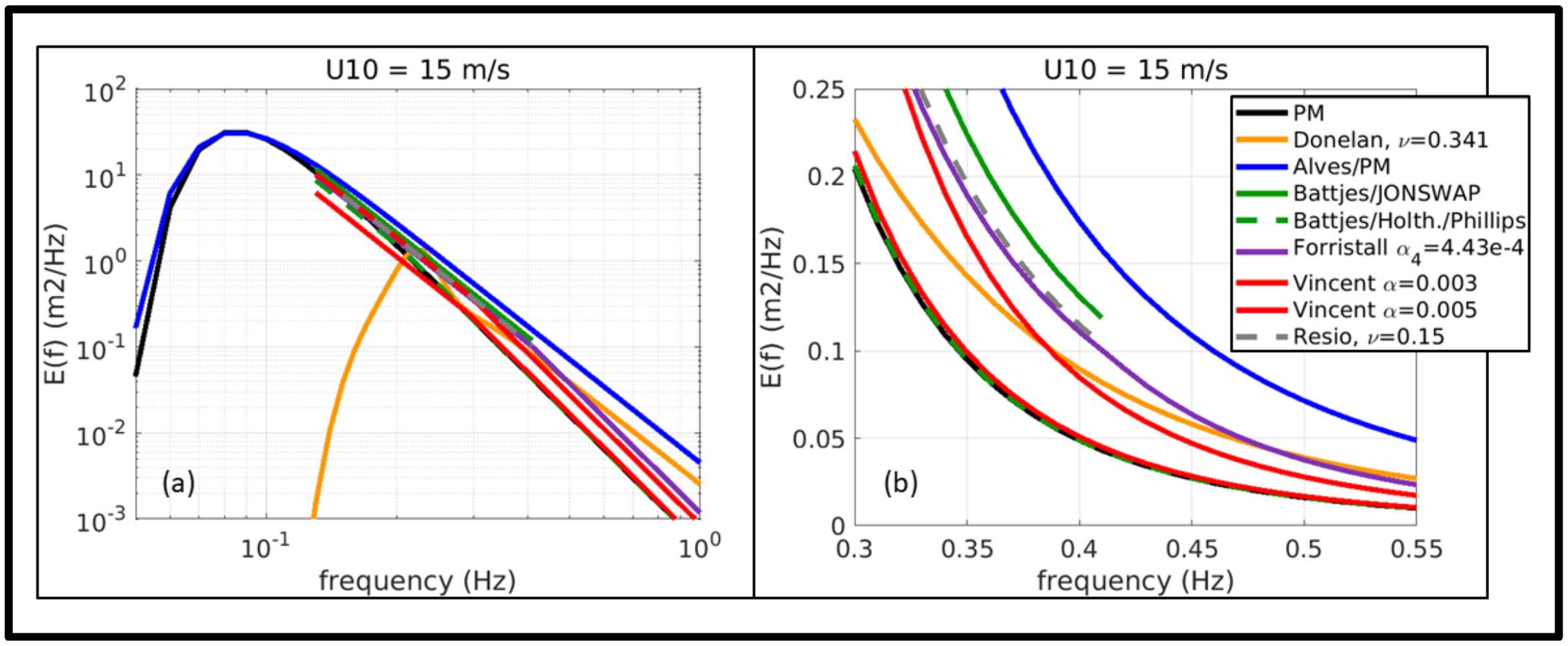

Figure 12. Comparisons of several parametric models for a near-gale force wind case, $U_{10} = 15$ m/s. Panel (a) full spectra, plotted on log-log scale. Panel (b): plotted on linear scale, showing only the high frequencies.

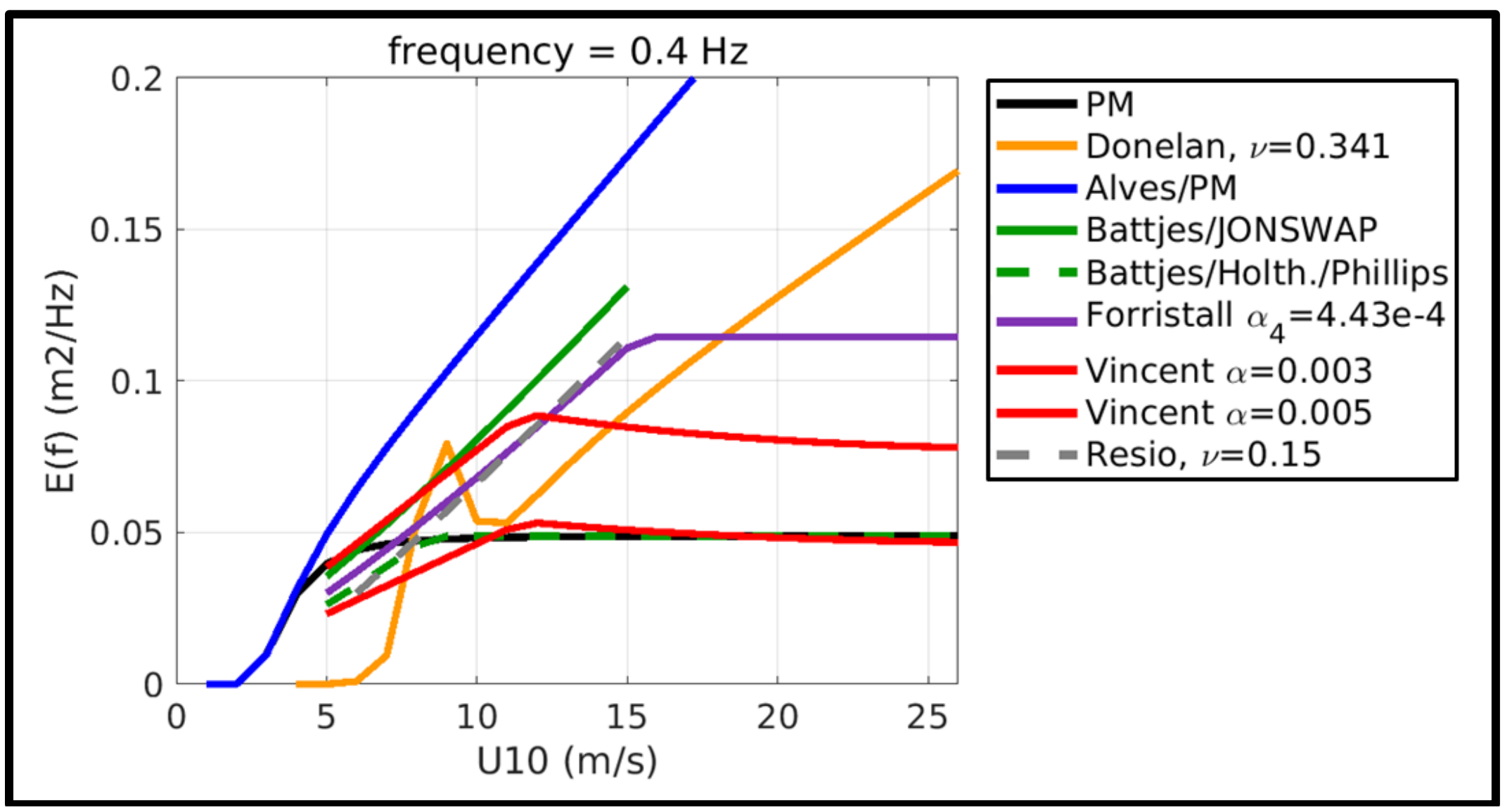

Figure 13. As in Figure 12, but plotting as a function of wind speed and only for a single frequency, 0.4 Hz.

## 5. Stokes drift predicted by candidate models and comparison to observational dataset

### 5.1. Formula for Stokes drift in non-deep water

The calculation of Stokes drift for a single wave train with amplitude $a$ is

$$\overline{u_L(z)} = g a^2 T(k, \omega, h, z), \text{ where}$$

$$T(k,\omega,h,z) = \frac{k^2 \cosh 2k(h+z)}{\omega \sinh 2kh}, \tag{9}$$

(Dean and Dalrymple 1993, equation 10.8 in Section 10.2.2, "Lagrangian Mass Transport"). [Here, $h$ is the distance from seafloor to mean sea level (MSL) and $z$ is the vertical coordinate from the MSL, positive up. So, for example, if the water depth is 100 m and one wants the Stokes drift one meter below the surface, one would use $h$ =100 m and $h + z = 99$ m in this formula.] This may be converted to spectral form by interpreting $a$ as $a_{rms} \approx H_{m0}/2\sqrt{2}$ for a spectrum of very small bandwidth $\delta f, \delta\theta$. With $m_0 = \iint E(f,\theta)dfd\theta$ and $H_{m0} = 4\sqrt{m_0}$, we have $a^2 = 2m_0$, and then

$$\overline{u_L(z)} = 2g \iint E(f,\theta)T(k,\omega,h,z)\, dfd\theta.$$

The general form for calculation of Stokes drift from a wave spectrum was given by Kenyon (1969) as

$$\vec{U}(z) = g \iint F(\vec{k}) \frac{\vec{k}}{\omega(k)} \left[\frac{2k \cosh 2k(z+h)}{\sinh 2kh}\right] d\vec{k}.$$

This can be rewritten as

$$\vec{U}(z) = 2g \iint F(k_x,k_y) \frac{\vec{k}}{k} T(k,\omega,h,z)\, dk_x dk_y,$$

or

$$U_x(z) = 2g \iint E(f,\theta)\cos(\theta)T(k,\omega,h,z)\, dfd\theta,$$
$$U_y(z) = 2g \iint E(f,\theta)\sin(\theta)T(k,\omega,h,z)\, dfd\theta.$$

Thus, we see that $\overline{u_L(z)}$ from Dean and Dalrymple (1993) and $\vec{U}(z)$ from Kenyon (1969) are consistent.

### 5.2. Stokes drift in deep water

In deep water (large $h$), the $T(k,\omega,h,z)$ term given by Kenyon (1969) is undefined, so in practice, we set a threshold, such as $kh = 6$, and use a simpler formula from Tang et al. (2007) in deep water:

$$T(k,\omega,z) = g^{-1}\omega k e^{2kz} \text{ or}$$
$$U_x(z) = 2\int_{f_{min}}^{f_{max}} E(f)\omega k e^{2kz} a_1(f)\, df, \tag{10}$$
$$U_y(z) = 2\int_{f_{min}}^{f_{max}} E(f)\omega k e^{2kz} b_1(f)\, df,$$

and of course, in deep water, wavenumber is simply $k = \sigma^2/g$, where $\sigma$ is the intrinsic frequency, and in absence of currents, the intrinsic frequency is the same as the frequency in a fixed frame of reference, $\sigma = \omega$.

### 5.3. Directionality in Stokes drift using buoy observations

The first order Fourier coefficients can be used to describe features of the directional distribution of the wave spectrum (e.g., Kuik et al. (1988)):

$$a_1(f) = \int_0^{2\pi} D(f,\theta)\cos(\theta)\, d\theta, \text{ and}$$
$$b_1(f) = \int_0^{2\pi} D(f,\theta)\sin(\theta)\, d\theta,$$

where $D(f,\theta)$ is the normalized directional distribution, defined using $E(f,\theta) = D(f,\theta)E(f)$ and $\int_0^{2\pi} D(f,\theta)d\theta = 1$. These Fourier coefficients, along with their second-order counterparts, $a_2(f)$ and $b_2(f)$, can be computed from buoy motion (e.g., Steele et al. 1985), so the waves community has long history of using them in calculations.

Following an unpublished document (David Wang, personal communication), we can rewrite the $x$-component:

$$U_x(z) = 2g \iint D(f,\theta)E(f)\cos(\theta)T(k,\omega,h,z)\, df d\theta, \text{ or}$$

$$U_x(z) = 2g \int_{f_{min}}^{f_{max}} E(f)T(k,\omega,h,z) \int_0^{2\pi} D(f,\theta)\cos(\theta)d\theta\, df, \text{ or}$$

$$U_x(z) = 2g \int_{f_{min}}^{f_{max}} E(f)T(k,\omega,h,z) \left[\int_0^{2\pi} D(f,\theta)\cos(\theta)d\theta\right] df, \text{ or}$$

$$U_x(z) = 2g \int_{f_{min}}^{f_{max}} E(f)T(k,\omega,h,z)a_1(f)\, df.$$

Similarly, for the $y$-component we get

$$U_y(z) = 2g \int_{f_{min}}^{f_{max}} E(f)T(k,\omega,h,z)b_1(f)\, df.$$

### 5.4. Stokes drift at the surface in deep water

At the surface, the $x$-component is:

$$U_{s,x} = U_x(z=0) = 2 \int_{f_{min}}^{f_{max}} E(f)\omega k a_1(f)\, df, \text{ or}$$

$$U_{s,x} = 2g^{-1} \int_{f_{min}}^{f_{max}} E(f)\omega^3 a_1(f)\, df, \text{ or}$$

$$U_{s,x} = 16\pi^3 g^{-1} \int_{f_{min}}^{f_{max}} E(f)f^3 a_1(f)\, df,$$

where '$s$' denotes "surface", and if we assume that all waves are travelling in the same direction, with no directional spread, then:

$$U_{s,s} = 16\pi^3 g^{-1} \int_{f_{min}}^{f_{max}} E(f)f^3\, df.$$

where '$s, s$' denotes "surface, simple". Taking the definition of spectral moments,

$$m_n = \int_{f_{min}}^{f_{max}} E(f)f^n\, df,$$

we have

$$U_{s,s} = 16\pi^3 g^{-1} m_3.$$

In the ocean waves literature, one encounters statements like "the third spectral moment is proportional to the surface Stokes drift in deep water". We emphasize that this assumes zero directional spread. See also Ardhuin et al. (2009), Appendix C.

### 5.5. Directionality in Stokes drift using WAVEWATCH III spectral files

The ocean wave model WAVEWATCH III ("WW3", Tolman 1991, WW3DG 2019) can provide directional spectra over a region, and as a forecast. Though we do not use WW3 spectra in this report, we explain here how its spectra could be used if one is working from files stored by the model. (In a separate effort, we are using this methodology for computation of particle drift.)

WW3 provides full directional spectra in its restart files. Provided one has a restart "reader", this can be used to compute Stokes drift profiles at every computational grid point. These files are very large (e.g., 3 GB per time field) and inconvenient (being in WW3-native binary on the three

WW3 computational grids). Thus, it is more common to instead store spectra in NetCDF files on an 'application grid' that has uniform grid spacing. The data in the restart files are the "two-dimensional" $(f, \theta)$ spectra, with one number per $(f, \theta)$. With $\Delta\theta = 10°$, this implies 36 numbers per frequency bin. Instead of storing this information in the NetCDF files, we store either three or five numbers per frequency bin that provide the energy level and directional moments at each frequency. In WW3 notation, the five parameters stored are $E(f)$, $\sigma_1(f)$, $\theta_1(f)$, $\sigma_2(f)$, $\theta_2(f)$, with the last two being second-order terms. In the WW3 instruction files, they are denoted as `EF`, `TH1M`, `STH1M`, `TH2M`, and `STH2M`, respectively. The $\theta_1$ parameter is the "mean direction for each frequency" and $\sigma_1$ is the "first directional spread for each frequency". Both are provided in degrees. Since there are three or five numbers per frequency, instead of 36, we refer to these data files at representing the "1.5-dimensional spectra". A representation using $E(f)$, $a_1(f)$, $b_1(f)$, $a_2(f)$, $b_2(f)$ is an alternative type of "1.5-dimensional spectra". The conversion between $\langle a_n, b_n \rangle$ and $\langle \sigma_n, \theta_n \rangle$ is effectively a conversion between Cartesian and polar representation of the same data. We find that it is more convenient to work with the Cartesian parameters. This also involves conversions between degrees and radians and between nautical and Cartesian angle convention. We use the following formulae to convert:

$$\theta_{1,cd} = (270 - \theta_1)\,(\pi/180)$$
$$r_1 = 1 - 0.5\left[\sigma_1\left(\frac{\pi}{180}\right)\right]^2$$
$$a_1 = r_1 \cos(\theta_{1,cd})$$
$$b_1 = r_1 \sin(\theta_{1,cd})$$

### 5.6. Comparisons of Stokes drift profiles

In this section, we compare Stokes drift profiles computed using our parametric models against those computed using the CDIP observational dataset, binned by wind speed. We also present the parametric models for $E(f)$ again, but most of them differ from their counterparts illustrated in Section 4: specifically, the "tail only" models that do not include a prediction of the spectral peak are modified. This includes: the Vincent et al. model, the Forristall model, the BHP model, the Battjes et al. model, and the Resio model. They are modified by attaching the PM spectrum to the left of the equilibrium range. This is denoted by adding "+PM" to the label. For example, the modified BHP model is "BHP+PM". In many cases, this attachment results in a discontinuity in $E(f)$.

The minimum and maximum frequencies used in the Stokes drift calculations (10) are determined by the buoy frequency limits: $f_{min}$ =0.025 and $f_{max}$ =0.56 Hz.

All calculations use a simple assumption for directionality, in WW3 notation: $\theta_1(f)$ =270° and $\sigma_1(f) = 0°$. The latter implies zero directional spread, resulting in a high estimate of the Stokes drift (e.g., Webb and Fox-Kemper 2011). Since the simplifying assumption is applied to both types of spectra, it does not bias our comparison.

We acknowledge here that we are only quantifying the impact of the method of estimating non-directional spectra on the accuracy of a Stokes drift prediction. We do not investigate how sensitive the accuracy of prediction is to the method used to estimate directional spread. In principle, we could use a frequency-dependent directional spread model such as Donelan et al. (1985) and compare it to results using frequency-dependent directional spread observed by the buoys, binned by wind speed. This is left for possible future analysis.

The twelve models listed in Table 2 ($U_{10} = 6$ m/s) and Table 3 ($U_{10} = 12$ m/s) are evaluated in this fashion, and results are summarized therein. For both wind speed examples, the analysis is broken into three frequency bands. In both cases, band 1 corresponds to the low-frequency peak in the observations, and band 3 corresponds to the high frequency tail. In the case of $U_{10} = 6$ m/s, the low-frequency peak is clearly a swell peak, and band 2 includes frequencies near the peak of the PM spectrum, i.e., the modeled "full development" wind-based spectrum. In the case of $U_{10} = 12$ m/s, the peak of the PM spectrum is in band 1, while band 2 includes frequencies near the peak of a hypothetical $\nu = 0.2$ (i.e., $U_{10}/C_p = 1.26$) wind sea spectrum, which only exists for the two "immature wind sea" parametric models that accept $\nu$ as an input parameter (Donelan et al. model and JONSWAP model).

The "band bias" values in columns #2, #3, and #4 in Table 2 and Table 3 refers to the bias in the wave energy spectrum. The last column (#5) on the two tables summarizes the outcome with respect to the surface Stokes drift (SSD). Since the SSD is roughly proportional to $m_3$ (i.e., weighted by $f^3$), the impact of higher frequencies (e.g., band 3) is disproportionately stronger than that of the lower frequencies (e.g., band 1). However, band 1 tends to contain much more energy than band 3. The integration of the bias across the bands (with the $f^3$ weighting) determines the bias outcome for SSD. By breaking the contributions into three parts, we illustrate that favorable SSD outcomes are almost always a result of fortuitous cancellation of errors, and poor SSD outcomes occur when either there is no cancellation of errors (less common), or when there is cancellation of errors, but they are not well balanced (most common).

The two tables do not report on the outcomes of the Stokes drift at depth. However, because they summarize the bias of the wave spectra, the tables can guide our expectations for the bias in Stokes drift at depth. Since the latter has the extra dimension ($z$), it is shown in figures (Figure 14 through Figure 17) rather than tables. At depth, the Stokes drift is reduced by the $e^{2kz}$ factor. Thus, while high frequencies are very important in SSD, their role rapidly disappears as depth increases. Conversely, low frequencies dominate the Stokes drift (SD) at depths.

We make the following observations about the $U_{10} = 6$ m/s case:

1. Swell has an enormous impact on the observed spectrum. The PM spectrum for mature wind sea, by its nature, does not include the swell peak. Thus, <u>all models</u> are low relative to the observed spectrum in band 1.
2. The non-prediction of swell means that <u>all models</u> underpredict SD at depth (see the $U_{10} = 6$ m/s panels in Figure 14 through Figure 17).
3. The SSD outcome depends on how bias in latter two frequency bands (frequencies near PM peak, and frequencies in tail) balance, reinforce, or overcorrect the negative bias in the first band (swell frequencies). This is illustrated in Table 2. The non-prediction of swell increases the likelihood that SSD will be underpredicted, but this doesn't happen for all models, since an overprediction of high frequencies can cancel this error (which happens for several models: PM, Alves/PM, Vincent $\alpha = 0.005$ + PM, Battjes/JONSWAP + PM). In one case (Alves/PM), the overprediction of high frequencies is so large that SSD is overpredicted.

We make the following observations about the $U_{10} = 12$ m/s case:

1. The PM "mature wind sea" peak frequency and the observed low-frequency peak both fall into band 1. For the models that rely on PM to provide the wind sea peak, the low-frequency peak is overpredicted (roughly twice the energy of the observed peak, overprediction in band 1). These models overpredict SD at depth.
2. The models of "immature wind sea" ($\nu$ =0.2) do not have the low-frequency peak in band 1 (underprediction). These models underpredict SD at depth.
3. As with the lower wind speed case, the SSD outcome depends on how bias in latter two frequency bands balance, reinforce, or overcorrect the bias in the first band (Table 3).

Table 2. Summary of outcomes for the $U_{10} = 6$ m/s case. The three bands refer to the outcome of prediction of spectral energy levels, $E(f)$. Band 1 contains the frequencies near the swell peak in the observations. Band 2 contains the frequencies near the peak of the PM spectrum. Band 3 is the high frequency tail. SSD is the surface Stokes drift based on integration of all frequencies. Gray shading indicates models that are not presented in the later plots.

| model | band 1 bias | band 2 bias | band 3 bias | SSD (i.e., integrated) outcome |
|---|---|---|---|---|
| PM | low | high | good | good |
| Alves/PM | low | high | very high | high |
| Vincent $\alpha = 0.003$ + PM | low | high | low | low |
| Vincent $\alpha = 0.005$ + PM | low | high | high | good |
| Forristall $\alpha_4$ =4.00× $10^{-4}$ + PM | low | high | slightly low | low |
| Forristall $\alpha_4$ =4.43× $10^{-4}$ + PM | low | high | good | slightly low |
| BHP + PM | low | high | slightly low | low |
| Battjes/JONSWAP + PM | low | high | good | good |
| Battjes/Mitsuyasu + PM | low | high | low | low |
| Resio + PM | N/A: non-unique result (largely reverts to PM) | | | |
| Donelan et al. wind sea | low | low | good | low |
| JONSWAP wind sea | low | low | low | very low |

Table 3. Summary of outcomes for the $U_{10} = 12$ m/s case. The three bands refer to the outcome of prediction of spectral energy levels. Band 1 is the frequencies near the peak in the observations, which in this case also corresponds to the peak of the PM spectrum, i.e., the modeled "full development" wind-based spectrum. Band 2 is the frequencies near the peak of the wind sea spectrum, i.e., the wind-based model spectrum for $\nu = 0.2$. Band 3 is the high frequency tail. SSD is the surface Stokes drift based on integration of all frequencies. Gray shading indicates models that are not presented in the later plots.

| model | band 1 bias | band 2 bias | band 3 bias | SSD (i.e., integrated) outcome |
|---|---|---|---|---|
| PM | high | N/A | low | good |
| Alves/PM | high | N/A | very high | very high |
| Vincent $\alpha = 0.003$ + PM | high | N/A | low | slightly low |
| Vincent $\alpha = 0.005$ + PM | high | N/A | good | high |
| Forristall $\alpha_4 = 4.00 \times 10^{-4}$ + PM | high | N/A | good | slightly high |
| Forristall $\alpha_4 = 4.43 \times 10^{-4}$ + PM | high | N/A | good | high |
| BHP + PM | high | N/A | low | good |
| Battjes/JONSWAP + PM | high | N/A | slightly high | high |
| Battjes/Mitsuyasu + PM | high | N/A | slightly low | good |
| Resio + PM | high | N/A | good | slightly high |
| Donelan wind sea | low | good | good | good |
| JONSWAP wind sea | low | good | low | low |

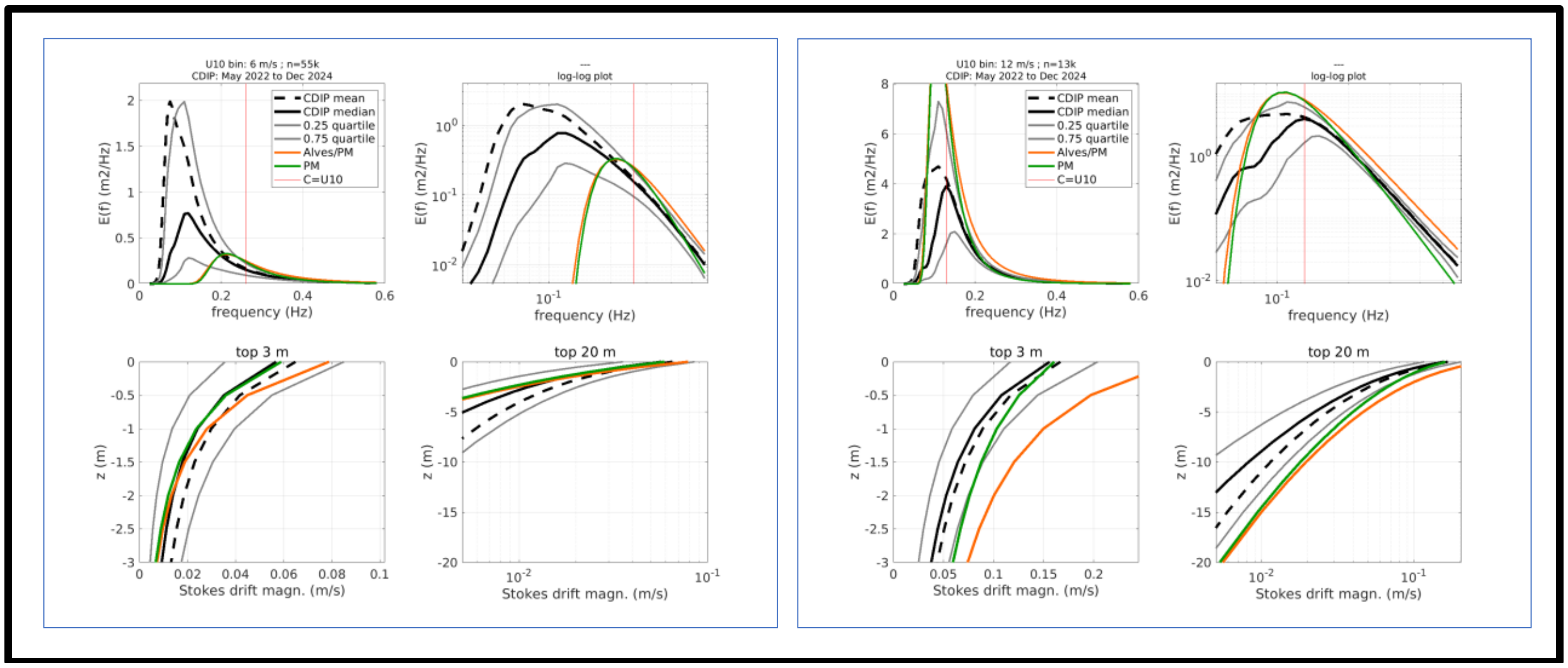

Figure 14. Comparison of wave spectra (top four plots) and Stokes drift profiles (bottom four plots). The four plots on the left correspond to $U_{10} = 6$ m/s, and the four plots on the right correspond to $U_{10} = 12$ m/s. For each set of Stokes drift profiles, one sub-plot is for the top 3 m of the water column (noted in the title of the sup-plot), plotted in linear scale, and a second sub-plot shows the top 20 m of the water column (again, noted in the title of the sub-plot), where Stokes drift is plotted in log scale. In this figure, the computations using the observed CDIP spectra are compared against the PM and Alves/PM parametric models. The vertical red lines indicate $U_{10}/C = 1$.

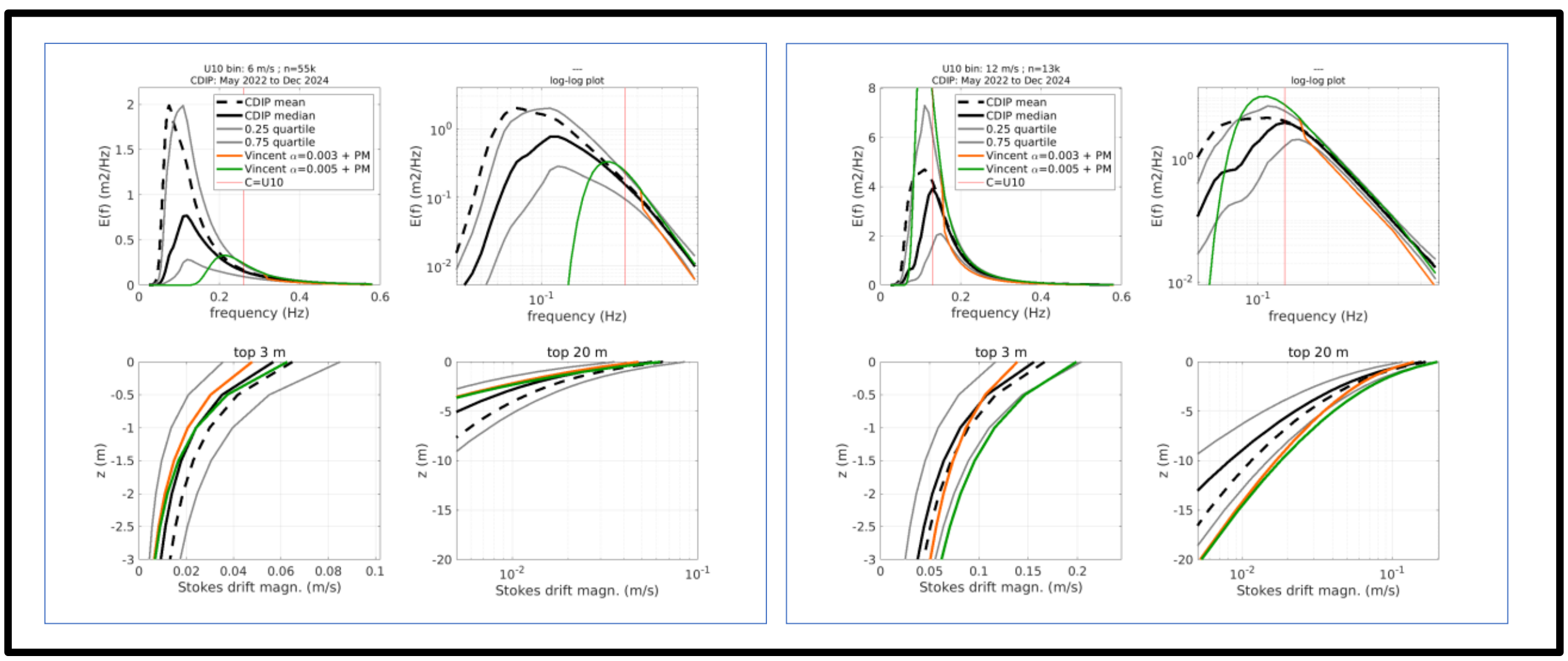

Figure 15. Like Figure 14, but comparing with the Vincent et al. model results.

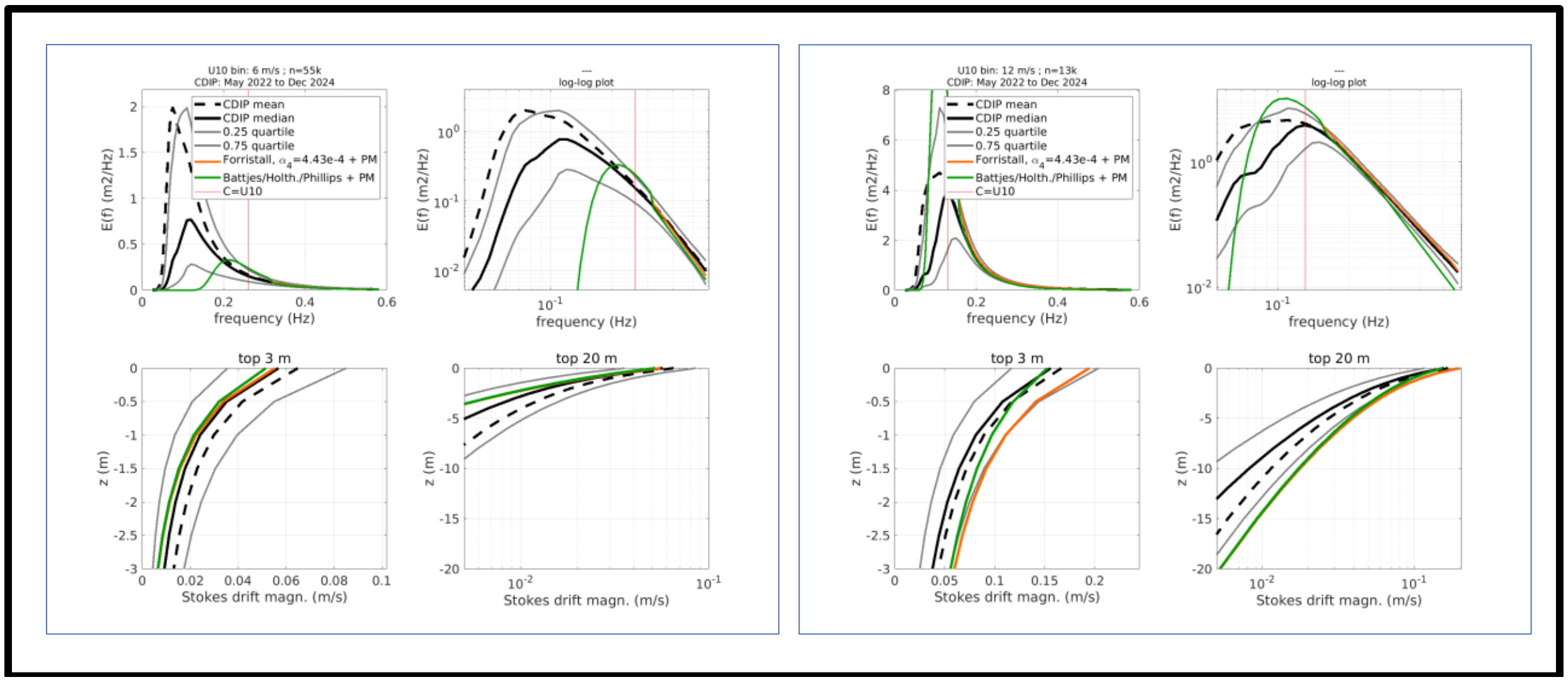

Figure 16. Like Figure 14, but comparing with the Forristall and BHP model results.

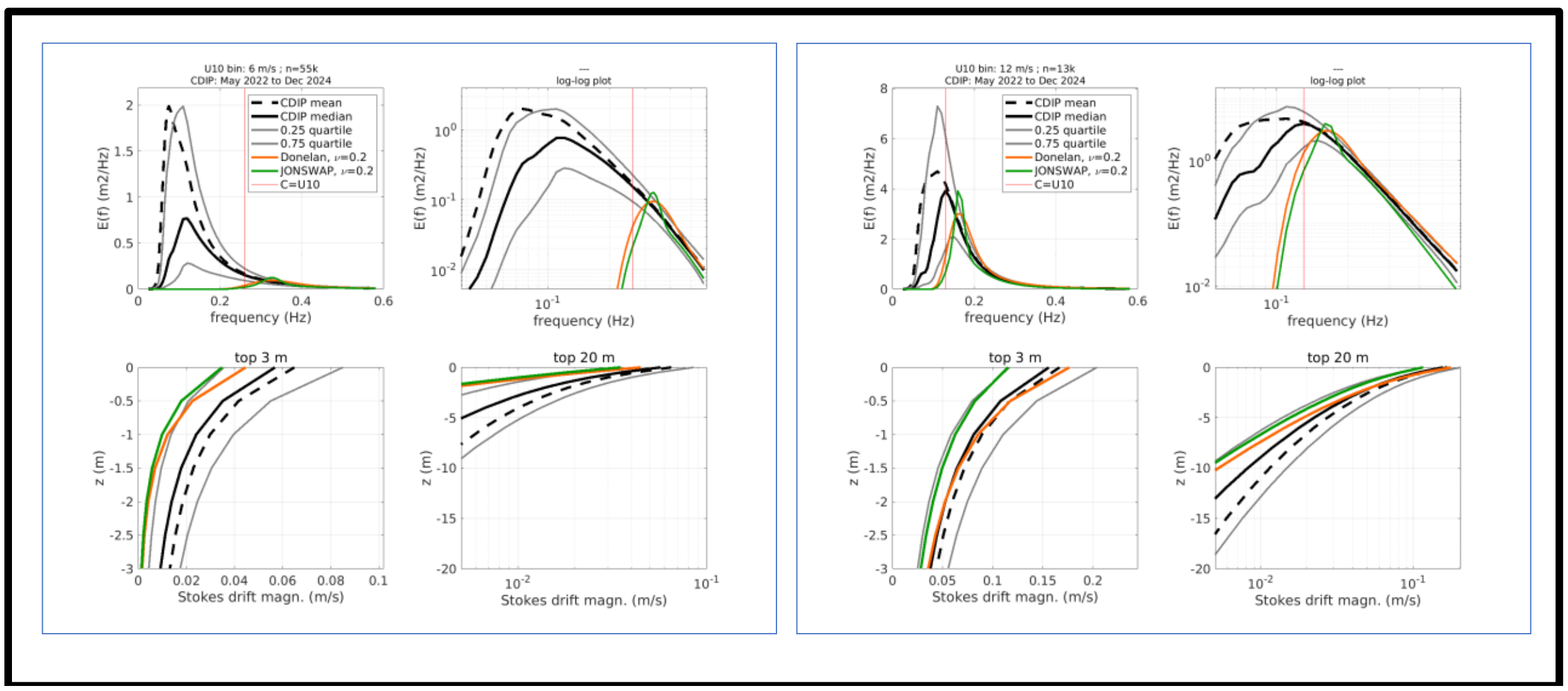

Figure 17. Like Figure 14, but comparing with the Donelan et al. and JONSWAP model results.

### 5.7. Sensitivity to maximum frequency

As noted already, maximum frequency used in the Stokes drift profiles, Figure 14 to Figure 17, was $f_{max}$ =0.56 Hz. This was used to be consistent with the buoy frequency limits. In practice, however, the maximum frequency can be higher than frequencies measured by the CDIP buoys. In the case of the USN's operational global WW3 (both standalone and coupled), $f_{max}$ =0.73 Hz at time of writing. This can be extended by attaching a diagnostic tail with $f^{-5}$ slope, which is a defensible approximation until frequencies around 50 Hz, where the decline becomes more rapid (Elfouhaily et al. 1997, Figure 5a).

In Figure 18, we explore the sensitivity of Stokes drift at four different depths (0, 5, 20, and 50 cm), five different values of $f_{max}$ (0.73, 0.97, 1.42, 9.6, and 94 Hz), and wind speeds up to $U_{10}$ =16 m/s are shown. The "Forristall, $\alpha_4 = 4.43 \times 10^{-4}$+PM" model is used. We find significant sensitivity of surface Stokes drift $U_{ss}$ to the $f_{max}$ setting (unsurprising, given the rough proportionality, $U_{ss} \propto m^3$). However, even at 5 cm depth, the sensitivity to $f_{max}$ is weak, and by 50 cm, it is hardly visible.

In the case of predicting the drift of small items, films, etc., on the ocean surface, Figure 18 implies that $f_{max}$ is a major consideration. In the case of larger objects, e.g., a person in the water with center of mass 30 to 45 cm below the surface, the figure implies that $f_{max}$ is of minor importance. This is also true when providing Stokes drift to an ocean model, unless the ocean model has very high vertical resolution near the ocean surface.

To estimate the contribution of frequencies beyond $f_{max}$ to Stokes drift, one could use equation (11) from Breivik et al. (2014), replacing $\omega_p$ with $2\pi f_{max}$. This assumes a Phillips spectrum level in the tail.

See also Section 4 in Breivik et al. (2014): "4. High-frequency contribution to the profile".

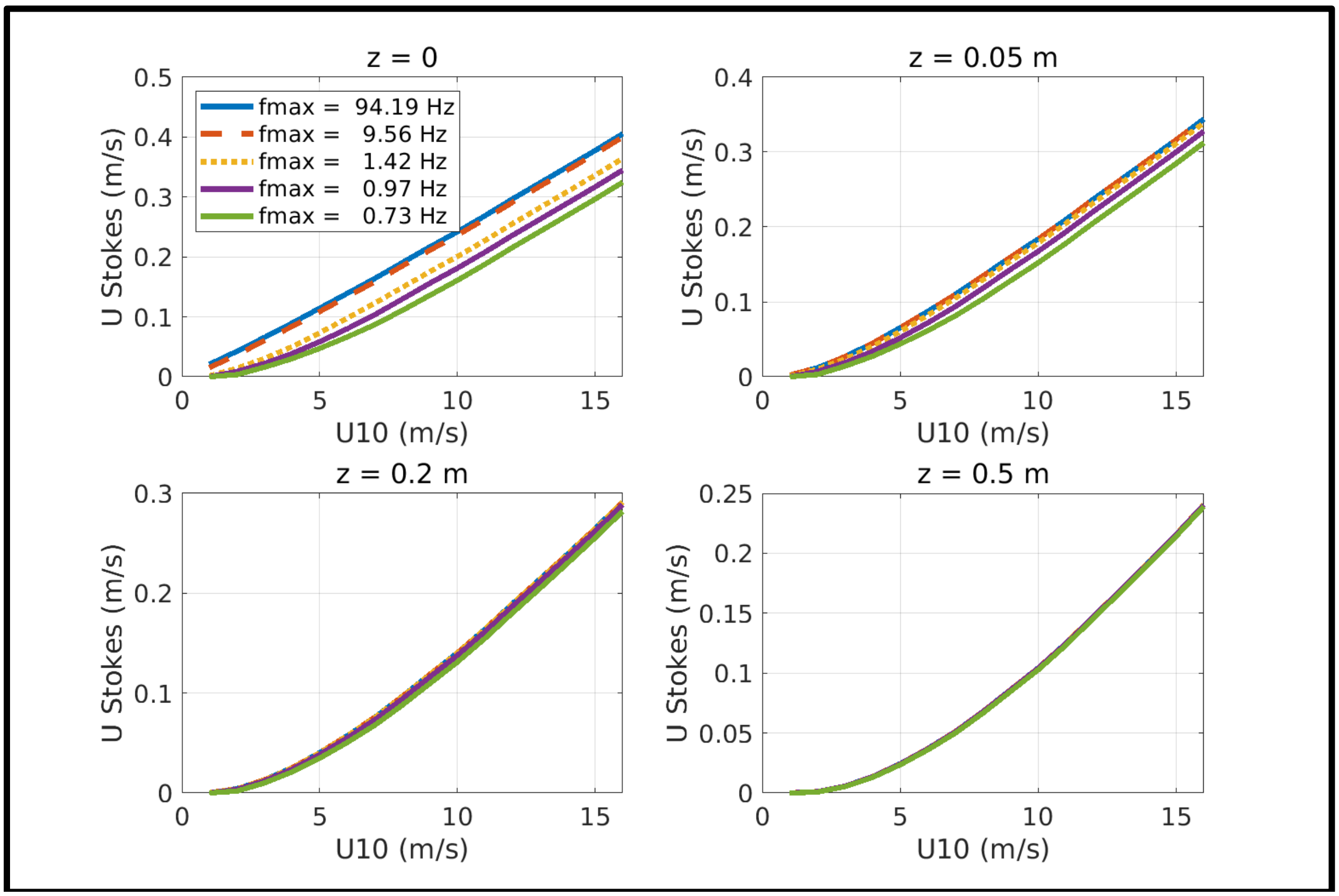

Figure 18. Dependence of Stokes drift on vertical position, maximum frequency, and wind speed. The "Forristall, $\alpha_4 = 4.43 \times 10^{-4}$+PM" parametric model is used to compute $E(f)$, which is then used to compute Stokes drift.

## 6. Background: Prior testing of a global ocean model with parameterized Stokes drift

### 6.1. Methods

The HYbrid Coordinate Ocean Model (HYCOM, Chassignet et al. 2003) includes five different embedded ocean mixed layer sub-models that can be used to simulate the variability of the mixed layer. One of those models is the K-Profile Parameterization (KPP) (Large et al., 1994). Zhitao Yu (2024, personal communication, henceforth denoted as "ZY") incorporated a theoretical Stokes drift (wind wave) in KPP to explore the impact of the Stokes drift in the ocean mixed layer using only wind speed, i.e., lacking any information from a dynamical wave model. Wind speed is readily available to HYCOM, as it already uses it for other components of its governing equations.

#### *6.1.1. Parametric wave model and wind-speed-based Stokes drift calculation*

The Phillips parametric model is used as a starting point, $F(\omega) = \alpha_P g^2 \omega^{-5}$. As noted in Section 4.1, this spectrum has no dependence on wind speed. For uni-directional waves in deep water, the Stokes drift is: $U(z) = 2\int_{\omega_{min}}^{\omega_{max}} F(\omega)\omega k e^{2kz}\, d\omega$.
Thus, $U(z) = 2\int_{\omega_{min}}^{\omega_{max}} \alpha_P g^2 \omega^{-4} k e^{2kz}\, d\omega = 2\int_{\omega_{min}}^{\omega_{max}} \alpha_P g \omega^{-2} e^{2\omega^2 z/g}\, d\omega$.
Following Breivik et al. (2014) (henceforth, "BJB"), the spectral peak is used for the lower bound of the integration,

$$U(z) = 2\int_{\omega_p}^{\omega_{max}} \alpha_P g \omega^{-2} e^{2\omega^2 z/g}\, d\omega \tag{11}$$

and $\alpha_P = 0.0083$. Insofar as $\omega_p$ depends on wind speed, this makes $U(z)$ depend on wind speed.

BJB used $\omega_{max} = \infty$ in equation (11) (this is their equation 10) and integrated it analytically. They did not, however, apply this result directly. They use it to develop an approximate "exponential integral profile" that matches the shape of (11). This approximation requires information from a wave model (the Stokes transport and surface Stokes drift), so it was not an option for ZY. ZY's solution was to use equation (11), and construct an estimate for the $\omega_p$ that it requires. ZY started with the equations for non-dimensional fetch and frequency as given in the JONSWAP report (Hasselmann et al. 1973, page 24a): $\tilde{x} = \frac{gF}{U_{10}^2}$ and $\tilde{f} = \nu = \frac{f_p U_{10}}{g}$ where $F$ is fetch in meters. On page 29 of that report, $\tilde{f}_p = 3.5\tilde{x}^{-0.33}$. Thus, $f_p \approx 3.5\left(\frac{g^2}{U_{10}F}\right)^{1/3}$ or $\omega_p \approx 22\left(\frac{g^2}{U_{10}F}\right)^{1/3}$. Unfortunately, this introduces a new variable, $F$. ZY assumed a constant value: $F = 100$ km. ZY used $\omega_{max} = (2\pi)1.173$ Hz and integrated (11) numerically. ZY's method is discussed further in Section 6.3.

## 6.2. Results: Impact of wave-enhanced mixing on HYCOM via KPP

### *6.2.1. Uncoupled model results with wind-speed-based Stokes drift*

Here, results from 3-months (April-June 2023) of HYCOM simulations with and without Stokes drift are presented.

Mixed layer depth (MLD) is defined by a positive density difference of 0.15 kg/$m^3$ between the surface and a given depth if both temperature and salinity profiles are available, or a negative temperature difference of 0.5°C between the surface and a given depth if only temperature profiles are available. Global maps of MLD, for June 2023, with and without Stokes drift effect, and the difference between the two maps are shown in Figure 19. The impact of the Stokes drift in the deepening (yellow and red colors) of the MLD along the Antarctic Circumpolar Current, around the 30°N, and some coastal regions in the Western Pacific Marginal Seas are distinctly depicted in Figure 19c. Equally notable is the shallowing (light and dark blue colors) impact of Stokes drift on the MLD along some coastal regions in the Antarctic (Figure 19c). Correspondingly, time series of daily globally averaged MLD shows a clear deepening of the MLD from the inclusion of the Stokes drift in KPP (Figure 20). Then, how does the modeled MLD (with and without Stokes drift) compare with the observed MLD? An error analysis (model results vs. observations) examining MLD was performed for the eight different regions of Figure 21. Histograms of MLD mean error (ME) and root mean square error (RMSE) during May-June 2023 show that the simulation with Stokes drift generally has lower ME and RMSE than the simulation without Stokes drift (Figure 22 and Figure 23). Conceivably, from the results presented here, the inclusion of the Stokes drift on KPP produces a deepening of the MLD.

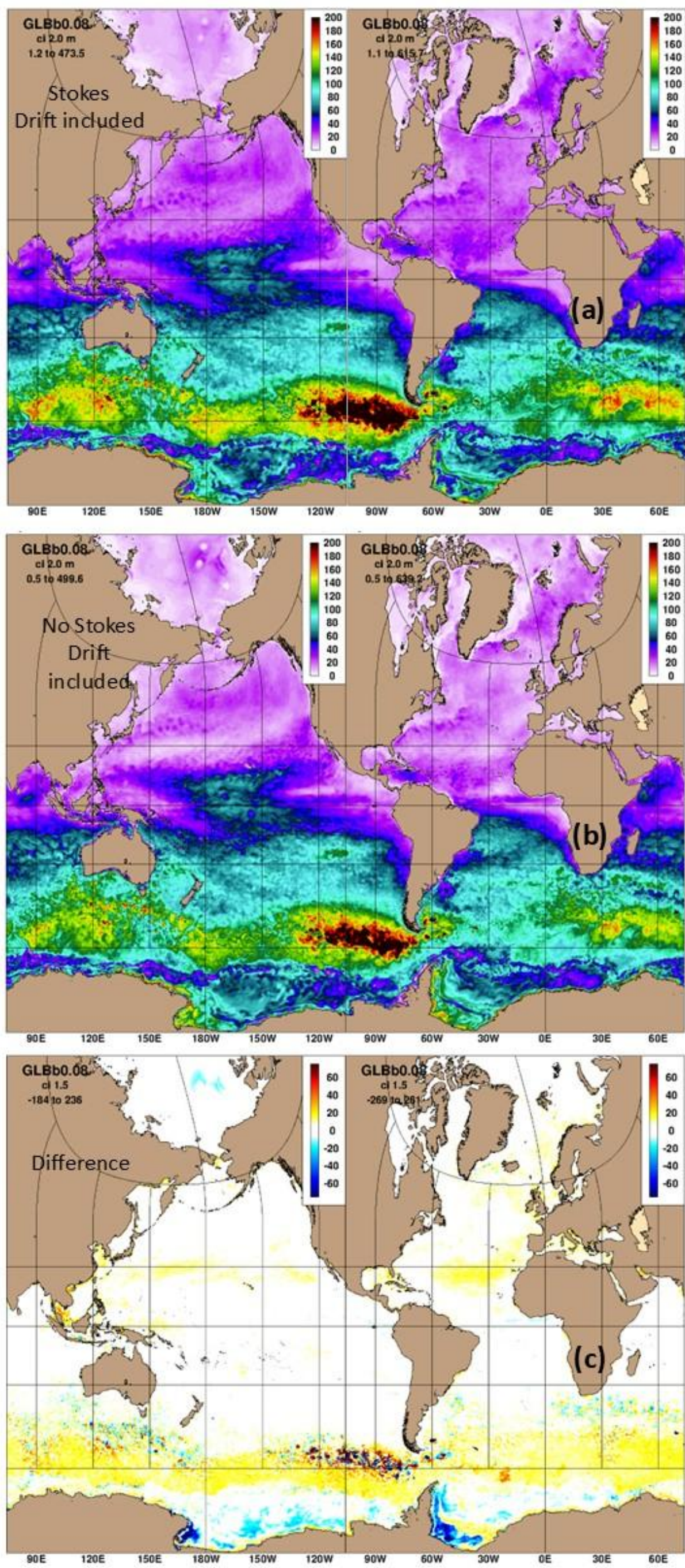

Figure 19. Mean mixed layer depth (color contours in meters) of June 2023, (a) with, and (b) without Stokes drift effect, and (c), the difference ((a) minus (b)).

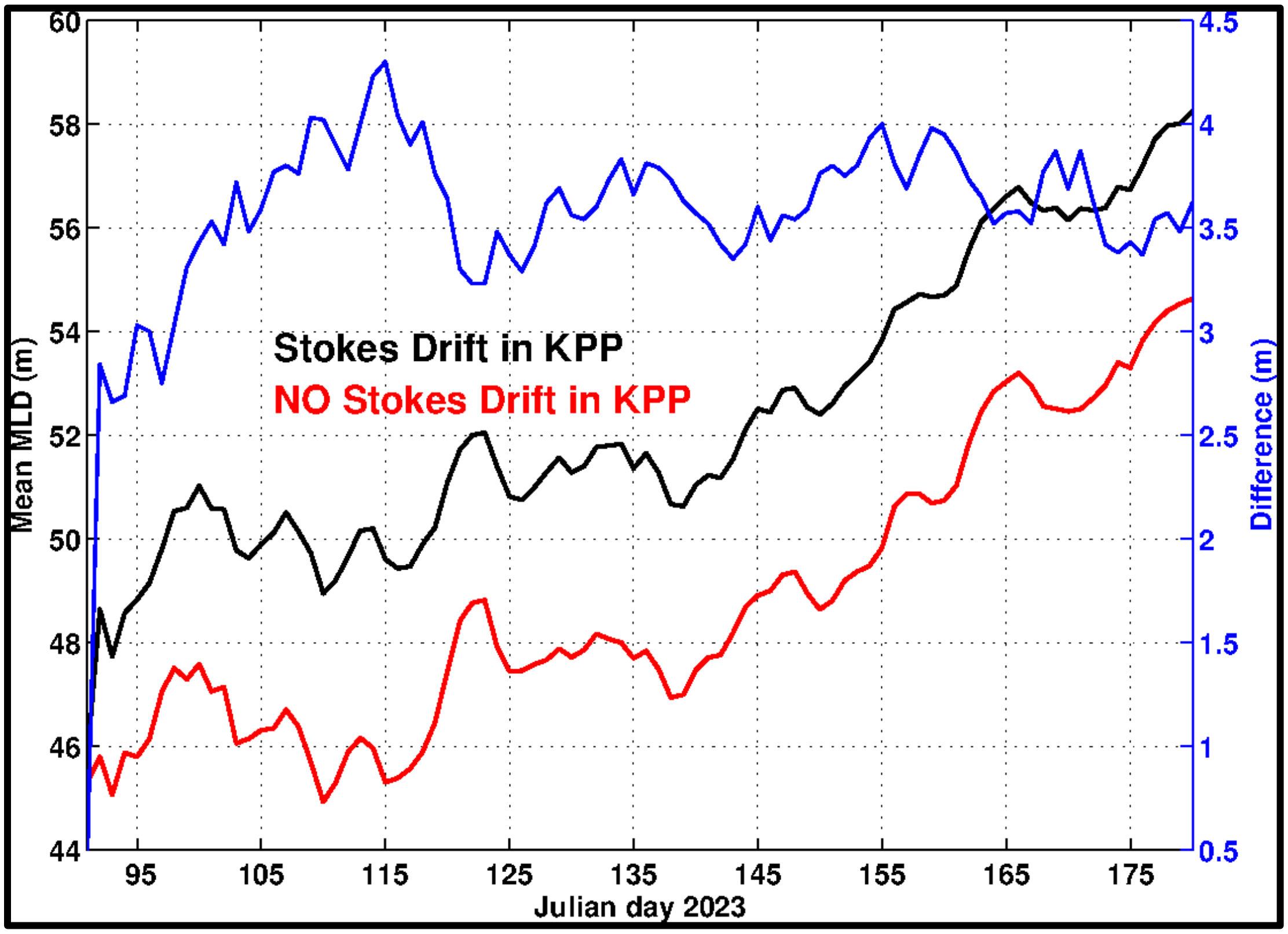

Figure 20. Globally averaged mixed layer depth (MLD) with (black line), and without (red line) Stokes drift, and the difference (blue line).

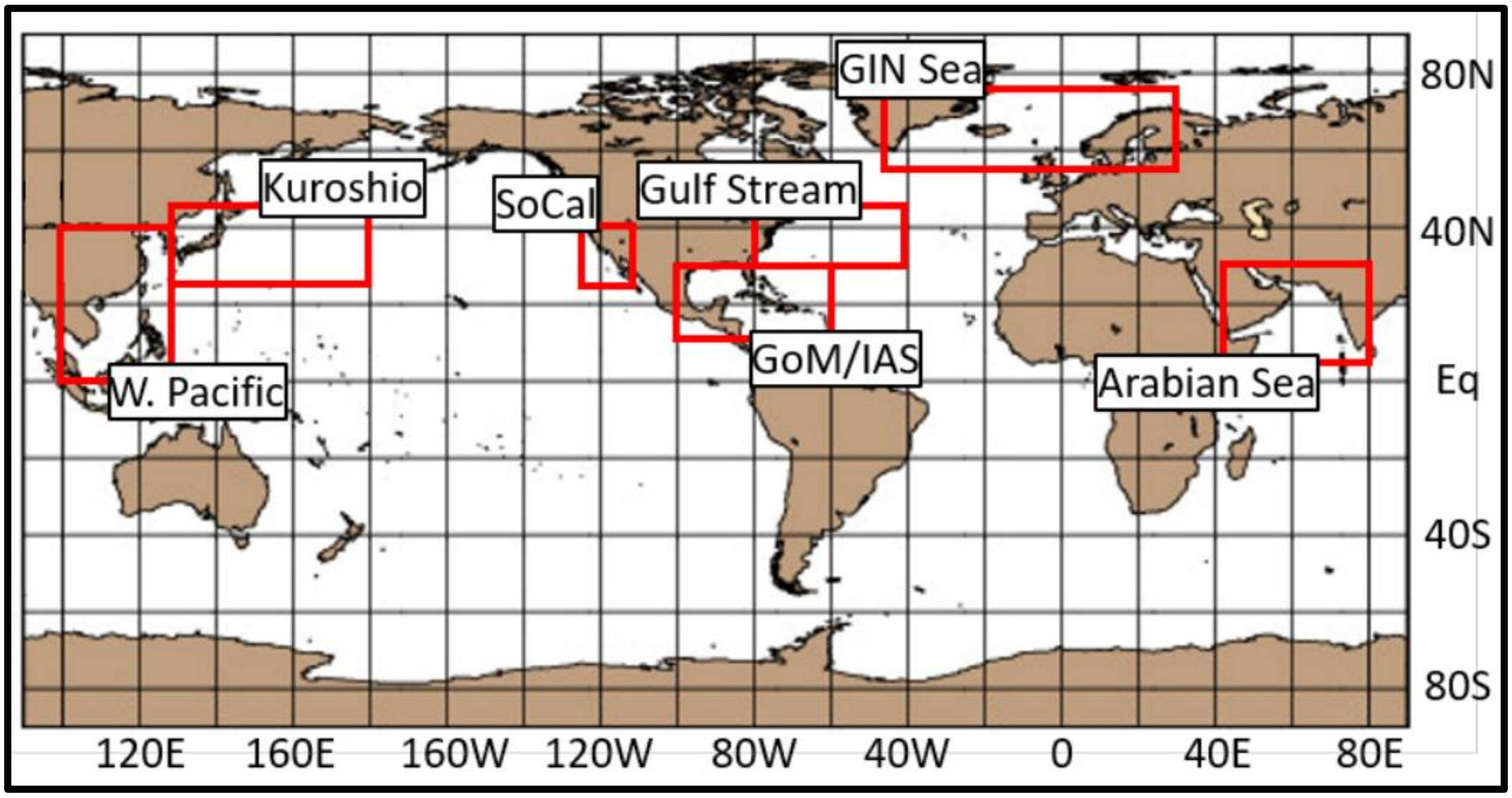

Figure 21. The ocean analysis regions used in this section that include the Globe (180°W-180°E, 90°S-90°N), Western Pacific marginal seas (100-130°E, 0-40°N), Kuroshio Extension (130°E-180°, 25-45°), Arabian Sea/Persian Gulf/Gulf of Aden (44-80°E, 5-30°N), Gulf Stream (80-40°W, 30-45°N), Southern California (SoCal) (125-110°W, 25-40°N), Greenland/Iceland/Norwegian (GIN) Seas (45°W-30°E, 55-75°N), and Gulf of America/Mexico (GoM)/Intra-Americas Sea (IAS) (100-60°W, 10-30°N).

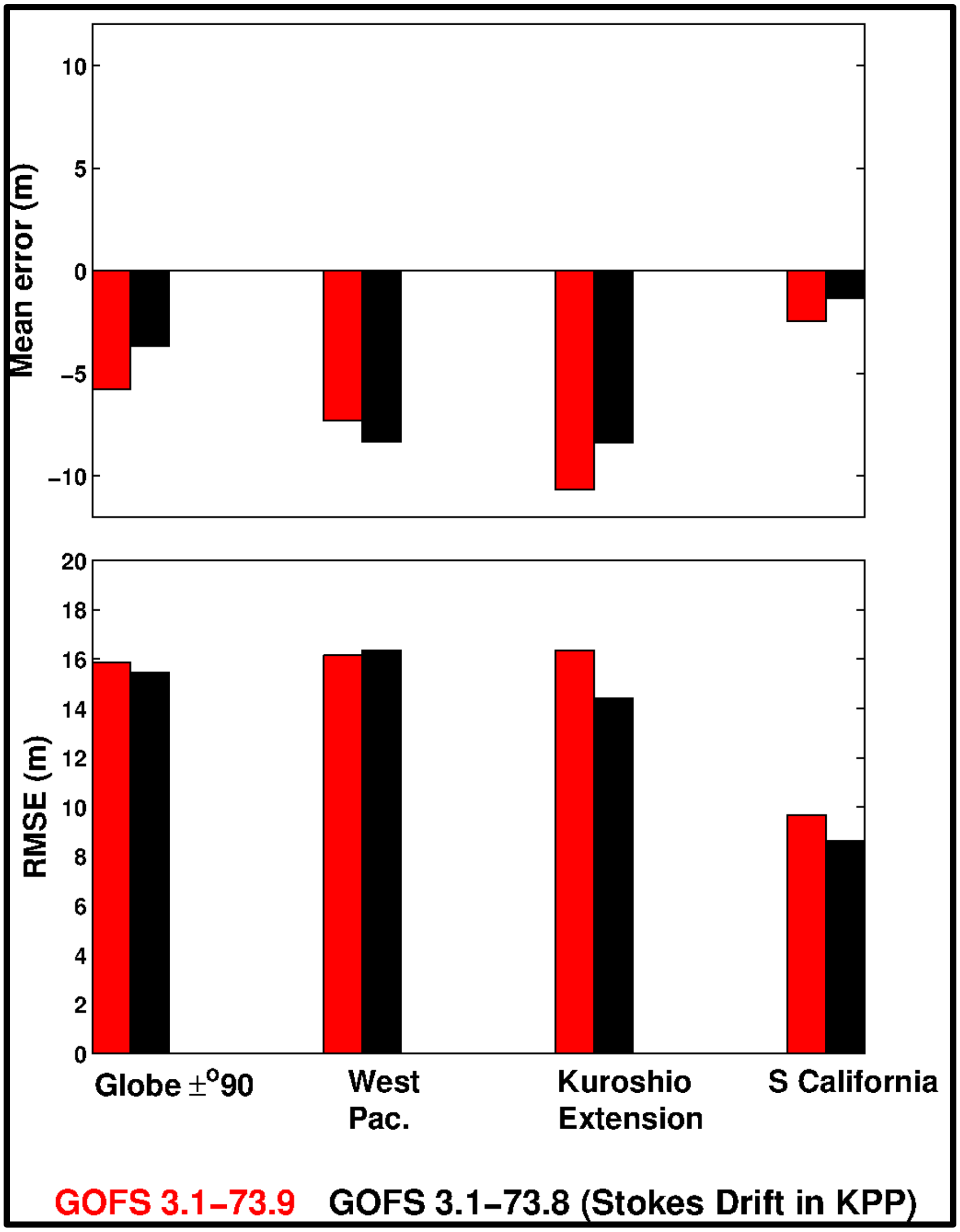

Figure 22. Histograms of mixed layer depth mean error (top panel) and RMSE (bottom panel) (both in m) of profile observations against modeled profiles for the indicated regions during May-June 2023. The "GOFS" referred to here is the "Global Ocean Forecasting System" (Metzger et al. 2017) for ocean circulation and sea ice.

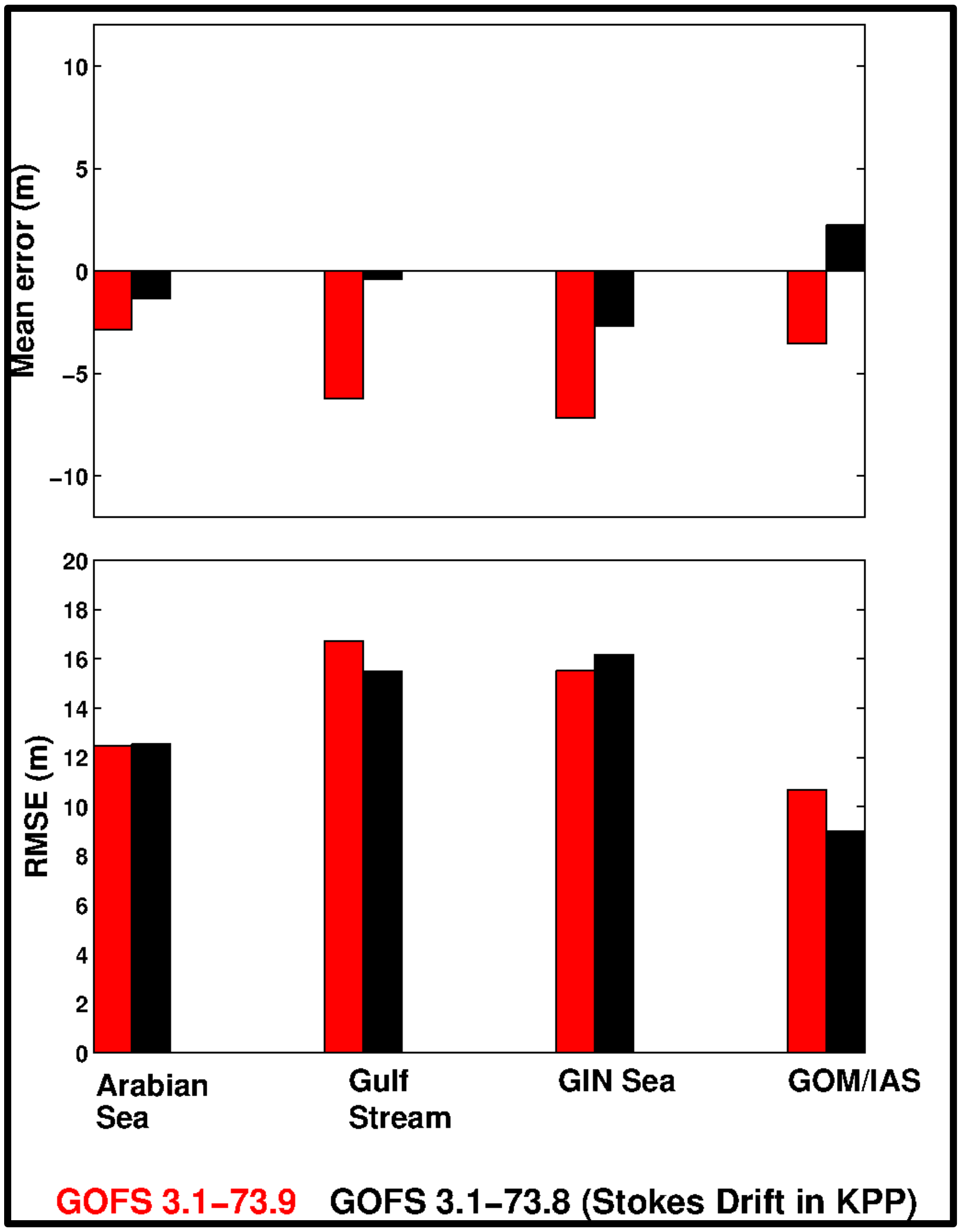

Figure 23. Histograms of mixed layer depth mean error (top panel) and RMSE (bottom panel) (both in m) of profile observations against modeled profiles for the indicated regions during May-June 2023.

#### *6.2.2. Coupled model results with wind-speed-based Stokes drift*

In the Section 6.2.1, we presented results from experiments with an uncoupled HYCOM model. That analysis was on mixed layer depth (MLD). The sea surface temperature (SST) was never plotted. Based on the apparent success of these experiments in MLD performance, the method was applied to a coupled HYCOM model: the ocean component of the Navy Earth System Prediction Capability (NESPC) global modeling system, in cycling runs for November 2023, performed for validation purposes. Unfortunately, when SST was first plotted for these cycling runs, anomalies were apparent in the equatorial regions of the east Pacific and the Atlantic (Figure 24). Differences are also visible in some warm regions at higher latitudes (15°-30°N), such as the Indian Ocean and the South China Sea.

Given the strange and unexpected appearance of the equatorial features, these runs were discontinued (reverting to the “no Stokes drift” configuration), pending further investigation. At time of writing, that investigation has not yet been carried out. However, one (#1) hypothesis was that the anomalies are caused by non-optimal Stokes drift profiles, i.e., a result of shortcomings/assumptions of the wind-speed-based Stokes drift. This motivated the study documented in this report.

Other possible explanations exist, of course: #2: that the anomalies are caused by some unknown technical problem with the implementation of the wind-speed-based Stokes drift, or #3: that there is some unknown problem using the wave-enhanced KPP in coupled mode.

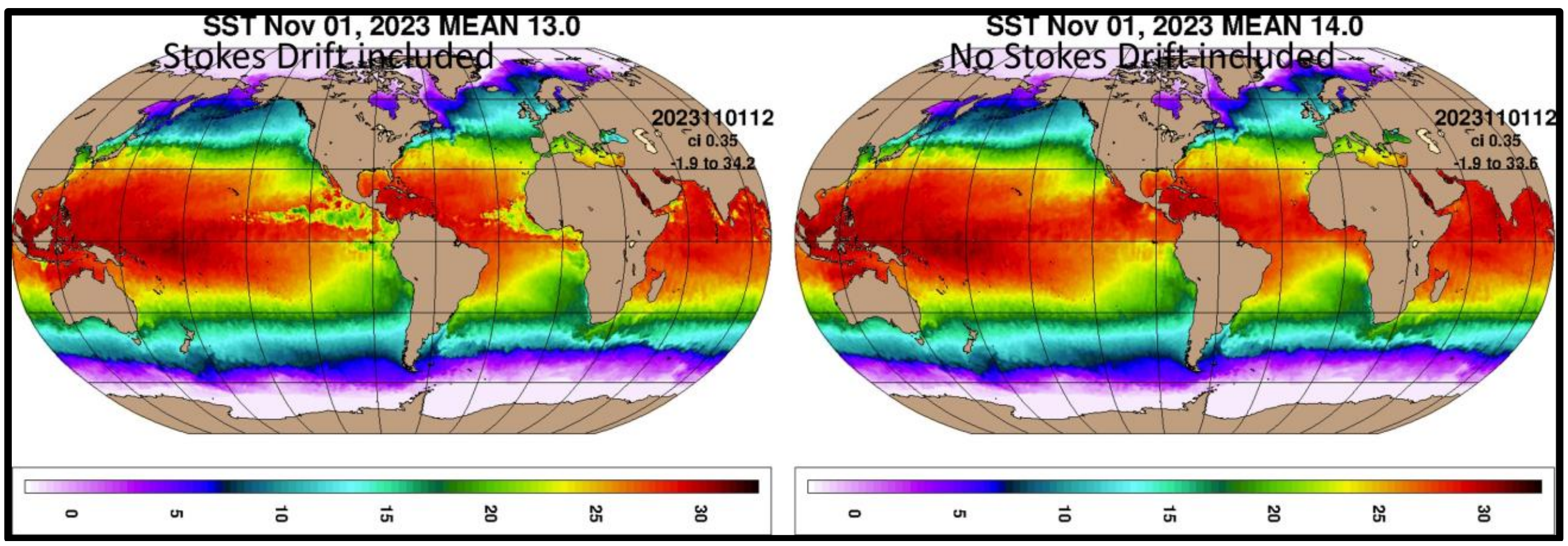

Figure 24. Average values of SST (°C) from HYCOM coupled experiments, November 01, 2023. Left panels: With wind-speed-based Stokes drift in KPP. Right panels: Without Stokes drift in KPP.

### *6.2.3. Revisiting the uncoupled model results with wind-speed-based Stokes drift*

To test hypothesis #3, we revisited the results from the uncoupled case with wind-speed-based Stokes drift. If the problem is associated with running in coupled mode, then we would expect to find no temperature anomalies in the uncoupled model results. An example plot of mixed layer temperature (MLT) is shown in Figure 25. The anomalies are clearly visible. This appears to eliminate hypothesis #3.

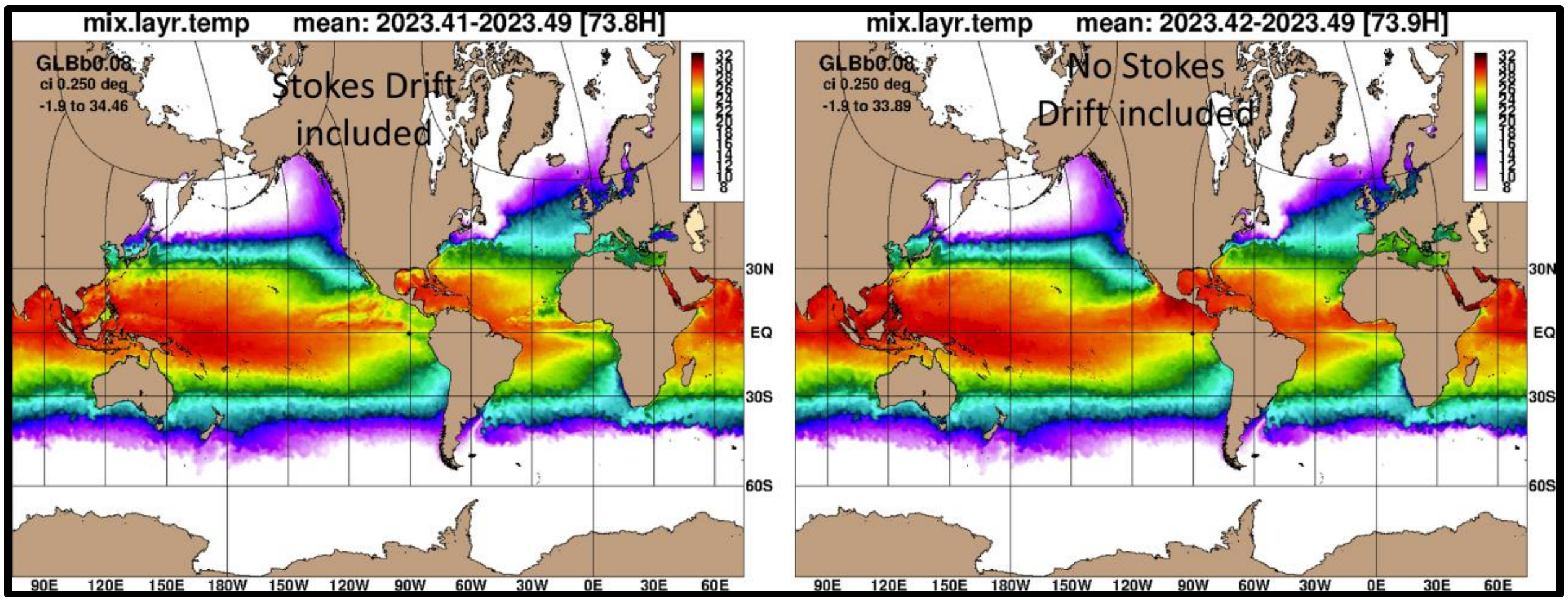

Figure 25. Average values of Mixed Layer Temperature (°C), June 2023, from uncoupled HYCOM experiments. Left panels: With wind-speed-based Stokes drift in KPP. Right panels: Without Stokes drift in KPP.

#### *6.2.4. Uncoupled model results with wave-model-based Stokes drift*

A one-year reanalysis (calendar year 2017) was conducted using the GOFS framework with data assimilation (Fan et al. 2026). In this run, a unified upper ocean mixing parameterization with the effect of Langmuir enhanced turbulence is used in HYCOM. Stokes drift profiles on the HYCOM horizontal grid are provided to HYCOM using files created by post-processing of WW3 output (Stokes frequency spectra on the WW3 application grid). Model results are analyzed in the eight domains shown in Figure 21. The new run has led to significant error reduction in both temperature and salinity predictions compared to standard GOFS 3.1, including an overall 24.4% reduction in global temperature RMSE (0.266 ºC vs. 0.352 ºC) and reaching 47.7% and 47.4% in the W. Pacific and Kuroshio domain respectively; and overall 26.4% reduction in global salinity RMSE (0.053 psu vs. 0.072 psu) and reaching 53.1% and 51.6% RMSE reduction in W. Pacific and Kuroshio respectively.

Systematic biases in temperature and salinity were also substantially reduced globally by 44.9% and 45.5% respectively. The W. Pacific (Kuroshio) domain again shows higher improvements with 68.7% (58.1%) and 70.3% (55.2%) respectively on temperature (salinity) predictions. Overall, the addition of Langmuir enhanced turbulence is not only able to adequately mix the warm/fresh water from the surface with the entrained cold/salty pycnocline water in the mixed layer, but its impact can also reach all the way into the deep ocean and beyond the depth of the pycnocline and thus reduce the cold/fresher bias in the model in the entire water column.

An example of a monthly mean mixed layer temperature is shown in Figure 26.

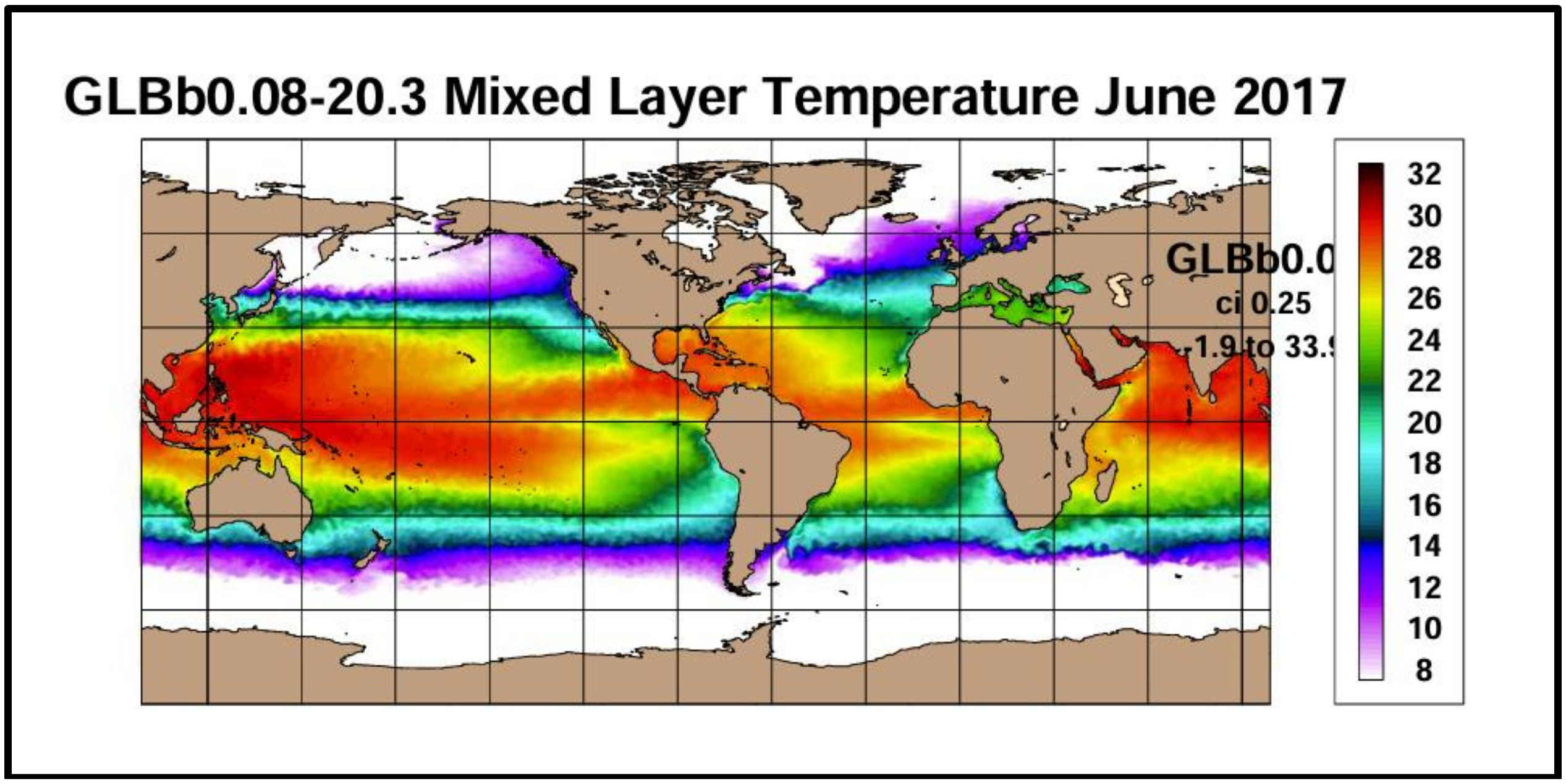

Figure 26. Monthly mean mixed layer temperature, in ºC, for June 2017, from the uncoupled model run with wave-model-based Stokes drift.

### *6.2.5. Discussion of SST and MLT anomalies*

The uncoupled HYCOM experiments shown in Figure 26 performed using Stokes drift profiles taken from the wave model WW3 do not show temperature anomalies like those in Figure 24 and Figure 25. This is consistent with the hypothesis that the anomalies are linked to shortcomings/assumptions of the wind-speed-based Stokes drift, which we will discuss in Section 6.3. However, the second hypothesis—that the anomalies are caused by some unknown technical problem with the implementation of that scheme—is still possible.

## 6.3. Discussion of Stokes drift approximation

As described in Section 6.1.1, the wind-speed-based Stokes drift estimate uses the JONSWAP relation $f_p \approx 3.5\left(\frac{g^2}{U_{10}F}\right)^{1/3}$, with fetch $F$ set to a fixed $100 \times 10^3$ m. A shortcoming of this approach is that $f_p$ becomes unbounded. JONSWAP is intended for growing seas, e.g., $\tilde{f} = \nu = f_p U_{10}/g$ values from 0.2 to 0.7, while $\nu = 0.13$ for the fully-developed PM seas. The relation $f_p \approx 3.5\left(\frac{g^2}{U_{10}F}\right)^{1/3}$ implies $\nu = 3.5\left(\frac{U_{10}^2}{gF}\right)^{1/3}$, and so there is a cross-over with PM at $U_{10} = 7.16 \times 10^{-3}\sqrt{gF}$, and with $F$=$100 \times 10^3$ m, that cross-over is at $U_{10} = 7.1$ m/s. For lower wind speeds, the $\nu$ is smaller (so wave age is larger) than that of PM.

The outcome of this shortcoming is that the wind-speed-based Stokes drift will be overestimated when $U_{10} < 7.1$ m/s. This has a simple fix: set a lower limit on $f_p$ that corresponds to the $f_p$ of PM for that wind speed.

A second shortcoming in the implemented approach is that it does not account for directional spread. Directional spread is expected to reduce Stokes drift by 16 to 19% (according to findings of BJB, page 2441) or by ~20% (Clarke and van Gorder (2018), page 4684). For a simple wind-speed based approximation, a 20% reduction would be appropriate, following the latter reference.

A third shortcoming is that it uses uses $E(f) \propto f^{-5}$ in the equilibrium range, even though the equilibrium range is well-known to be $E(f) \propto f^{-4}$. A fourth shortcoming is that it starts the equilibrium range at $f_p$, which is lower than the $1.5f_p$ cited by Tsagareli (2009) and others. Figure 27 illustrates how this results in a large overestimation of the spectrum in the region $f_p < f < 1.5f_p$. Both of these matters are discussed in Section 4 of this report. A potential fix for this would be to use a model such as the "Forristall, $\alpha_4 = 4.43 \times 10^{-4}$+PM" parametric model, but we do not wholeheartedly endorse this approach, since as we showed in Section 5.6, such a parametric model tends to provide accurate Stokes profiles only when it benefits from fortuitous cancellation of errors.

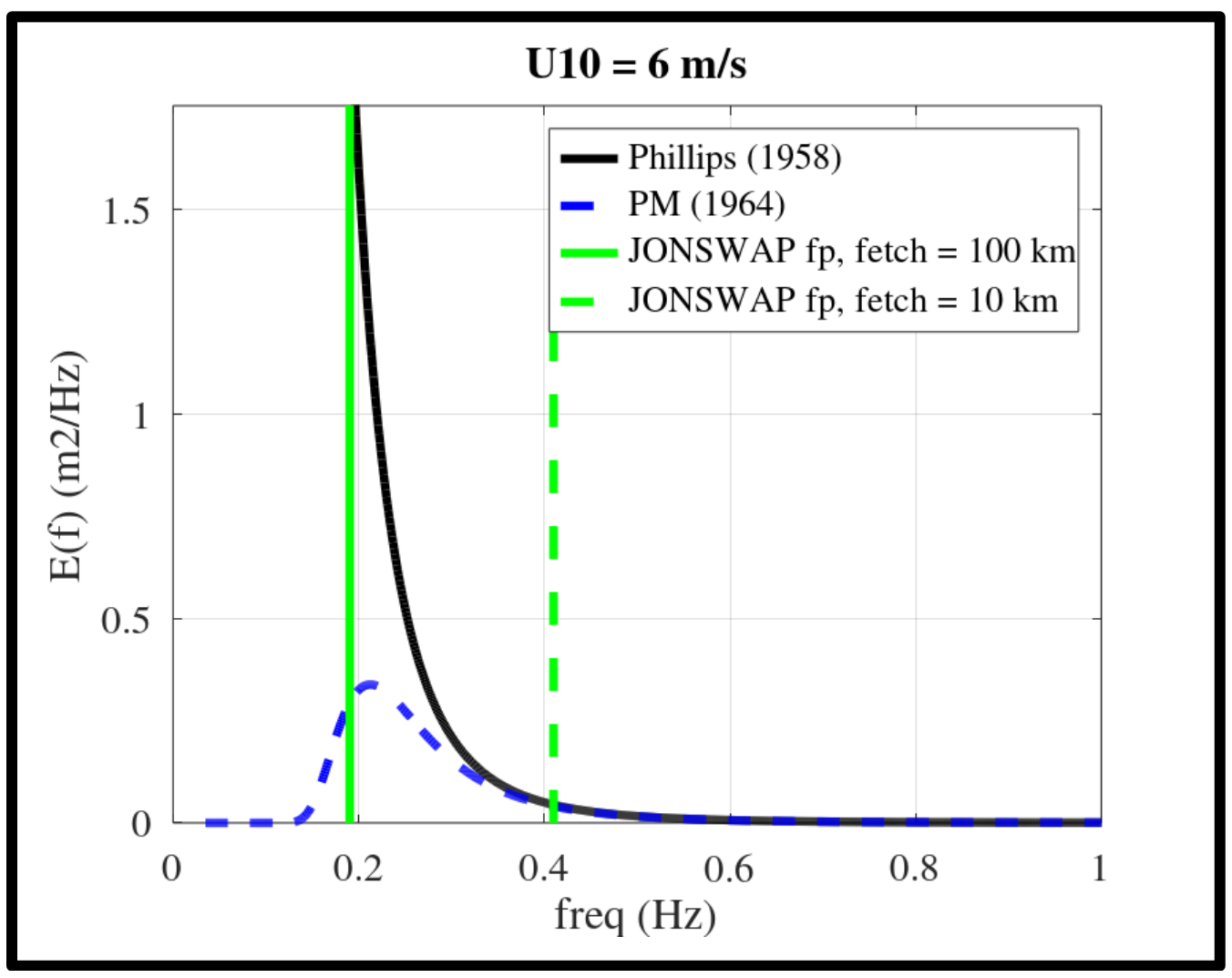

Figure 27. Demonstration of how extending the Phillips spectrum to the spectral peak results in a high estimate of $E(f)$, relative to that of PM. The black line is the Phillips spectrum, with a minimum frequency $f_{min}$ at the $f_p$ predicted by JONSWAP for $U_{10} = 6$ m/s and $F = 100$ km (this $f_p$ is indicated with a solid vertical green line). The blue dashed line is the PM spectrum, which is effectively the Phillips spectrum, with an exp() term added (see Section 4.7). To illustrate sensitivity of the JONSWAP $f_p$ to $F$, the dashed vertical green line indicates the $f_p$ predicted by JONSWAP for $U_{10} = 6$ m/s and $F = 10$ km. The spectrum indicated with the black line is an example of what would be used in the wind-speed-based Stokes drift as implemented in HYCOM prior to this study.

## 7. Stokes drift in a coupled regional model

In this section, we discuss the use of Stokes drift in the USN's regional coupled modeling system, COAMPS (Coupled Ocean Atmosphere Mesoscale Prediction System, Campbell et al. (2010), Smith et al. (2013), Doyle et al. (2014)). The USN global coupled modeling system,

ESPC ("Earth System Prediction Capability") described in Metzger et al. (2023) does not include two-way coupling between its wave component, WW3, and the other modeling components. This is primarily due to the technical complexity of exporting fields from multiple WW3 grids via the "Earth System Modeling Framework" (ESMF). This technical limitation was very recently addressed by Dr. Tim Campbell (NRL 7322). In contrast, two-way coupling has been a feature of the USN regional coupled modeling system for many years, e.g., Smith et al. (2013). The COAMPS permits export of data from one wave model grid, though the nesting can include multiple grids. Further, more than one ocean model grid can receive data from the exporting wave model grid. Like in ESPC, this is done via ESMF. COAMPS users can select between two wave models, SWAN ("Simulating WAves Nearshore", Booij et al. 1999) and WW3. The former is preferred for higher resolution applications (e.g., 500 m spacing), while the latter is preferred for larger regions (e.g., with 6 km spacing); see Rogers et al. (2025) for examples of uncoupled wave models run within COAMPS.

Figure 28 shows examples of surface Stokes drift from a coupled COAMPS reanalysis (D'Addezio et al. 2025). The left panel shows the "surface Stokes drift" (sic) from the ocean model component, NCOM ("Navy Coastal Ocean Model", Martin (2000)), after being imported from the wave model and re-gridded by COAMPS/ESMF. This, however, is not actually an estimate of the SSD. Rather, it is the Stokes drift at a vertical level corresponding to the middle of the top layer of the ocean model, so perhaps it should be called the "top layer Stokes drift" (TLSD?) instead. Since the top layer here is 1 m thick, the plot shows the Stokes drift at 0.5 m. The right panel shows the actual SSD exported from WW3.

The stark difference between TLSD and SSD has important implications. As we discussed in Section 5.7, inclusion/omission of high frequency wave components has a strong impact on SSD, but very little impact on Stokes drift at depth, even shallow depths like 0.5 m (Figure 18).

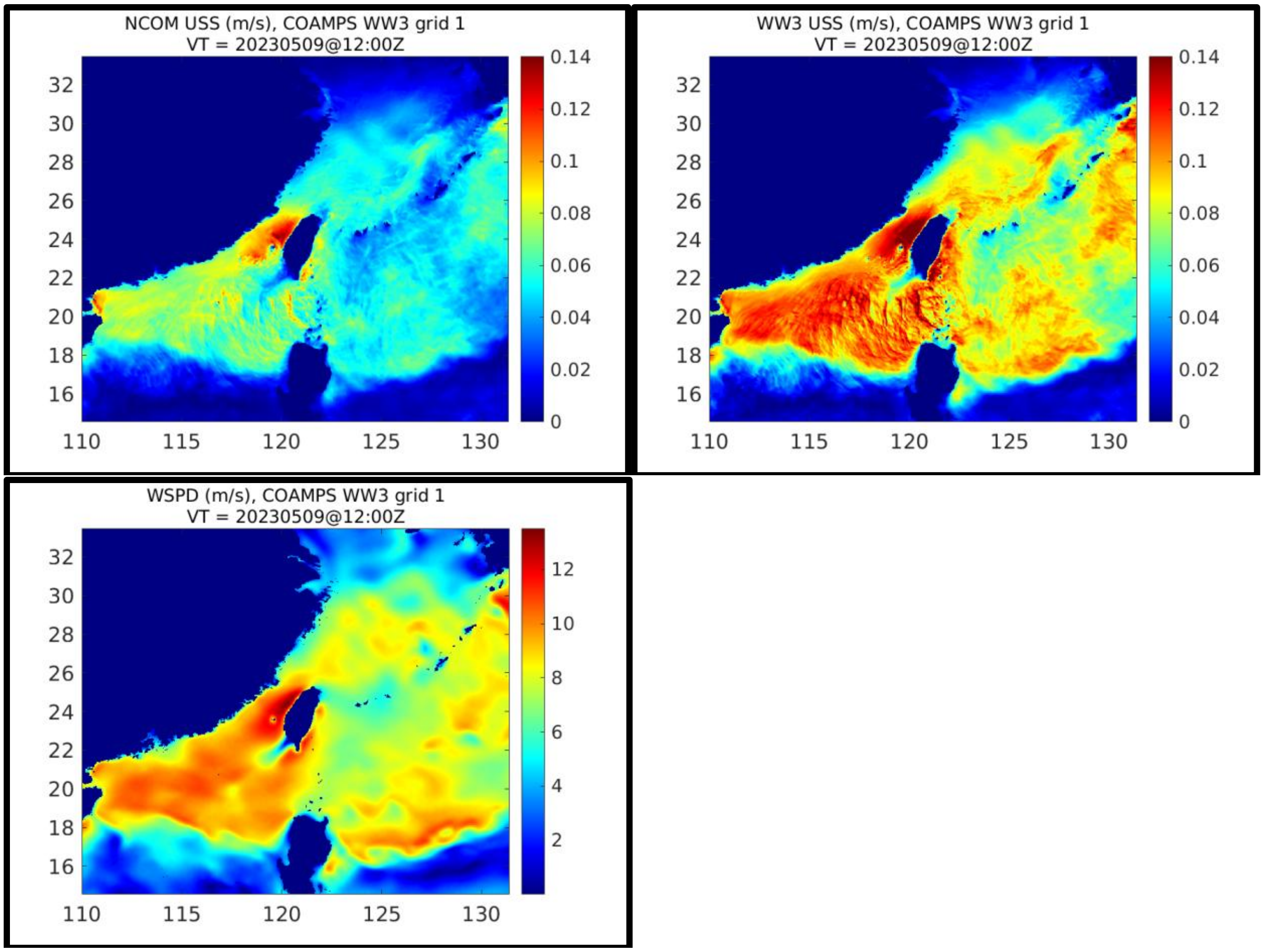

Figure 28. Example of surface Stokes drift (SSD) in a regional coupled model. Upper left: SSD used by regional ocean model, NCOM. Upper right: SSD computed by regional WW3. Lower figure: the wind speed used as forcing to WW3. The simulated time and date are 1200 UTC 9 May 2023.

In Figure 29, we take the SSD from the WW3 field shown in the upper right panel of Figure 28, plot against the wind speed shown in the lower panel of the same figure, and compare with SSD computed using simplified methods.

The first simplified method is from Clarke and van Gorder (2018), henceforth denoted as CG18, near the bottom of their page 4688:

$$U(z=0) = 0.01U_{10}$$

The second simplified method is from CG18 (their equation 23):

$$U(z=0) = 4.4u_* \ln(0.0074U_{10}/u_*)$$

This equation only predicts SSD. We use the Large et al. (1994) formula to calculate $u_*$, for consistency with CG18.

The third simplified method is from equation 10 in BJB, which we discussed in Section 6.1.1:

$$U(z) = 2\int_{\omega_p}^{\omega_{max}} \alpha_P g\omega^{-2} \mathrm{e}^{2\omega^2 z/g} d\omega,$$

Except that we use $f_{max} = \omega_{max}/(2\pi) = 0.729$ Hz, where BJB used $\omega_{max} = \infty$. Thus,

$$U(z=0) = 2\int_{\omega_p}^{\omega_{max}} \alpha_P g\omega^{-2} d\omega,$$

Here peak frequency $\omega_p$ is determined from the JONSWAP formula as we described in Section 6.1.1, using two illustrative values for fetch: 10 km and 100 km. In cases where the JONSWAP-peak frequency is lower than that of the fully developed PM spectrum, the estimate is considered invalid.

The fourth simplified method is to use the PM spectrum given in Section 4.7, in combination with the simplified SSD formula given in Section 5.4.

We computed correlation between the SSD predicted using three of the simplified methods (#1, #2, and #4)[5] and the WW3 predictions. Rounded to two significant digits, all three are the same: 0.94 Pearson correlation coefficient and 0.92 Spearman correlation coefficient. This correlation is quite good, implying that these three simple methods have skill for computing Stokes drift *at the surface*.

However, two things are obvious from this plot. First, there are large differences between different estimates of SSD. Second, by design, the simple models each give one SSD value for each wind speed value, while WW3 gives a histogram of many SSD values. WW3 dynamically accounts many physical processes that affect Stokes drift. At each wind speed, the variability in the histogram of WW3's SSD shown here is a result of physics implemented in the model. The simplified methods cannot account for these effects.

Examples of the physical processes represented by WW3 which affect Stokes drift are:

---

[5] Correlation was not calculated for the third simplified method, because the JONSWAP-based estimates of peak frequency are not valid for the full range of wind speeds.

1) Time- and space- varying wind speed, and direction. Thus, the “history” of the wind field and the fetch are dynamically integrated.
2) Contributions from one or more swell systems, each with different frequency and directional distribution. To give a canonical example: if two swell systems are identical and travelling in the opposite direction, the net contribution to Stokes drift is zero. WW3 can represent this.
3) Horizontally sheared currents. These produce effects analogous to shoaling and refraction by bathymetry and are responsible for much of the spatial variability of SSD shown in the upper panels of Figure 28.

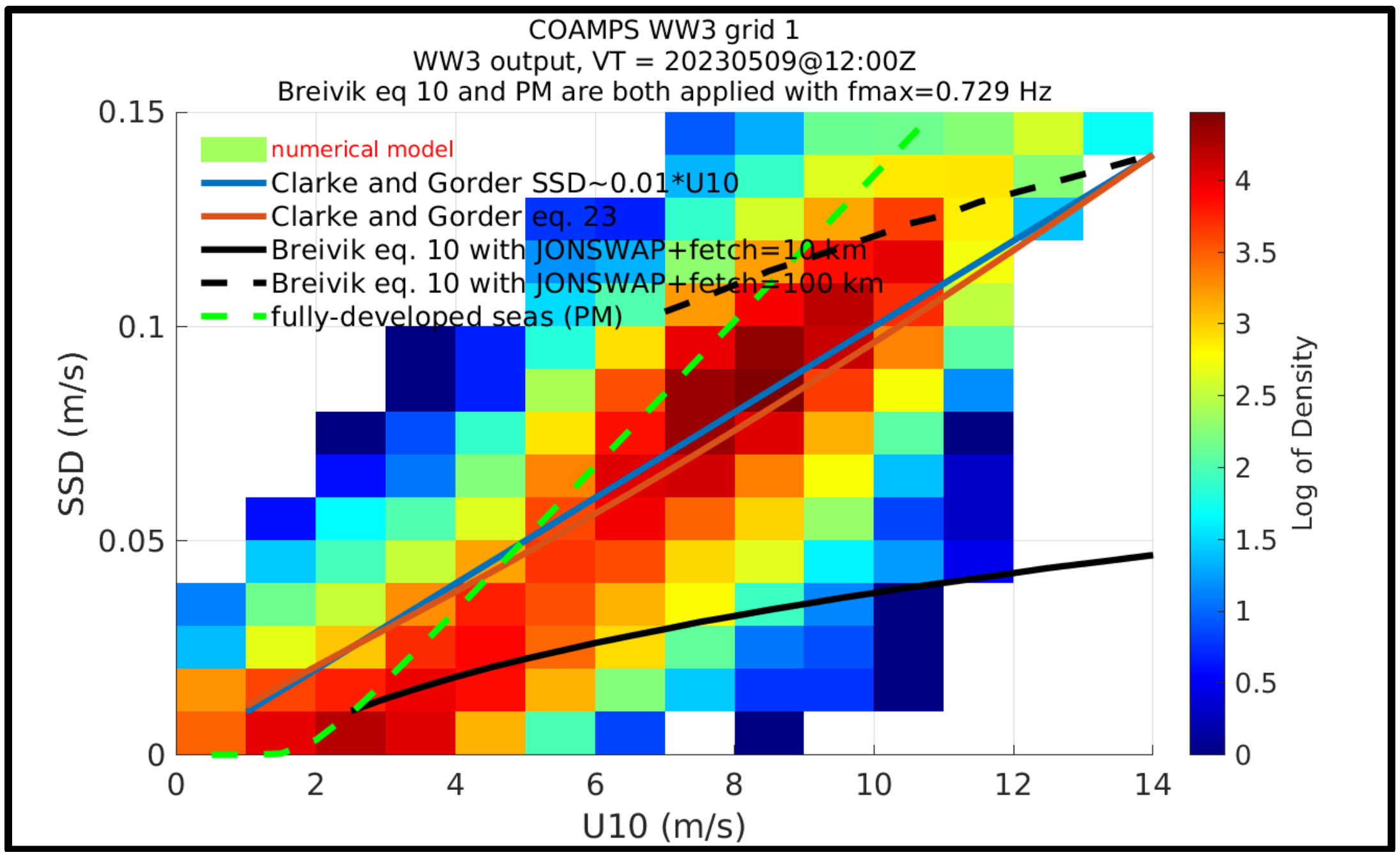

Figure 29. SSD as a function of wind speed. The color scale is the SSD predicted by WW3 for the region and time shown in Figure 28. The lines are the simple models for SSD described in the text.

## 8. Evaluations of surface Stokes drift (SSD) using Sofar drifting buoys

In this section, we use data from drifting Spotter buoys deployed by Sofar Oceans, provided by Dr. P. Smit of Sofar. These buoys measure wave spectra over much of the world’s oceans. The period is September 2021 to May 2022. This dataset was previously used in Rogers and Thomson (2026) and Rogers et al. (2026). For a graphic of the drifter positions, the reader is referred to Figure 2 of Rogers and Thomson (2026). The data set is very large: over 3 million wave spectra are used here.

The data is used in two ways here. First, we compute the SSD from the buoy directional spectra. This calculation accounts for the effects of directional spread by using the buoy moments, as described in Section 21. Second, we use the time sequence of drifter positions to compute the total drift of the buoys, previously described in Rogers et al. (2026).

The Sofar data provided to NRL fall into one of two types. The first "buoy type" represents 82% of the data and provides spectra at frequencies up to 0.65 Hz. The frequency grid is very irregular: with the last three bins being [0.469 0.498 0.654] Hz. The second "buoy type" represents the remaining 18% of the data and provides spectra at frequencies up to 1.25 Hz at a regular spacing, $\Delta f = 0.0098$ Hz, with the last three bins (rounded to three decimals) being [1.226 1.235 1.245] Hz. In the analysis that follows, we present results from the two buoy types separately.

The results are presented as two-dimensional histograms (i.e., point-density scatter plots) and we provide a few error metrics in the diagrams: "CC" is the Pearson correlation coefficient, "bias" is the mean error, and "rmse" is the RMS error. The number of co-locations ("n") is also indicated for each histogram.

### 8.1. Comparison to simple estimate of SSD

Clarke and van Gorder (2018, "CG18" below), propose three empirical formula for calculation of SSD. The simplest method assumes that the SSD is 1% of the 10-m wind speed. Here, we compare that simple method against SSD computed from the buoy spectra. As with the CDIP buoys, the Sofar buoys do not measure the wind speed, so we co-locate the ERA5 $U_{10}$ values with the buoys.

Comparisons are shown in Figure 30 (SSD magnitude), Figure 31 (eastward, i.e., $x$-component, SSD), and Figure 32 (northward, i.e., $y$-component, SSD).

Generally, the simple empirical model for SSD is quite accurate for this global scenario (CC=0.88 and higher), especially the $x$-component of SSD (CC=0.93 and 0.94). The high correlation is consistent with the correlation that we reported in Section 7 (Pearson CC=0.94) for the regional scenario.

Though we do not present them here, we also applied two other empirical models proposed by CG18. One is $U(z=0) = 7.9u_{*,w}$, with the water-side friction velocity computed as $u_{*,w} = u_*\sqrt{\rho_a/\rho_w} \approx 0.035u_*$, and the air-side friction velocity $u_*$ computed using the Hwang formula. The third was already described in Section 7; it is from equation 23 in CG18, $U(z=0) = 4.4u_*\ln(0.0074U_{10}/u_*)$. We found that these two models were not significantly more accurate than the simplest model $U(z=0) = 0.01U_{10}$ shown in the figures below. We also tested a recalibration of the simplest model to the dataset, in the form of $U(z=0) = mu_* + b$, but again, the accuracy was not significantly improved relative to the simple, uncalibrated formula shown.

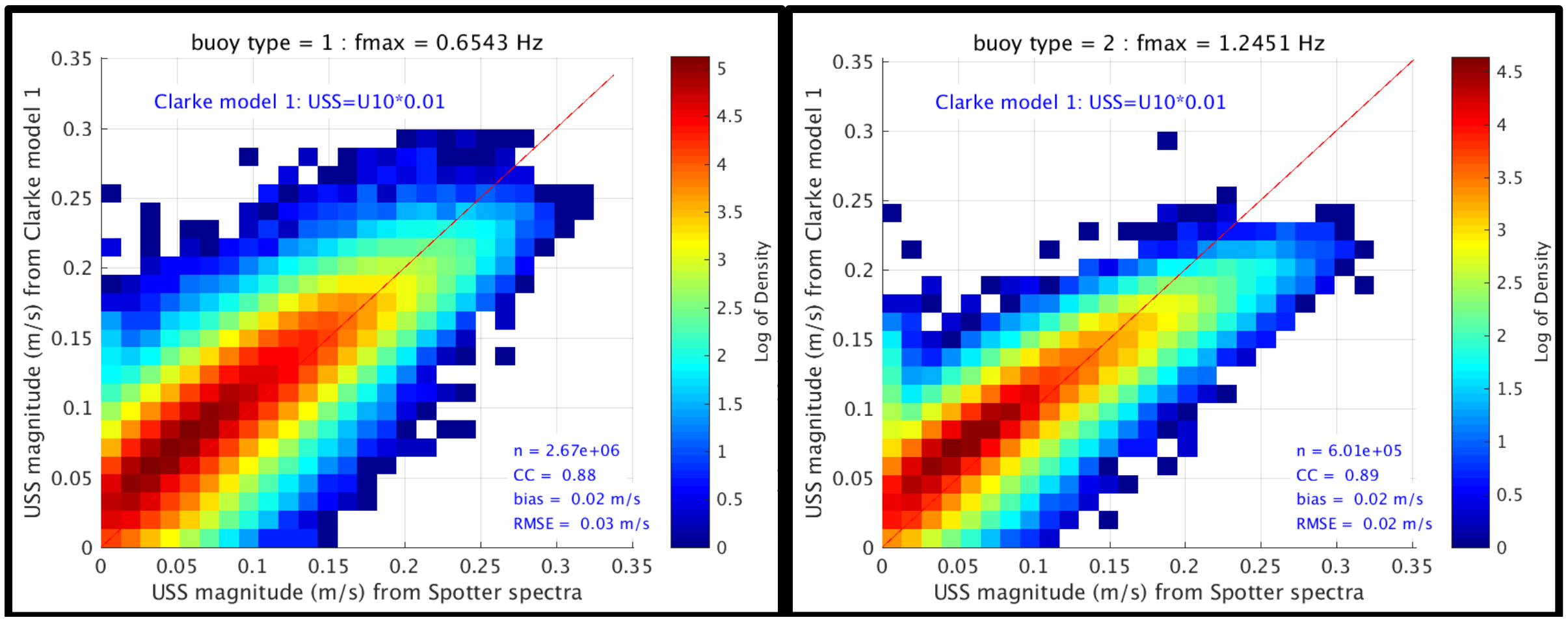

Figure 30. Comparison, via two-dimensional histograms, of the magnitude of SSD computed using Spotter spectra (horizontal axis) against SSD computed using the simplest empirical formula from CG18. Left panel: Sofar buoy type 1. Right panel: Sofar buoy type 2.

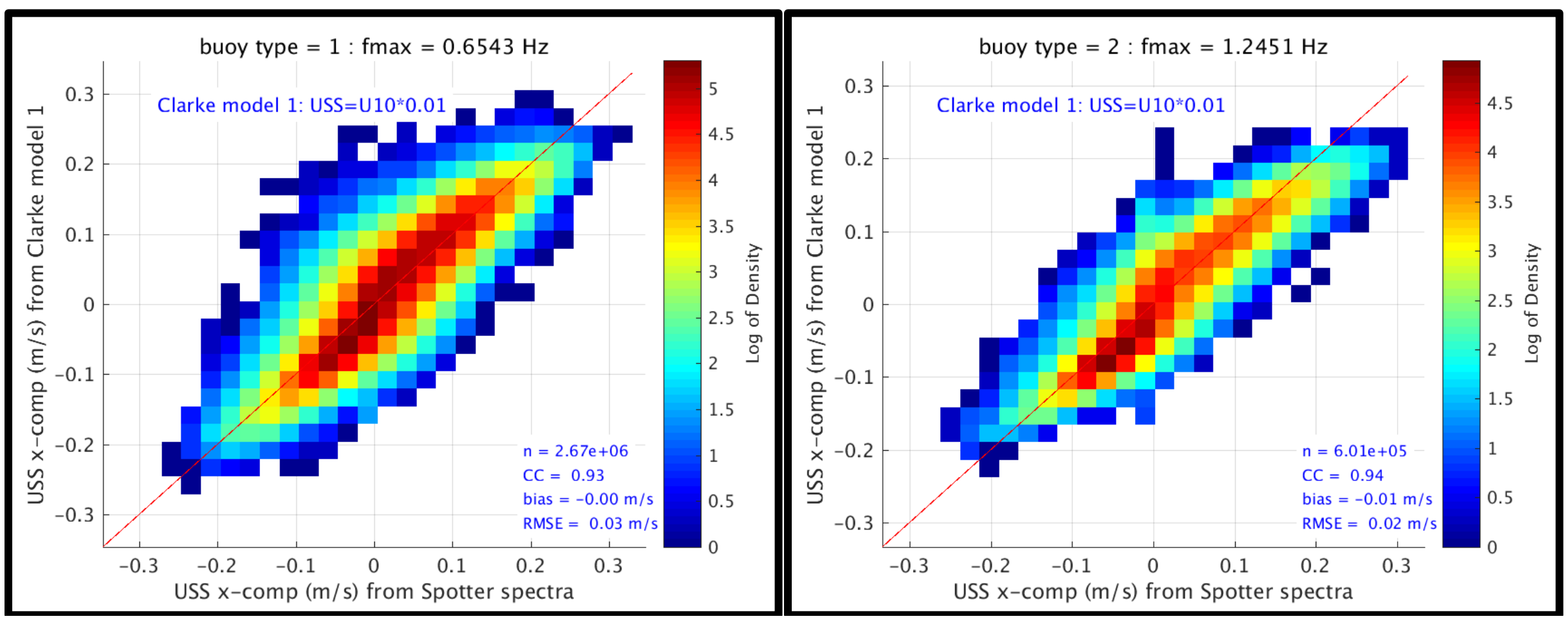

Figure 31. Like Figure 30, but showing the eastward component of SSD.

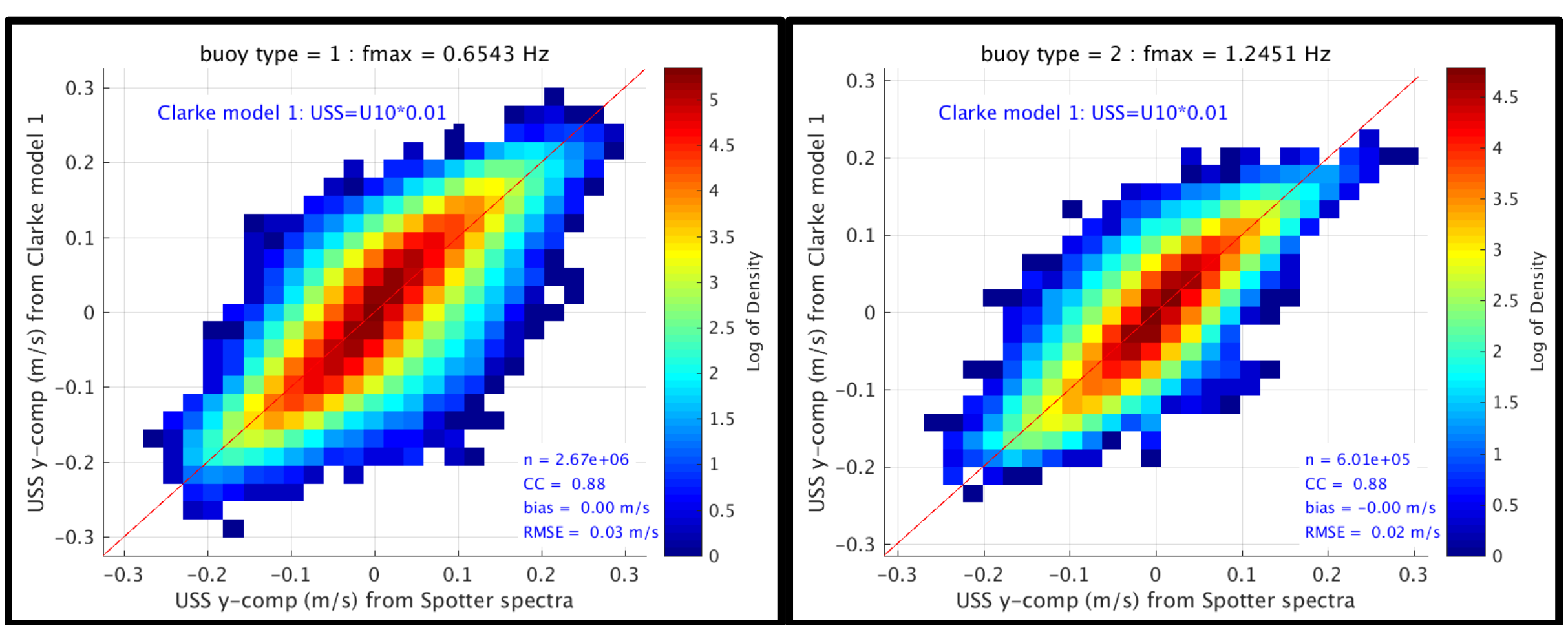

Figure 32. Like Figure 30, but showing the northward component of SSD.

## 8.2. Comparison to observed total drift

Here, we use total drift computed from the time sequence of buoy positions, previously described in Rogers et al. (2026). This is compared against the SSD computed from the buoy spectra.

To scale the former to the latter, we invoke the "3% of wind rule", which is that the Lagrangian mean current is $U_L = 0.03U_{10}$; see Samelson (2022). If the "wind-slip" or "windage" (i.e., the drift caused by the direct transfer of momentum from the wind to the Spotter buoys) is 1% of the wind speed (Houghton et al. (2021), Davis et al. (2025)), then the total drift can be approximated as $U_T \approx 0.04U_{10}$. With surface Stokes drift $U(z = 0) \approx 0.01U_{10}$, we can anticipate that $U(z = 0) \approx 0.25U_T$.

The comparisons are shown in Figure 33 (SSD magnitude), Figure 34 (eastward, i.e., $x$-component, SSD), and Figure 35 (northward, i.e., $y$-component, SSD).

Note that the $0.25$ factor in $U(z = 0) \approx 0.25U_T$ does not affect correlation. We would recover the same correlation values by plotting $U(z = 0)$ against $U_T$. In this context (linear correlation), the invocation of the "3% of wind rule" and the wind slip approximation are unnecessary: effectively, we are merely computing the linear correlation between the SSD and the position-based drift speed.

In general, there is much more scatter than was seen in the prior section, with CC of only 0.32 for the prediction of the magnitude of total drift. But again, the best accuracy is seen with the eastward component (CC=0.57 and 0.67). We speculate that much of the scatter is associated with contributions to buoy drift $U_T$ from Eulerian currents that are not connected to the local winds (boundary currents, eddies, tidal currents, etc.).

While the relatively low bias provides some confidence in the "rule of thumb" approximation $U(z = 0) \approx 0.25U_L$, the large scatter implies low skill.

The comparison provides insight about the contribution of the surface Stokes drift (SSD) to the total drift (Lagrangian current plus windage) of a Spotter buoy. For further reading, see Rascle et al. (2008), who study the relative magnitude of SSD and Eulerian current.

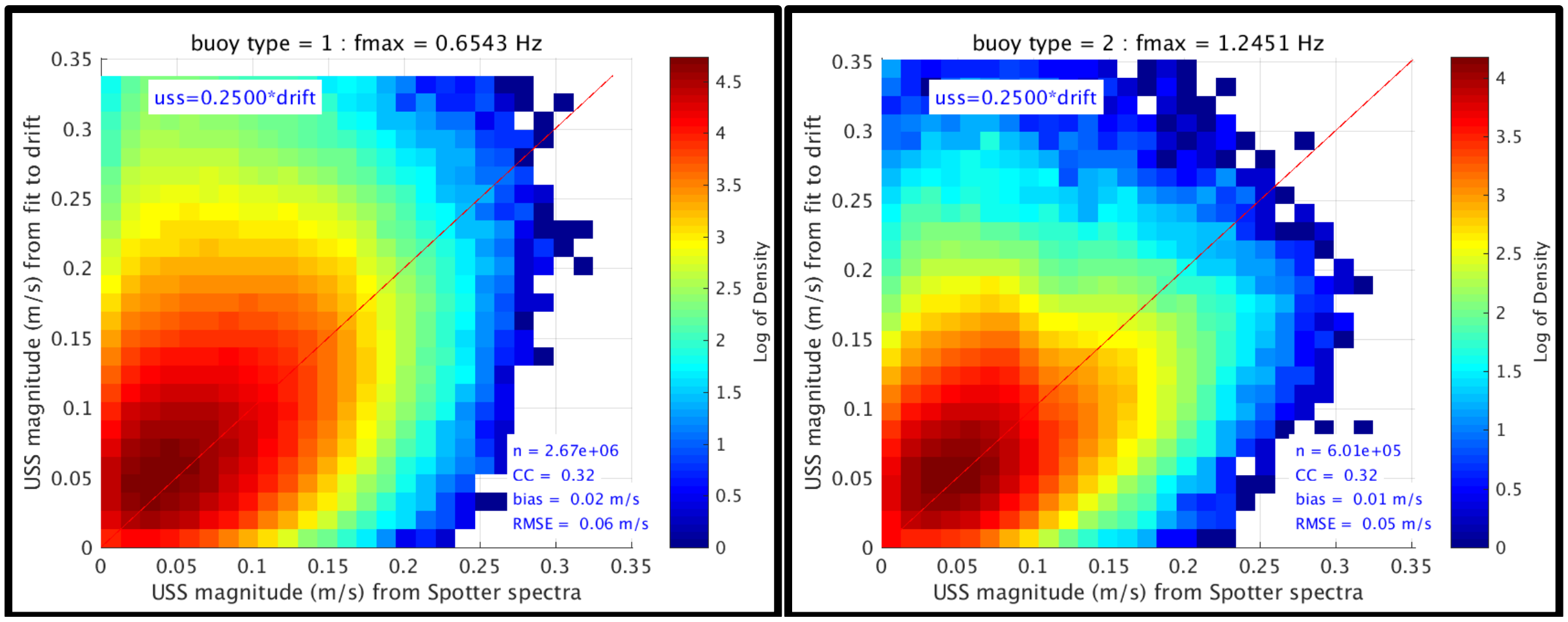

Figure 33. Comparison, via two-dimensional histograms, of the magnitude of SSD computed using Spotter spectra (horizontal axis) against the total drift computed using the buoy positions, after scaling the latter by 25%. Left panel: Sofar buoy type 1. Right panel: Sofar buoy type 2.

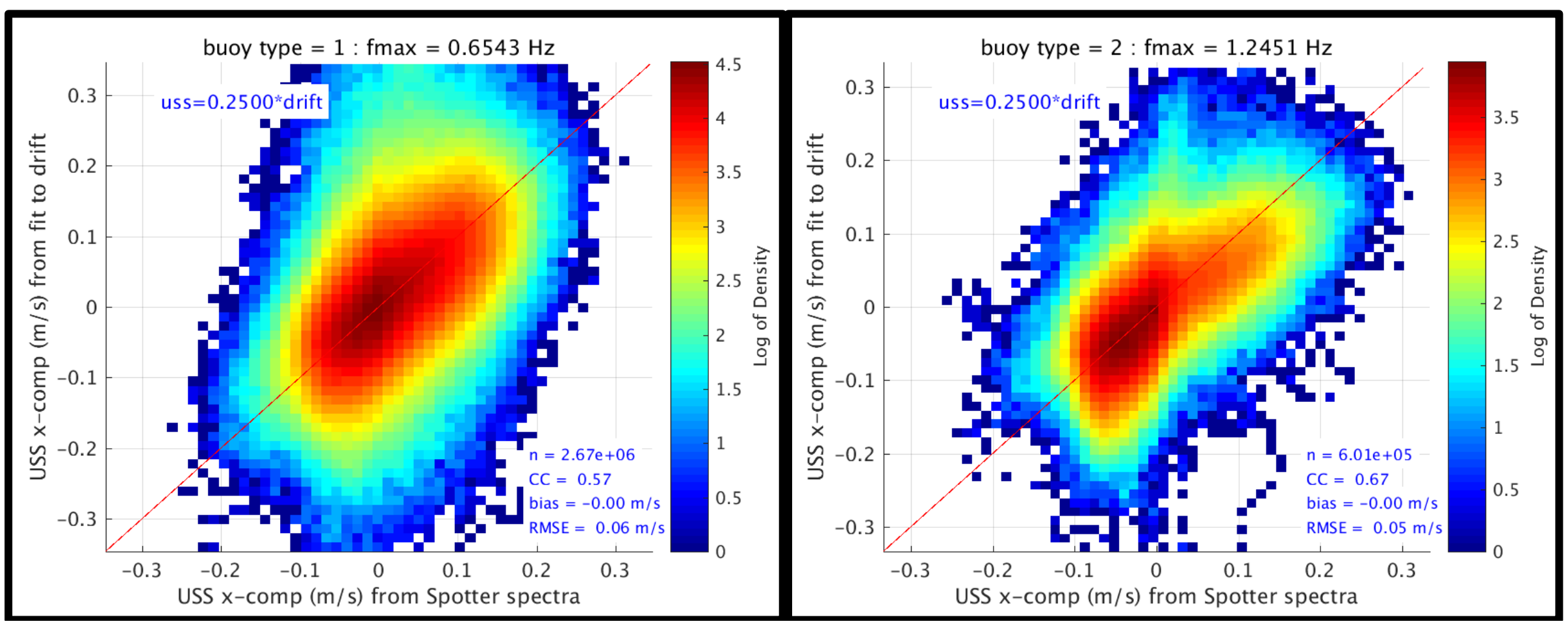

Figure 34. Like Figure 33, but showing the eastward component of SSD.

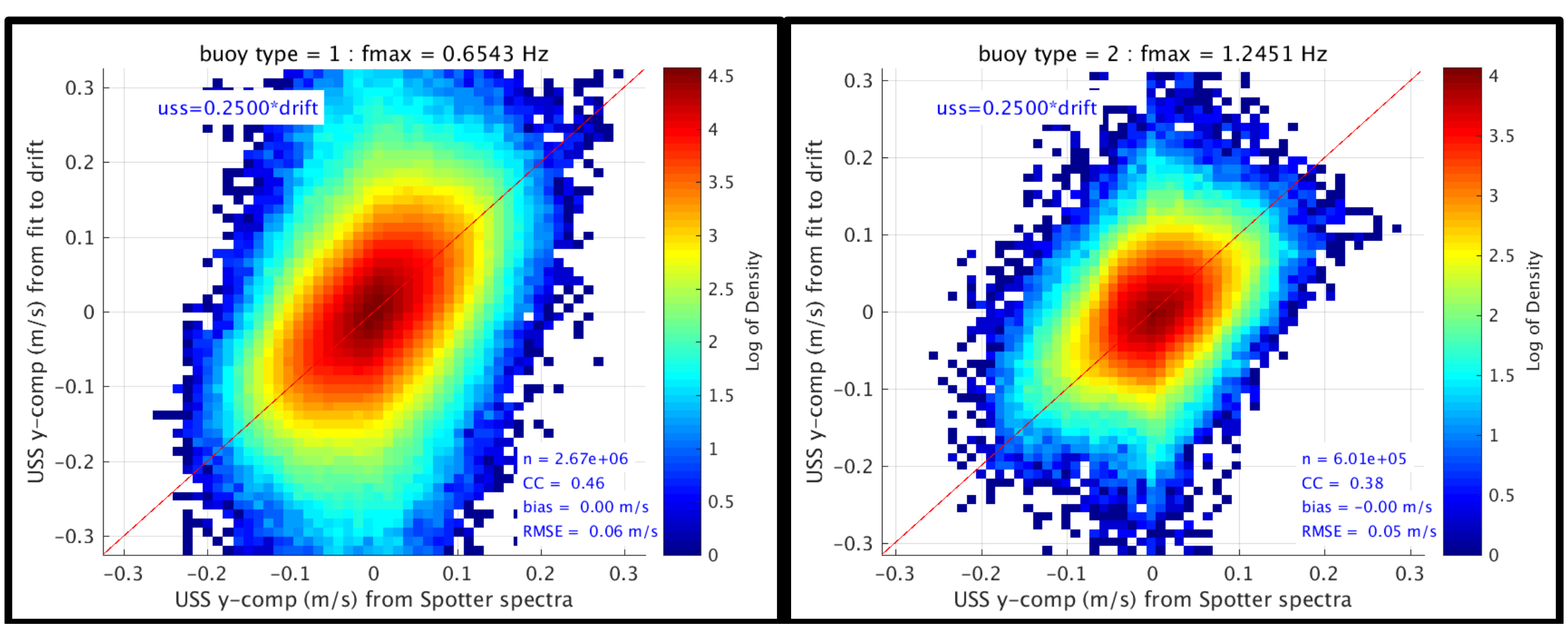

Figure 35. Like Figure 33, but showing the northward component of SSD.

## 9. Findings and Recommendations

Here, we summarize our findings and recommendations, split into two categories.

Findings and recommendations based primarily on prior knowledge

- Estimation of Stokes drift using a representative wavenumber should be avoided, since it will underpredict the vertical shear of Stokes drift near the surface. (Section 2)
- The impact of directional spread on Stokes drift should be accounted for (Section 2). This is a straightforward process when using Stokes spectra (Section 2) from a wave model, or 1.5-dimensional (1.5-D) spectra from a buoy (Section 5.3) or wave model (Section 5.5).
- There is no practical reason to use 2D spectra from a wave model to compute Stokes drift. This can be done with much lower storage and computational resources using Stokes spectra or 1.5-D spectra from a wave model. (Section 2)
- If there is a requirement to estimate Stokes drift without using an ocean wave model, then it should be forced only by wind speed or friction velocity. We know of no practical method for providing a 'wave age' or fetch parameter to a parametric model that requires this information, without using a wave model, so parametric models that require such parameters should not be used (Section 6.3).
- If there is a requirement to estimate Stokes drift using an ocean wave model, but it is not possible to pass Stokes spectra or 1.5-D spectra (e.g., due to storage or compute limitations), then the approximate method of Breivik et al. (2014), equation (17), which uses only three numbers (surface Stokes drift (SSD), Stokes transport, and mean wave direction) per time/location can be considered[6]. (Section 2)
- Our evaluations of parametric models use observations of wave spectra from fixed frames of reference (Section 3). In the context of the high-frequency tail, our conclusions are valid only for that frame of reference. For applications requiring an intrinsic frame of reference (i.e., applications dependent on wavenumber spectra), Doppler-shift must be considered (Rogers et al. 2026).

Findings and recommendations based primarily on new analysis presented in this report

- If there is a requirement to estimate Stokes drift without using an ocean wave model,
    a. one can expect a fairly accurate SSD, and a parametric wave model is not needed (Section 8.1),
    b. one can expect an inaccurate Stokes drift profile (Section 5.6), and
    c. the Forristall model (Section 4.12) is a "least bad" choice (Section 4.15) for estimating a Stokes profile using a parametric wave model.
- We recommend against estimating Stokes drift without using an ocean wave model. Parametric models such as the Forristall model, even if heavily calibrated to minimize absolute error, still have many shortcomings from missing physics. As a general rule, they are only accurate for representing very simple wave states, or when benefitting from fortuitous cancellation of errors (Section 5.6).

[6] Also, see later works developing these methods: Breivik et al. (2016), Li et al. (2017), Wu et al. (2019), and Brevik and Christensen (2020).

- If an ocean wave model can be used to provide Stokes drift, we suggest that the wave model provides Stokes spectra or 1.5-D spectra to the ocean model (Sections 2 and 5.6).
- Except in (unusual) cases where the ocean model is computing the top one or two meters of the water column with fine vertical resolution, the truncation of the ocean wave model spectra $E(f)$, e.g. at $f = 0.73$ Hz, is not consequential for wave-to-ocean coupling. Having said that, analytical calculation of an approximate contribution from these frequencies is not difficult. Moreover, the truncation does have a strong impact on SSD, which could be important in other applications (Section 5.7).
- Prior global ocean model experiments with wave-enhanced mixing using only information about the wind (i.e., wind-based-Stokes drift) were inspected. Anomalies were observed in SST and mixed layer temperature (MLT), in coupled and uncoupled model results, respectively. The cause of the anomalies has not been determined, but assumptions and approximations associated with the wind-based estimation methods used provides a credible explanation (Section 6.2).
- Global ocean model experiments with wave-enhanced mixing using Stokes drift from the ocean wave model WW3, run in uncoupled mode were conducted (Fan et al. 2026). A spot-check of MLT from that experiment (Section 6.2.4) does not indicate anomalies such as those found with the experiments with wave-enhanced mixing using only information about the wind, mentioned above.
- Simple parametric models for *surface* Stokes drift (SSD) have adequate or even good skill for predicting the SSD from a large (over 3 million wave spectra) global dataset from a drifting buoy network (CC=0.88 to 0.94) (Section 8.1) and for matching SSD estimates from WW3 for a regional case (CC=0.94) (Section 7). However, as stated above, we do not recommend using any simple model for the *sub-surface* Stokes drift, since they suffer greatly from missing physics (Sections 5.6 and 7).
- The SSD computed from the 3+ million wave spectra measured by drifting buoys were compared with the total drift calculated using positional change of the same buoys. The SSD is roughly 25% of the total drift, consistent with expectations based on "rules of thumb". However the scatter is large (e.g., CC=0.32 for SSD magnitude), indicating that other contributions to total drift (windage and Eulerian currents) obey these rules only in the gross, mean sense (Section 8.2).

**Acknowledgments**

This work was funded by the Office of Naval Research through the NRL Core Program, Program Element 0602435N, work units 62A1B6 and 62A1P9. It is approved for public release. We are grateful for the observational data used in this study, and acknowledge the significant effort and resources required to make this available. We thank James Behrens and others at CDIP for the UCSD/SIO/CDIP datasets, which are provided freely to the research community. We thank Pieter Smit and others at Sofar Oceans for sharing the Sofar Spotter dataset. We thank ECMWF and the European Commission for sharing their ERA5 wind and ice products through C3S of the Copernicus Earth Observation Programme.